\documentclass[aps,prb,twocolumn,showpacs,amsmath,amssymb,superscriptaddress,floatfix]{revtex4-2}

\usepackage[english]{babel}
\usepackage{blindtext}
\usepackage{latexsym}
\usepackage{amssymb}
\usepackage{physics}
\usepackage{braket}
\usepackage{amsmath}
\usepackage{bm}
\usepackage{amsfonts}
\usepackage{relsize}
\usepackage{xcolor}
\usepackage{verbatim}
\usepackage{bbold}
\usepackage{slashed}
\usepackage{appendix}
\usepackage{graphicx}
\usepackage{color}
\usepackage[normalem]{ulem}
\usepackage[colorlinks = true,
            linkcolor = blue,
            urlcolor  = blue,
            citecolor = blue,
            anchorcolor = blue]{hyperref}
\usepackage{graphicx} 
\usepackage{dcolumn} 
\usepackage{here}
\usepackage{array}
\usepackage{tabularray}
\newcommand{\approptoinn}[2]{\mathrel{\vcenter{
  \offinterlineskip\halign{\hfil$##$\cr
    #1\propto\cr\noalign{\kern2pt}#1\sim\cr\noalign{\kern-2pt}}}}}

\newcommand{\up}{\uparrow}
\newcommand{\down}{\downarrow}

\definecolor{ao(english)}{rgb}{0.0, 0.5, 0.0}
\definecolor{amaranth}{rgb}{0.9, 0.17, 0.31}
\definecolor{green(html/cssgreen)}{rgb}{0.0, 0.5, 0.0}

\newcommand\greensout{\bgroup\markoverwith{\textcolor{green(html/cssgreen)}{\rule[0.5ex]{2pt}{1.0pt}}}\ULon}

\begin{document}

\title{
Type II Lifshitz invariant and optically active Higgs mode in time-reversal symmetry broken superconductors
}

\author{Raigo Nagashima}
\thanks{These authors contributed equally to this work.}
\affiliation{Institute for Theory of Condensed Matter, Karlsruhe Institute of Technology, Karlsruhe 76131, Germany}
\affiliation{Department of Physics, University of Tokyo, Bunkyo-ku, Tokyo 113-8656, Japan}

\author{Chihiro Mamiya}
\thanks{These authors contributed equally to this work.}
\affiliation{Department of Physics, University of Tokyo, Bunkyo-ku, Tokyo 113-8656, Japan}

\author{Naoto Tsuji}
\affiliation{Department of Physics, University of Tokyo, Bunkyo-ku, Tokyo 113-8656, Japan}
\affiliation{RIKEN Center for Emergent Matter Science (CEMS), Wako, Saitama 351-0198, Japan}
\affiliation{Trans-Scale Quantum Science Institute, University of Tokyo, Bunkyo-ku, Tokyo 113-8656, Japan}

\date{\today}

\begin{abstract}
Lifshitz invariant is a symmetry-allowed term in the Ginzburg-Landau free energy of an ordered phase, involving the order parameters and a single spatial derivative, which serves as a source of unusual optical responses.
Here we introduce a ``type II" Lifshitz invariant for superconductors, which changes its sign under the particle-hole transformation and can be distinguished from the ordinary particle-hole even ``type I" Lifshitz invariant.
We show that the type II Lifshitz invariant appears only in superconductors that break time-reversal symmetry and allows the Higgs mode to be visible in the optical conductivity spectrum.
We provide a classification of all pairs of irreducible corepresentations of order parameters in the magnetic point groups that admit a type II Lifshitz invariant.
We also numerically calculate the optical conductivity for various models of time-reversal symmetry broken multiband superconductors, finding agreement with the group-theoretical analysis.
Our results establish a universal class of time-reversal symmetry broken superconductors hosting an optically active Higgs mode.
\end{abstract}

\maketitle

\section{Introduction}
Dynamical properties of superconductors provide valuable insights into the underlying microscopic characteristics of superconducting states.
In particular, the spectrum of collective modes contains numerous pieces of information on pairing symmetries, fluctuations, and spatial configurations of superconducting orders.
Conventional superconductors~\cite{BCS_1957} typically show a collective amplitude mode, or the Higgs mode, which corresponds to the massive excitation of an amplitude fluctuation of the order parameter~\cite{Anderson1958, Schmid1968, Littlewood_Varma1981, Littlewood_Varma1982, Pekker_Varma2015, Higgs_mode_review2020, Tsuji-Ency2024}.
On the other hand, the phase mode, or the Nambu-Goldstone (NG) mode~\cite{Goldstone1961, Nambu_JonaLasinio1961}, couples to electromagnetic fields in superconductors, and is lifted up to high energy, a phenomenon known as the Anderson-Higgs mechanism~\cite{Anderson1963, Englert1964, Higgs1964, Guralnik1964}.
Hence the Higgs mode is one of the low-energy surviving modes in superconductors, whose energy gap corresponds to the superconducting gap $2\Delta$, 
i.e., the lower bound of the quasiparticle excitation continuum.

The difficulty of observing the Higgs mode stems from the fact that it does not linearly couple to gauge fields in ordinary situations. 
Thus, previous studies have mainly focused on nonlinear responses of superconductors~\cite{Tsuji2015, Kemper2015, Cea2016, Tsuji2016, Jujo2018, Silaev2019, Schwarz2020, Tsuji_and_Nomura2020, Haenel2021, Udina2022, Derendorf2024}, such as Raman spectroscopy for superconductors with charge density wave order~\cite{Sooryakumar1980, Sooryakumar1981, Measson2014, Grasset2018, Majumdar2020}, THz pump-probe spectroscopy, and third harmonic generations~\cite{Matsunaga2013, Matsunaga2014, Matsunaga2017, Katsumi2018, Chu2020}.
A few exceptions that allow for the linear response of the Higgs mode include the case of supercurrent injection~\cite{Moor2017, Nakamura2019}, coupling to an LC circuit~\cite{Lu2023}, and spatially modulated superconducting states~\cite{Nagashima2025}.

Superconductors that host multiple order parameters exhibit an even wider variety of collective phenomena. Such systems arise in multiband superconductors, including iron-based superconductors~\cite{Iron_SC_review}, niobium-based superconductors~\cite{Niobium_SC_review}, $\text{MgB}_{2}$~\cite{MgB2_review}, and Kagome superconductors~\cite{Kagome_SC_review}.
Since each order parameter carries two degrees of freedom, i.e., its amplitude and phase, a relative phase mode between two different order parameters, known as the Leggett mode~\cite{Leggett1966}, emerges in multiband superconductors, in addition to Higgs modes. The Leggett mode has been previously studied in the nonlinear response regime in multiband superconductors~\cite{Cea2016, Balatsuky2000, Burnell2010, Ota2011, Lin2012, Marciani2013, Bittner2015, Krull2016, Murotani2017, Murotani2019, Giorgianni2019, Fiore2022, Seibold2021}.
Experimental studies of Raman spectroscopy have observed the Leggett mode in a multiband superconductor, $\text{MgB}_{2}$~\cite{Blumberg2007, Giorgianni2019}.

Recently, the Leggett mode has been proposed to arise in the linear-response regime~\cite{Kamatani2022} in a multiband superconducting system with a Lifshitz invariant~\cite{Nagashima2024}.
In the Ginzburg-Landau (GL) theory, the Lifshitz invariant appears as a linear derivative term, $\sum_{\alpha\beta\lambda}d_{\alpha\beta}^{\lambda}\psi^{*}_{\alpha}D_{\lambda}\psi_{\beta}$, where $d_{\alpha\beta}^{\lambda}$ is a constant, $\psi_{\alpha}$ is a superconducting order parameter of $\alpha$-th band, and $D_{\lambda}$ is the covariant derivative in $\lambda$-direction~\cite{LandauLifshitz}.
Although the Lifshitz invariant has often been considered in the context of, for example,  noncentrosymmetric superconductors~\cite{Mineev1994, Mineev2008, Samokhin2013, Fuchs2022, Kochan2023}, parity- and time-reversal (TR) symmetry broken superconductors~\cite{Kanasugi2022, Kitamura2023}, commensurate-incommensurate phase transition~\cite{Kopsky1977, Ishibashi1978}, and liquid crystals~\cite{Sparavigna2009}, the Lifshitz invariant can exist even in systems preserving inversion symmetry~\cite{Nagashima2024}.
There are also theoretical studies to observe collective modes, including the Leggett modes, in the linear response regime~\cite{Levitan2024, Levitan2025}.
A higher-order generalization of the Lifshitz invariant is proposed to be relevant to the nonlinear Hall effect with a geometric phase~\cite{Takasan2025} and the Raman response~\cite{Yamazaki2026} in superconducting systems.

In previous studies of the Lifshitz invariant in multiband superconductors, the TR symmetry is assumed to be preserved, thereby allowing the Leggett mode to be excited in the linear-response regime.
In this paper, we consider TR symmetry-broken superconductors \cite{Ghosh2021}, which attract much attention in the context of, e.g., unconventional superconductors \cite{Luke1998, Xia2006, Schemm2015, Grinenko2020} and Kagome superconductors \cite{Mielke2022, Deng2024, Elmers2025}.
In superconductors, the particle-hole (PH) symmetry emerges as an approximate symmetry at low energies.
Depending on how the coefficient of the Lifshitz invariant $d_{\alpha\beta}^{\lambda}$ behaves under the PH transformation, 
we can classify the Lifshitz invariant into two types: One is the PH even term that does not change the sign under the PH transformation (type I Lifshitz invariant), and the other is the PH odd term (type II Lifshitz invariant). The latter requires the breaking of the TR symmetry. We find that when the type II Lifshitz invariant is present in TR-symmetry broken superconductors, the Higgs mode couples to electromagnetic fields linearly, and thus can become optically active in the linear response regime. 
We summarize the classification in Table \ref{tab: type I, II}.

\begin{table}[t]
    \centering
    \caption{Classification of type I and type II Lifshitz invariants ($\sum_{\alpha\beta\lambda} d_{\alpha\beta}^\lambda \psi_\alpha^\ast D_\lambda \psi_\beta$) in superconductors. The third column indicates whether the Lifshitz invariant changes the sign under the particle-hole transformation. The fourth column shows optically active collective modes in superconductors associated with each type of Lifshitz invariant.}
    \label{tab: type I, II}
    \begin{tabular}{cccc}
        \hline\hline
         & coefficient & particle-hole & $\;$ optically active mode $\;$  \\
        \hline
        type I & $d_{\alpha\beta}^\lambda=-d_{\beta\alpha}^\lambda\in \mathrm{i}\mathbb R$ & even & Leggett mode \\
        type II & $d_{\alpha\beta}^\lambda=d_{\beta\alpha}^\lambda\in \mathbb R$ & odd & Higgs mode \\
        \hline\hline
    \end{tabular}
\end{table}

To determine under what conditions the type II Lifshitz invariant is allowed to emerge in systems without TR symmetry, we perform group-theoretical classification based on the magnetic point group,
which is an extension of the previous classification of the Lifshitz invariant based on the crystallographic point groups \cite{Nagashima2024}.
We list all possible combinations of irreducible corepresentations of order parameters in the magnetic point groups that support the existence of the Lifshitz invariant.
Due to the anti-unitary TR operation being absent in the crystallographic point groups, the classification becomes more enriched than that of the point groups.

We numerically demonstrate that the Higgs mode indeed appears as a peak in the optical conductivity spectrum in several lattice models of TR-symmetry-broken multiband superconductors.
The examples include a Kagome lattice model that breaks TR symmetry via a quantum geometric phase (Aharonov-Bohm flux)~\cite{Bohm2003, Xiao2010, Cohen2019}, which is partly motivated by a loop current order~\cite{Hsu1991, Varma1997, Chakravarty2001, Sun2008} that has been proposed to be relevant for Kagome materials~\cite{Park2021, Lin2021, Denner2021, Tazai2023}.

This paper is organized as follows.
In Sec.~\ref{Sec.II}, we consider the Lifshitz invariant in the GL theory of multiband superconductors.
We also see that the Higgs mode can appear in the linear-response regime when the coefficient of the Lifshitz invariant $d_{\alpha\beta}^{\lambda}$ changes its sign under the PH transformation, i.e., the Lifshitz invariant belongs to type II.
We perform the group-theoretical classification of the Lifshitz invariant by the magnetic point group, and show applications to several lattice models without TR symmetry in Sec.~\ref{Sec.III}.
In Sec.~\ref{Sec.IV}, we numerically calculate the optical conductivity for those models studied in the previous section using an imaginary-time path-integral approach in the clean limit.
Section~\ref{Sec.V} summarized the paper.
We set $\hbar=1$ throughout the paper.

\section{Ginzburg-Landau theory}
\label{Sec.II}

Let us first consider a phenomenological GL theory for general $N$-band superconductors with a Lifshitz invariant~\cite{Kamatani2022, Nagashima2024}. For the single-band case, we refer to the review article~\cite{Higgs_mode_review2020}.

In the GL theory, one can write down the free energy density of superconductors in terms of a complex order parameter $\psi_\alpha$ ($\alpha=1,2,\dots, N$), corresponding to the $\alpha$-th band.
For simplicity, we assume that the equilibrium state has the $s$-wave spin-singlet pairing and is uniform in space.
We are interested in the effect arising from the Lifshitz invariant; the single spatial-derivative terms in the form of~\cite{Kamatani2022, Nagashima2024, LandauLifshitz}
\begin{equation}
    \sum_{\alpha,\beta=1}^{N}\sum_{\lambda}d_{\alpha\beta}^{\lambda}\psi^{*}_{\alpha}D_{\lambda}\psi_{\beta},
    \label{GL_free_energy}
\end{equation}
where $D_{\lambda}=-\mathrm{i}\nabla_{\lambda} - e^{*}A_{\lambda}$ ($\lambda=x,y,z$) is the covariant derivative 
with an effective charge $e^{*}=2e$, 
and $A_{\lambda}$ is an electromagnetic vector potential.
This term works similarly to the Dzyaloshinskii-Moriya interaction in magnets~\cite{Dzyaloshinsky1958, Moriya1960}.
The Lifshitz invariant term can appear in multiband systems even when the system preserves the inversion symmetry~\cite{Nagashima2024}.

As a basic premise, the free energy must be hermitian, and so is the Lifshitz invariant.
By taking the hermitian conjugate of Eq.~(\ref{GL_free_energy}) and neglecting the total derivative terms, we can see that the coefficient $d_{\alpha\beta}^\lambda$ should satisfy
\begin{align}
    (d_{\alpha\beta}^\lambda)^\ast
    &=
    d_{\beta\alpha}^\lambda.
    \label{d hermiticity}
\end{align}
The Lifshitz invariant (\ref{GL_free_energy}) is invariant under the gauge transformation, $A_\lambda \to A_\lambda+\partial_\lambda\chi$ and $\psi_\alpha\to e^{ie^\ast\chi}\psi_\alpha$ ($\chi$ is an arbitrary real function in space), due to the form of the covariant derivative. 

On top of these, we consider the PH and TR symmetries.
Let us define the PH (charge conjugation) and the TR transformations as
\begin{align}
&\text{PH: } \psi_{\alpha}\to\psi^{*}_{\alpha}, \ A_{\lambda}\to-A_{\lambda}, \ D_{\lambda}\to -D^{*}_{\lambda}, \\
&\text{TR: } \psi_{\alpha}\to\psi_{\alpha}, \ A_{\lambda}\to-A_{\lambda}, \ D_{\lambda}\to-D_{\lambda},
\end{align}
where we adopt the standard convention for complex scalar fields~\cite{WeinbergQFT1}.
Since $\psi_\alpha$ corresponds to $\langle c_{\alpha\downarrow}c_{\alpha\uparrow}\rangle$ ($c_{\alpha\sigma}$ is the annihilation operator of electrons of $\alpha$-th band with spin $\sigma$), the PH operation exchanges $\psi_\alpha$ and $\psi_\alpha^\ast$, while $\psi_\alpha$ remains unchanged under the TR operation (here we consider non-chiral $s$-wave pairings). Note that $A_\lambda$ changes the sign under the PH transformation, since it is equivalent to the change of the sign of $e^\ast$ ($e^\ast$ itself is not a dynamical degree of freedom, and hence does not change under the PH transformation).

As we have mentioned in the introduction, the PH symmetry is an emergent approximate symmetry that appears at low energies in superconductors.
Thus, we suppose that the GL theory respects the PH symmetry. From this, we can claim that the coefficient $d_{\alpha\beta}^\lambda$ of the Lifshitz invariant (\ref{GL_free_energy}) belongs to an irreducible representation of the PH symmetry. Since the PH operation is involution, there are two choices: 
$d^{\lambda}_{\alpha\beta}$ either remains unchanged or changes the sign under the PH transformation:
\begin{align}
    d^{\lambda}_{\alpha\beta} &\xrightarrow{\text{PH}}d^{\lambda}_{\alpha\beta}, \quad d^{\lambda}_{\alpha\beta}=-d^{\lambda}_{\beta\alpha} \quad (\text{Type I}), 
    \label{type I}
    \\
    d^{\lambda}_{\alpha\beta} &\xrightarrow{\text{PH}}-d^{\lambda}_{\alpha\beta}, \quad d^{\lambda}_{\alpha\beta}=d^{\lambda}_{\beta\alpha} \quad (\text{Type II}).
    \label{type II}
\end{align}
In the former case, Eq.~(\ref{GL_free_energy}) becomes invariant under the PH transformation if the coefficient is antisymmetric with respect to $\alpha$ and $\beta$ ($d_{\alpha\beta}^\lambda=-d_{\beta\alpha}^\lambda$). We call this case the type I Lifshitz invariant. In the latter case, Eq.~(\ref{GL_free_energy}) becomes invariant under the PH transformation if the coefficient is symmetric with respect to $\alpha$ and $\beta$ ($d_{\alpha\beta}^\lambda=d_{\beta\alpha}^\lambda$). We call this the type II Lifshitz invariant (see Table \ref{tab: type I, II}).

By combining the relation (\ref{type I}) with the hermiticity condition (\ref{d hermiticity}), we can see that for the Lifshitz invariant of type I, the coefficient $d_{\alpha\beta}^\lambda$ is purely imaginary ($d_{\alpha\beta}^\lambda\in \mathrm{i}\mathbb R$). Since $(d_{\alpha\beta}^\lambda)^\ast=-d_{\alpha\beta}^\lambda$, the type I Lifshitz invariant is compatible with the TR symmetry, and has already been considered for systems preserving the TR symmetry. The group-theoretical classification of the type I Lifshitz invariant for all the crystallographic point groups has been given in Ref.~\cite{Nagashima2024}.

For the Lifshitz invariant of type II, on the other hand, the coefficient becomes real ($d^{\lambda}_{\alpha\beta} \in \mathbb R$) due to the relation (\ref{type II}) and hermiticity (\ref{d hermiticity}).
This type II Lifshitz invariant can appear if the coefficient $d^{\lambda}_{\alpha\beta}$ is an odd function of an external magnetic field.
This type of Lifshitz invariant has been discussed, for example, in the context of magnetoelectric effects in non-centrosymmetric superconductors \cite{BauerSigrist2012}.
The type II Lifshitz invariant also appears if $d^{\lambda}_{\alpha\beta}$ is an odd function of an Aharonov-Bohm flux.
In fact, we can confirm from a microscopic argument that $d_{\alpha\beta}^\lambda$ is an odd function of a flux $\phi$ in lattice models used in later sections (see Appendix~\ref{Appendix.B}).
For the type II Lifshitz invariant to exist, the system must break the TR symmetry, since the Lifshitz invariant transforms under the TR operation as
\begin{equation}
    d^{\lambda}_{\alpha\beta}\psi_{\alpha}^{*}D_{\lambda}\psi_{\beta}\xrightarrow{\text{TR}}d^{\lambda}_{\alpha\beta}\psi_{\alpha}^{*}(-D_{\lambda})\psi_{\beta}.
\end{equation}

Since $d_{\alpha\beta}^\lambda$ is symmetric ($d_{\alpha\beta}^\lambda=d_{\beta\alpha}^\lambda$), the type II Lifshitz invariant can, in principle, be realized with a single order parameter ($N=1$). 
This situation has been discussed in the context of noncentrosymmetric superconductors under an external magnetic field~\cite{Smidman2017}.
Assuming a spatial dependence of the order parameter of the form $\psi_{\alpha} \propto e^{\mathrm{i}\bm{q}\cdot\bm{r}}$, one finds that, in addition to the kinetic energy term proportional to $q^2$, a term linear in $\bm{q}$ appears in the GL free energy.
This indicates that, for $N=1$, the type II Lifshitz invariant drives an instability toward a non-uniform state, resulting in a spatially modulated phase called the helical state.
To obtain a stable uniform state, it is therefore necessary to consider cases with $N\ge 2$, where additional terms such as the Josephson coupling between different order parameters may stabilize the uniform solution (see the discussion in Ref.~\cite{Nagashima2025}).

Let us examine how the order parameter couples to electromagnetic fields when the Lifshitz invariant is present in the GL free energy.
To this end, we decompose each order parameter into amplitude and phase fluctuations ($H_{\alpha}$ and $\theta_{\alpha}$) as $\psi_{\alpha}=(\psi_{\alpha0} + H_{\alpha})e^{\mathrm{i}\theta_{\alpha}}$, where $\psi_{\alpha0}$ is the equilibrium value that is assumed to be a real constant.
We focus on linear coupling terms between the fluctuations ($H_{\alpha}$ and $\theta_{\alpha}$) and the vector potential $A_{\lambda}$ that contribute to the linear response.
For the type I Lifshitz invariant, the coupling term reads
\begin{equation}
    -2\mathrm{i}e^{*}A_{\lambda}\psi_{\alpha0}\psi_{\beta0}d^{\lambda}_{\alpha\beta}(\theta_{\alpha} - \theta_{\beta}),
\end{equation}
which is real because $d^{\lambda}_{\alpha\beta}$ is purely imaginary.
This term describes the linear Leggett-light coupling~\cite{Kamatani2022, Nagashima2024}.
On the other hand, the coupling term for the type II Lifshitz invariant has the following form:
\begin{equation}
    -2e^{*}A_{\lambda}d^{\lambda}_{\alpha\beta}(\psi_{\alpha0}H_{\beta} + H_{\alpha}\psi_{\beta0}).
\end{equation}
This term indicates that the Higgs mode linearly couples to light through the type II Lifshitz invariant (see Table~\ref{tab: type I, II}), which arises from the multiband nature.
We note that terms linear in $A_{\lambda}$ without fluctuations should vanish, since otherwise the system would become unstable against perturbations.
Let us also comment that this result is consistent with the recent theoretical study of the linear optical conductivity of multiband $s$-wave superconductors in a magnetic field with vertex corrections~\cite{Tanaka2025}.

If we further impose the symmetry of a crystal structure, we can obtain additional constraints on the emergence of the type II Lifshitz invariant.
Since the system must break TR symmetry, it is natural to include the TR operation among the point-group operations that form magnetic point groups.
In Sec .~\ref{Sec.III}, we will discuss symmetry classification of the type II Lifshitz invariant based on magnetic point groups.

\section{Group-theoretical classification of type II Lifshitz invariant}
\label{Sec.III}

To classify the crystal structure in which the type I Lifshitz invariant is allowed to exist, Ref.~\cite{Nagashima2024} has employed the representation theory of crystallographic point groups.
To extend this analysis beyond the crystallographic point groups to include TR operation, we generalize the classification from the 32 crystallographic point groups to the 122 magnetic point groups based on the theory of corepresentations.
The corepresentations introduced by Wigner~\cite{wigner2012} allow one to represent groups containing anti-unitary operations (such as TR) in terms of matrices.
This framework is essential for classifying the existence of type-II Lifshitz invariants in TR-symmetric broken-symmetry superconductors. 

The magnetic point groups are categorized into three types: 32 crystallographic point groups, 32 gray magnetic point groups, and 58 black-and-white magnetic point groups~\cite{Bradley1972}.
Both gray and black-and-white magnetic point groups are generated from a parent crystallographic point group $G$ by adding an anti-unitary operation $A$,
such that $M=G + AG$ (coset decomposition).
If one can choose $A$ as the TR operator itself ($A=\theta$ and $\theta$ is the TR operator), the system is TR-symmetric, and the corresponding group is referred to as a gray magnetic point group. Otherwise, when $A$ is of the form 
$A=\theta R_0$ (where $R_0\not\in G$ is a unitary operation), the group is called a black-and-white magnetic point group, which describes a TR-symmetry-broken system.
We note that, unlike the gray magnetic point groups, there is generally no simple one-to-one correspondence between a crystallographic point group $G$ and a black-and-white magnetic point group $M$. 
This is because the number of possible black-and-white groups derived from a single $G$ depends on the choice of the spatial operation $R_0$ combined with $\theta$ to form $A=\theta R_0$.

\subsection{Corepresentation theory}
\label{sec.IIIA}
In this subsection, we briefly review the corepresentation theory for magnetic point groups \cite{Bradley1972}.
We begin by recalling the conventional representation theory of a point group $G$, in which one defines a map from each element $R \in G$ to a transformation matrix $\bm\Delta(R)$ acting on a complex vector space.
The set of such matrices satisfying
\begin{equation}\label{eq:rep_def}
    \{\bm \Delta(R)|\bm \Delta(R)\bm \Delta(S)=\bm \Delta(RS)\, (R, S\in G)\},
\end{equation}
constitutes a (complex) representation of $G$. 
In general, a representation can be decomposed into irreducible representations, and the corresponding matrices can be simultaneously block-diagonalized by a unitary transformation.
Since each term in the GL free energy must be invariant under every $R\in G$, it should transform according to the trivial representation 
($\bm\Delta(R)\equiv1$).

For magnetic point groups $M=G+AG$, one must employ a corepresentation rather than an ordinary representation in order to account for the anti-unitary nature of TR symmetry operations. 
A corepresentation is defined as a map from $M$ to a set of matrices $\bm D$ and the complex conjugate operator $K$, namely $R\in G \mapsto \bm D(R)$ and $B\in AG\mapsto \bm D(B)K$, satisfying the following properties:
\begin{align}
    \bm D(R)\bm D(S)
    &=
    \bm D(RS),
    \\
    \bm D(R)[\bm D(B)K]
    &=
    \bm D(RB)K,
    \\
    [\bm D(B)K]\bm D(R)
    &=\bm D(B)\bm D^\ast(R)K=
    \bm D(BR)K,
    \\
    [\bm D(B)K][\bm D(C)K]
    &=\bm D(B)\bm D^\ast(C)K^2=
    \bm D(BC),
\end{align}
for $R, S\in G$ and $B, C\in AG$. 
The anti-unitary nature of the symmetry operations also affects how the matrices transform under a change of basis. In particular, for $B\in AG$, one finds
$\bm D^\prime(B)K=U^\dagger [\bm D(B)K] U=U^\dagger\bm D(B)U^*K$,
whereas for $R\in G$, the matrices 
obey the ordinary unitary transformation,
$\bm D^\prime(R)=U^\dagger \bm D(R) U$.

There exists a systematic procedure for constructing corepresentations and their matrices from irreducible representations $\Gamma$ of $G$ and the corresponding matrices $\bm\Delta^\Gamma(R)$.

Let $\ket{\psi_i}$ ($i=1,2,\cdots n$) be a basis of the representation space.
Acting with an anti-unitary operator $A$ generates another set of basis vectors $\ket{\phi_i}:=A\ket{\psi_i}$. As a result,
the corepresentation matrices are generally doubled in dimension compared to those of the original representation.

In general, irreducible corepresentations of a magnetic point group $M$ can be derived from irreducible representations of the corresponding point group $G$.
They are classified into three types, (a), (b), and (c), according to the Wigner criterion,
\begin{equation}
    \frac{1}{|G|}\sum_{B\in AG}\chi^\Gamma(B^2)=
    \begin{cases}
    +1&\text{for type (a),}\\
    -1&\text{for type (b),}\\
    0&\text{for type (c),}
    \end{cases}
    \label{wigner_criterion}
\end{equation}
where $\chi^\Gamma(R)=\mathrm{Tr}(\bm\Delta^\Gamma(R))$ is the character of $R\in G$, and $|G|$ is the order of the group.
Since $B^2\in G$ (because $\theta^2=E$, $E$ is the identity operation), the above expression is well defined.
Note that a given representation of $G$ may belong to different types for each $M$.
It is known that types (a), (b), and (c) correspond to two-to-one, one-to-one, and one-to-two mappings, respectively, between irreducible representations of $G$ and irreducible corepresentations of $M$.

We denote an irreducible corepresentation of $M$ as $\mathrm{D}\Gamma$, with corepresentation matrices $\bm{D}(R)$ and $\bm{D}(B)$. 
For type (a), one can introduce a ``mixed" basis $\{\ket{\psi_{1}}+\ket{\phi_{1}},\cdots,\ket{\psi_{1}}-\ket{\phi_{1}}, \cdots\}$, in which the corepresentation matrices take a block-diagonal form,
\begin{gather}
        \!\bm{D}(R)\!\!=\!\!
            \begin{pmatrix}
                \!\bm{D}^{\mathrm{D}\Gamma}\!(R)\!&\bm{0}\\
                \bm{0}&\!\bm{D}^{\mathrm{D}\Gamma}\!(R)\!
            \end{pmatrix}\!\!=\!\!
            \begin{pmatrix}
                \!\bm{\Delta}^\Gamma(R)\!&\bm{0}\\
                \bm{0}&\!\bm{\Delta}^\Gamma(R)\!
            \end{pmatrix},\label{eq:corepmat_a_r}
            \\
            \begin{aligned}
            \bm{D}(B)=&
            \begin{pmatrix}
                \bm{D}^{\mathrm{D}\Gamma}(B)&\bm{0}\\
                \bm{0}& -\bm{D}^{\mathrm{D}\Gamma}(B)
            \end{pmatrix}\\=&
            \begin{pmatrix}
                \bm{\Delta}^\Gamma(BA^{-1})\bm{N}^{\mathrm{D}\Gamma}&\bm{0}\\
                \bm{0}&-\bm{\Delta}^\Gamma(BA^{-1})\bm{N}^{\mathrm{D}\Gamma}
            \end{pmatrix}.                
            \end{aligned}\label{eq:corepmat_a_b}
\end{gather}
Thus, the corepresentation decomposes into two irreducible corepresentations $2\mathrm {D\Gamma}$, corresponding to the upper-left and lower-right blocks. Although these blocks have opposite parity under $A$ (e.g., the TR parity in gray groups), they are unitarily equivalent due to the antiunitary nature of $B$. Indeed, the unitary transformation $U=\mathrm{i}\bm1$ leaves $\bm D(R)$ invariant while flipping the sign of $\bm D(B)$, i.e., $U^\dagger\bm D(R)U=\bm D(R)$ and $U^\dagger\bm D(B)U^\ast=-\bm D(B)$. The two blocks are sometimes denoted by $\mathrm{D}\Gamma^+$ and $\mathrm{D}\Gamma^-$ to distinguish their parity under $A$; however, such a distinction is meaningful only when additional constraints (e.g., gauge fixing) are imposed.
The matrix $\bm{N}^{\mathrm{D}\Gamma}$ becomes non-identity only if there exists an element $R\in G$ that does not commute with the anti-unitary generator $A\in M$.

For types (b) and (c), corepresentation matrices are not block diagonalizable and are instead expressed in the original basis $\{\ket{\psi_1},\cdots,\ket{\phi_1},\cdots\}$.
For type (b), one obtains
\begin{gather}
        \bm{D}^{\mathrm{D}\Gamma}(R)=
            \begin{pmatrix}
                \bm{\Delta}^\Gamma(R)&\bm{0}\\
                \bm{0}&\bm{\Delta}^\Gamma(R)
            \end{pmatrix},\label{eq:corepmat_b_r}
            \\
            \bm{D}^{\mathrm{D}\Gamma}(B)=
            \begin{pmatrix}
                \bm{0}&-\bm{\Delta}^\Gamma(BA^{-1})\bm{N}^{\mathrm{D}\Gamma}\\
                \bm{\Delta}^\Gamma(BA^{-1})\bm{N}^{\mathrm{D}\Gamma}&\bm{0}
            \end{pmatrix},\label{eq:corepmat_b_b}
\end{gather}
where $\bm N^{\mathrm{D}\Gamma}=\bm1$ for all single-valued corepresentations considered in this work. 
Here, ``single-valued'' means that the corepresentation is invariant under a $2\pi$ rotation, in constast to double-valued corepresentations, where a $2\pi$ rotation becomes a non-identity operation.

For type (c), the matrices are given by
\begin{gather}
        \bm{D}^{\mathrm{D}\Gamma}(R)=
            \begin{pmatrix}
                \bm{\Delta}^\Gamma(R)&\bm{0}\\
                \bm{0}&\bm{\Delta}^{\Gamma^\prime}(R)
            \end{pmatrix},\label{eq:corepmat_c_r}
            \\
            \bm{D}^{\mathrm{D}\Gamma}(B)=
            \begin{pmatrix}
                \bm{0}&\bm{\Delta}^{\Gamma}(BA)\\
                \bm{\Delta}^{\Gamma^\prime}(BA^{-1})&\bm{0}
            \end{pmatrix},\label{eq:corepmat_c_b}
\end{gather}
where $\Gamma$ and $\Gamma^\prime$ are different representations of $G$ and $\bm{\Delta}^{\Gamma^\prime}(R)=[\bm{\Delta}^{\Gamma}(A^{-1}RA)]^\ast$, which means there is no transformation between $\bm\Delta^\Gamma(R)$ and $[\bm{\Delta}^{\Gamma}(A^{-1}RA)]^\ast$ unlike type (a).

If one focuses only on the unitary part ($R\in G$), type (b) exhibits a one-to-one correspondence between representations and corepresentations, whereas type (c) corresponds to a one-to-two mapping.

Since the Lifshitz invariant has the form $\psi_\alpha^\ast D_\lambda \psi_\beta$,
one must consider the irreducible decomposition of Kronecker products of corepresentations.
It is known~\cite{Bradley1972} that such decompositions can be obtained from those of the corresponding representations.
For example, let us assume that an irreducible decomposition of the product between representations of type (a), $\Gamma_i$, and of type (c), $\Gamma_j\oplus\Gamma_j^\prime$, has a term of type (c), $\Gamma_k\oplus\Gamma_k^\prime$:
\begin{equation}
    \Gamma_i\otimes(\Gamma_j\oplus\Gamma_j^\prime)=c_{ij,k}(\Gamma_k\oplus\Gamma_k^\prime)\oplus\cdots.
\end{equation}
Considering the size of the corepresentation matrices, its corepresentation counterpart is given by
\begin{equation}
    \mathrm{D}\Gamma_i\otimes\mathrm{D}\Gamma_j=c_{ij,k}\mathrm{D}\Gamma_k\oplus\cdots
\end{equation}
with the same coefficient $c_{ij,k}$.

We show the product tables of corepresentations for all magnetic point groups in Appendix~\ref{Appendix A}.
The relation between decomposition coefficients for representations and those for corepresentations is provided in Table \ref{tab:transform_table} in Appendix~\ref{app: Kronecker product}.

\subsection{Real corepresentation theory}
So far, we have formulated the corepresentation theory of magnetic point groups in terms of complex matrices,
i.e., the \textit{complex} corepresentation theory.
However, this framework does not distinguish between whether a given corepresentation is even or odd under the antiunitary operation $A\in M$ (e.g., the time reversal combined with a spatial operation in black and white groups)~\cite{knoll2022classification,erb2020vector}, as mentioned above.
Let us recall that, for type (a) corepresentations, such a parity distinction can be glimpsed in the block-diagonalized form given in Eqs.~(\ref{eq:corepmat_a_r}) and (\ref{eq:corepmat_a_b}).
In some situations, it is desirable to keep track of such parity. For instance, the covariant derivative $D_\lambda$ changes the sign under time reversal, $D_\lambda\to -D_\lambda$, and hence its transformation properties can be treated explicitly in the GL free energy.
One way to address this issue is to fix the gauge by restricting the order parameters to be real,
and to formulate the classification in terms of {\it real} corepresentations.

The real corepresentations of a magnetic group $M = G+AG$ are equivalent to the real representations of the point group $H=G+R_0G$, since the anti-unitary nature of the time reversal operations can be disregarded in a real matrix representation. As a consequence, the Kronecker products of real corepresentations coincide with those of the representations of $H$. The corresponding product tables are provided in Appendix~\ref{Appendix A}.
To make the $A$-parity explicit, we denote the corepresentation as $\mathrm{D}\Gamma^\pm$, rather than adopting the standard notation of the representation of $H$ used in Ref.~\cite{erb2020vector}. However, there exist exceptional cases in which the $A$-parity is not well defined, since an orthogonal transformation interchanges the real corepresentation matrices $\bm D(B)$ and $-\bm D(B)$ while leaving $\bm D(R)$ unchanged. Moreover, the assignment of $A$-parity may depend on the choice of $R_0$. 
The specific choices of $R_0$ and the parity-indefinite corepresentations are listed in the second and third rows of Tables~\ref{selection_table1_r}-\ref{selection_table5_r}.

\subsection{Corepresentation of the Lifshitz invariant}
\label{3c}
\begin{table*}
    \centering
    \caption{
    Classification of the Lifshitz invariant based on complex corepresentations of magnetic point groups (MPGs) derived from triclinic, monoclinic, and orthorhombic crystallographic point groups (PGs).
   The first column lists the MPGs. 
    The second column explicitly shows the type-(b) and type-(c) representations (reps) 
    for each MPG, while all others are of type (a).
    The third column gives the corepresentation (corep) of the covariant derivative $D$.
    The fourth column lists all pairs of corepresentations of the order parameters $(\Psi_i,\Psi_j)$ that allow a Lifshitz invariant for each MPG.
    The fifth column indicates the parent PG in Hermann-Mauguin (Sch\"{o}nflies) notation.
    For each PG, gray MPGs are listed first, followed by black-and-white MPGs. 
    }
    \begin{tblr}{c|l|l|l|c}
    \hline\hline
        MPG   &   (b), (c) reps   &   Corep of $D$   &   Allowed coreps of $(\Psi_i, \Psi_j)$ &   PG  \\
         \hline
        $11^\prime$, $\bar{1}^{\prime}$, $2^\prime$, $m^\prime$   &      &   $3A$   &  $(A, A)$ &  $1\ (C_1)$    \\
        \hline[dashed]
        $\bar{1}1^\prime$, $2^\prime/m^\prime$   &      &   $3A_u$   &   $(A_{g}, A_{u})$ &    $\bar{1}\ (C_{i},S_{2})$  \\
        \hline[dashed]
        $21^\prime$, $2/m^\prime$, $2^\prime2^\prime2$, $m^\prime m^\prime2$   &      &   $A\oplus2B$   &   $(A, A), (A, B), (B, B)$ &   $2\ (C_2)$   \\
        $4^\prime$, $\bar{4}^\prime$   &   (b);$B$   &   $A\oplus B$   &   $(A, A), (A, B), (B,B)$ &      \\
        \hline[dashed]
        $m1^\prime$, $2^\prime/m$, $m^\prime m2^\prime$   &      &   $2A^{\prime}\oplus A^{\prime\prime}$   &   $(A^{\prime}, A^{\prime}), (A^{\prime}, A^{\prime\prime}), (A^{\prime\prime},A^{\prime\prime})$ &   $m\ (C_s,C_h)$   \\
        \hline[dashed]
        $2/m1^\prime$, $m^\prime m^\prime m$   &      &  $A_u\oplus2B_u$   &   $(A_{g}, A_{u}), (A_{g}, B_{u}), (B_{g}, A_{u}), (B_{g}, B_{u})$ &    $2/m\ (C_{2h})$  \\
        $4^\prime/m$   &   (b);$B_g,B_u$   &   $A_u\oplus B_u$   &   $(A_{g}, A_{u}), (A_{g}, B_{u}), (B_{g}, A_{u}), (B_{g}, B_{u})$ &      \\
        \hline[dashed]
        $2221^{\prime}$, $m^\prime m^\prime m^\prime$   &      &   $B_1\oplus B_2\oplus B_3$   &   $(A, B_{1}), (A, B_{2}), (A, B_{3}), $ &   $222\ (D_2)$   \\
           &     &      &   $(B_{1}, B_{2}), (B_{1}, B_{3}), (B_{2}, B_{3})$ &     \\
        $4^\prime22^\prime$, $\bar{4}^\prime2m^\prime$   &   (c);$(B_2,B_3)$   &   $B_1\oplus B_{23}$   &   $(A, B_{1}), (A, B_{23}), (B_1, B_{23}), (B_{23},B_{23})$ &      \\
        \hline[dashed]
        $mm21^\prime$, $mmm^\prime$   &      &   $A_1\oplus B_1\oplus B_2$   &   $(A_{1}, A_{1}), (A_{1}, B_{1}), (A_{1}, B_{2}), (A_{2}, A_{2}),$ &  $mm2\ (C_{2v})$    \\
           &&            &   $ (A_{2}, B_{1}), (A_{2}, B_{2}), (B_{1}, B_{1}), (B_{2}, B_{2})$ &     \\
        $4^\prime mm^\prime$, $\bar{4}^\prime m2^\prime$   &   (c);$(B_1,B_2)$   &   $A_1\oplus B_{12}$   &   $(A_{1}, A_{1}), (A_{2}, A_{2}), (A_{1}, B_{12}), $ &      \\
         && &  $(A_{2}, B_{12}),(B_{12},B_{12})$ & \\
        \hline[dashed]
        $mmm1^{\prime}$   &      &   $B_{1u}\oplus B_{2u}$   &   $(A_{g}, B_{1u}), (A_{g}, B_{2u}), (A_{g}, B_{3u}), $ &   $mmm$  \\
           &      &   ${}\oplus B_{3u}$   &   $(B_{1g}, A_{u}), (B_{1g}, B_{2u}), (B_{1g}, B_{3u}), $ &   $(D_{2h})$  \\
           &      &      &   $(B_{2g}, A_{u}), (B_{2g}, B_{1u}), (B_{2g}, B_{3u}),$ &     \\
           &&&$ (B_{3g}, A_{u}), (B_{3g}, B_{1u}), (B_{3g}, B_{2u})$&\\
        $4^\prime/mmm^\prime$   &   (c);$(B_{2g},B_{3g}),$   &   $B_{1u}\oplus B_{23u}$   &   $(A_g, B_{1u}), (A_g, B_{23u}), (B_{1g}, B_{23u}), $ &      \\
           &   $(B_{2u},B_{3u})$   &      &   $(A_u, B_{1g}), (A_u, B_{23g}), $  &     \\
           &&&$(B_{1u}, B_{23g}),(B_{23g},B_{23u})$&\\
        \hline\hline
    \end{tblr}
    \label{selection_table1_c}
\end{table*}

\begin{table*}
    \centering
   \caption{
   Classification of the Lifshitz invariant based on complex corepresentations of magnetic point groups (MPGs) derived from tetragonal crystallographic point groups (PGs). The notations and column definitions are the same as in Table~\ref{selection_table1_c}.
   }
    \begin{tblr}{c|l|l|l|c}
    \hline\hline
        MPG   &   (b), (c) reps   &   Corep of $D$   &   Allowed coreps of $(\Psi_i, \Psi_j)$ &   PG  \\
         \hline
        $41^{\prime}$, $4/m^\prime$   &   (c);(${}^{1,2}E$)   &   $A\oplus E$   &   $(A, A), (A, E), (B, B), (B, E), (E,E)$ &   $4$  \\
        $42^{\prime}2^{\prime}$, $4m^\prime m^\prime$   &      &   $A\oplus {}^1E$   &   $(A, A), (A, {}^{1,2}E), (B, B),$ &   $(C_{4})$  \\
           &      &   ${}\oplus{}^2E$   &   $(B, {}^{1,2}E), ({}^{1,2}E, {}^{1,2}E)$ &     \\
        \hline[dashed]
        $\bar{4}1^{\prime}$, $4^{\prime}/m^{\prime}$   &   (c);(${}^{1,2}E$)   &   $B\oplus E$   &   $(A, B), (A, E), (B, E), (E,E)$ &   $\bar{4}$  \\
        $\bar{4}2^\prime m^\prime$   &      &   $B\oplus {}^1E$   &   $(A, {}^{1,2}E), (A, B),$ &   $(S_4)$   \\
         & & ${}\oplus{}^2E$ & $ (B, {}^{1,2}E), ({}^{1,2}E, {}^{1,2}E) $ & \\
        \hline[dashed]
        $4/m1^{\prime}$   &   (c);(${}^{1,2}E_g$),   &   $A_u\oplus E_u$   &   $(A_g, A_u), (A_g, E_u), (A_u, E_g), (B_g, B_u), $ &   $4/m$  \\
           &   (${}^{1,2}E_u$)   &      &   $(B_g, E_u), (B_u, E_g), (E_g,E_u)$ &   $ (C_{4h})$  \\
        $4/mm^\prime m^\prime$   &      &   $A\oplus {}^1E_u$   &   $(A_{g}, A_{u}), (A_{g}, {}^{1,2}E_{u}), (B_{g}, B_{u}), (B_{g}, {}^{1,2}E_{u}), $ &      \\
           &      &   ${}\oplus{}^2E_u$   &   $({}^{1,2}E_{g}, A_{u}), ({}^{1,2}E_{g}, B_{u}), ({}^{1,2}E_{g}, {}^{1,2}E_{u})$ &     \\
        \hline[dashed]
        $4221^{\prime}$,    &      &   $A_2\oplus E$   &   $(A_{1}, A_{2}), (A_{1}, E), (A_{2}, E), (B_{1}, B_{2}), $ &    $422$  \\
        $4/m^\prime m^\prime m^\prime$   &      &      &   $(B_{1}, E), (B_{2}, E), (E, E)$ &  $(D_4)$   \\
        \hline[dashed]
        $4mm1^{\prime}$,    &      &   $A_1\oplus E$   &   $(A_{1}, A_{1}), (A_{1}, E), (A_{2}, A_{2}), $ &   $4mm$  \\
        $4/m^\prime mm$   &      &      & $(A_{2}, E), (B_{1}, B_{1})$, $(B_{1}, E),$ &   $ (C_{4v})    $  \\
           & & & $ (B_{2}, B_{2}), (B_{2}, E), (E, E) $ & \\
        \hline[dashed]
        $\bar{4}2m1^{\prime}$,     &      &   $B_2\oplus E$   &   $(A_{1}, B_{2}), (A_{1}, E), (A_{2}, B_{1}), (A_{2}, E), $ &   $\bar{4}2m$  \\
        $4^\prime/m^\prime m^\prime m$     &      &      &   $(B_{1}, E), (B_{2}, E), (E, E)$ &   $ (D_{2d})$  \\
        \hline[dashed]
        $4/mmm1^{\prime}$   &      &   $A_{2u}\oplus E_u$   &   $(A_{1g}, A_{2u}), (A_{1g}, E_{u}), (A_{2g}, A_{1u}), $ &   $4/mmm$  \\
            &      &      &   $(A_{2g}, E_{u}), (B_{1g}, B_{2u}),(B_{1g}, E_{u}), $ &   $ (D_{4h})  $  \\
           &      &      &   $(B_{2g}, B_{1u}), (B_{2g}, E_{u}), (E_{g}, A_{1u}),$ \\
           & & & $(E_{g}, A_{2u}),  (E_{g}, B_{1u}), (E_{g}, B_{2u}), (E_{g}, E_{u}) $ &  \\ \hline\hline 
    \end{tblr}
    \label{selection_table2_c}
\end{table*}

\begin{table*}
    \centering
        \caption{
     Classification of the Lifshitz invariant based on complex corepresentation of magnetic point groups (MPGs) derived from trigonal crystallographic point groups (PGs). The notations and column definitions are the same as in Table~\ref{selection_table1_c}.
        }
    \begin{tblr}{c|l|l|l|c}
    \hline\hline
        MPG   &  (b), (c) reps   &   Corep of $D$   &   Allowed coreps of $(\Psi_i, \Psi_j)$ &   PG  \\
         \hline
        $31^\prime$, $\bar{6}^\prime$, $6^\prime$, $\bar{3}^\prime$&   (c);(${}^{1,2}E$)   &   $A\oplus E$   &   $(A, A), (A, E), (E, E)$ &   $3$  \\
        $32^\prime$, $3m^\prime$  &      &   $A\oplus{}^1E\oplus{}^2E$   &   $(A, A), (A, {}^{1,2}E), ({}^{1,2}E, {}^{1,2}E)$ &   ($C_{3}$)   \\
        \hline[dashed]
        $\bar{3}1^{\prime}$, $6^\prime/m^\prime$   &   (c);(${}^{1,2}E_g$),    &   $A_u\oplus E_u$   &   $(A_g, A_u), (A_g, E_u),(A_u, E_g), (E_g, E_u)$ &   $\bar{3}$  \\
         & (${}^{1,2}E_u$) & &  & $ (S_6,C_{3i})$ \\
        $\bar{3}m^\prime$  &      &   $A_u\oplus {}^1E_u\oplus{}^2E_u$   &   $(A_{g}, A_{u}), (A_{g}, {}^{1,2}E_{u}), $ &    \\
         & & & $({}^{1,2}E_{g}, A_{u}), ({}^{1,2}E_{g}, {}^{1,2}E_{u})$ & \\
        \hline[dashed]
        $321^{\prime}$, $\bar{6}^\prime m^\prime 2$,    &      &   $A_2\oplus E$   &   $(A_{1}, A_{2}), (A_{1}, E), (A_{2}, E), (E, E)$ &   $32$  \\
        $\bar{3}^\prime m^\prime$, $6^\prime2^\prime2$   &      &      &    &   $(D_{3})$   \\
        \hline[dashed]
        $3m1^\prime$, $\bar{6}^\prime m2^\prime$   &      &   $A_1\oplus E$   &   $(A_{1}, A_{1}), (A_{1}, E), (A_{2}, E), (E, E)$ &   $3m$   \\
        $\bar{3}^\prime m$, $6^\prime m^\prime m$   &      &      &    &   $(C_{3v})$  \\
        \hline[dashed]
        $\bar{3}m1^\prime$,    &      &   $A_{2u}\oplus E_u$   &   $(A_{1g}, A_{2u}), (A_{1g}, E_{u}), (A_{2g}, A_{1u}), $ &   $\bar{3}m$   \\
        $6^\prime/m^\prime m^\prime m$    &      &      &   $(A_{2g}, E_{u}), (E_{g}, A_{1u}), (E_{g}, A_{2u}), (E_{g}, E_{u})$ &   $ (D_{3d})$  \\ \hline\hline
    \end{tblr}
    \label{selection_table3_c}
\end{table*}

\begin{table*}
    \centering
    \caption{
    Classification of the Lifshitz invariant based on complex corepresentations of magnetic point groups (MPGs) derived from hexagonal crystallographic point groups (PGs). The notations and column definitions are the same as in Table~\ref{selection_table1_c}.
    }
    \begin{tblr}{c|l|l|l|c}
    \hline\hline
        MPG   &   (b), (c) reps   &   Corep of $D$   &   Allowed coreps of $(\Psi_i, \Psi_j)$ &   PG  \\
         \hline
        $61^\prime$, $6/m^\prime$  &   (c);(${}^{1,2}E_1$),  &   $A\oplus E_1$   &   $(A, A), (A, E_{1}), (B, B), (B, E_{2}), $ &   $6$   \\
           &   (${}^{1,2}E_2$)   &      &   $(E_{1}, E_{1}), (E_{1},E_{2}), (E_{2},E_{2})$ &   $(C_{6})$  \\
        $62^\prime2^\prime$, $6m^\prime m^\prime$  &      &   $A\oplus{}^1E_1$   &   $(A, A), (A, {}^{1,2}E_{1}), (B, B), (B, {}^{1,2}E_{2}), $ &      \\
           &      &  {}${}\oplus{}^2E_1$    &   $({}^{1,2}E_{1}, {}^{1,2}E_{1}), ({}^{1,2}E_{1}, {}^{1,2}E_{2}), ({}^{1,2}E_{2}, {}^{1,2}E_{2})$ &     \\
        \hline[dashed]
        $\bar{6}1^\prime$, $6^\prime/m$  &   (c);(${}^{1,2}E^\prime$),  &   $A^{\prime\prime}\oplus E^{\prime}$   &   $(A^{\prime}, A^{\prime\prime}), (A^{\prime}, E^{\prime}), (A^{\prime\prime}, E^{\prime\prime}), $ &   $\bar{6}$   \\
            &  (${}^{1,2}E^{\prime\prime}$)    &      &   $(E^{\prime}, E^{\prime}), (E^{\prime}, E^{\prime\prime}), (E^{\prime\prime}, E^{\prime\prime})$ &   $ (C_{3h})$  \\
        $\bar{6}^\prime m2^\prime$   &      &   $A^{\prime\prime}\oplus {}^1E^{\prime}$   &   $(A^{\prime}, A^{\prime\prime}), (A^{\prime}, {}^{1,2}E^{\prime}), (A^{\prime\prime}, {}^{1,2}E^{\prime\prime}), $ &      \\
           &      &   ${}\oplus {}^2E^{\prime}$   &   $({}^{1,2}E^{\prime}, {}^{1,2}E^{\prime}), ({}^{1,2}E^{\prime}, {}^{1,2}E^{\prime\prime}), ({}^{1,2}E^{\prime\prime}, {}^{1,2}E^{\prime\prime})$ &     \\
        \hline[dashed]
        $6/m1^\prime$   &   (c);(${}^{1,2}E_{1g}$),   &   $A_u\oplus E_{1u}$   &   $(A_g, A_u), (A_g, E_{1u}), (B_g, B_u), $ &   $6/m$   \\
            &   (${}^{1,2}E_{2g}$),   &      &   $(B_g, E_{2u}),(A_u, E_{1g}), (B_u, E_{2g}), $ &   $(C_{6h})$  \\
           &  (${}^{1,2}E_{1u}$),  &      &   $(E_{1g}, E_{1u}), (E_{1g},E_{2u}),  (E_{1u},E_{2g}),$ &     \\
         & (${}^{1,2}E_{2u}$) & & $(E_{2g},E_{2u})$ & \\
        $6/mm^\prime m^\prime$   &      & $A_u\oplus {}^1E_{1u}$  &$(A_g, A_u), (A_g, {}^{1,2}E_{1u}), (A_u, {}^{1,2}E_{1g}), (B_g, B_u), $ &      \\
           &      &   ${}\oplus {}^2E_{1u}$   &  $(B_g, {}^{1,2}E_{2u}), (B_u, {}^{1,2}E_{2g}), ({}^{1,2}E_{1g}, {}^{1,2}E_{1u}),$   &     \\
           &    &   &$({}^{1,2}E_{1g},{}^{1,2}E_{2u}),  ({}^{1,2}E_{1u},{}^{1,2}E_{2g}), ({}^{1,2}E_{2g},{}^{1,2}E_{2u})$   &   \\
        \hline[dashed]
        $6221^\prime$,    &      &   $A_2\oplus E_1$   &   $(A_{1}, A_{2}), (A_{1}, E_{1}), (A_{2}, E_{1}), (B_{1}, B_{2}), (B_{1}, E_{2}), $ &   $622$   \\
         $6/m^\prime m^\prime m^\prime$   &      &      &   $(B_{2}, E_{2}), (E_{1}, E_{1}), (E_{1}, E_{2}), (E_{2}, E_{2})$ &   $(D_{6})$  \\
        \hline[dashed]
        $6mm1^\prime$,    &      &   $A_1\oplus E_1$   &   $(A_{1}, A_{1}), (A_{1}, E_{1}), (A_{2}, A_{2}), (A_{2}, E_{1}), $ &   $6mm$  \\
        $6/m^\prime mm$   &      &      &   $(B_{1}, B_{1}), (B_{1}, E_{2}), (B_{2}, B_{2}), (B_{2}, E_{2}), $ &   $(C_{6v})$  \\
           &      &      &   $(E_{1}, E_{1}), (E_{1}, E_{2}), (E_{2}, E_{2})$ &     \\
        \hline[dashed]
        $\bar{6}m21^\prime$,   &      &   $A_2^{\prime\prime}\oplus E^{\prime}$   &   $(A_{1}^{\prime}, A_{2}^{\prime\prime}), (A_{1}^{\prime}, E^{\prime}), (A_{1}^{\prime\prime}, A_{2}^{\prime}), $ &   $\bar{6}m2$   \\
        $6^\prime/mmm^\prime$   &      &      &   $(A_{1}^{\prime\prime}, E^{\prime\prime}), (A_{2}^{\prime}, E^{\prime}), (A_{2}^{\prime\prime}, E^{\prime\prime}), $ &   $(D_{3h})$   \\
           &      &      &   $(E^{\prime}, E^{\prime}), (E^{\prime}, E^{\prime\prime}), (E^{\prime\prime}, E^{\prime\prime})$ &     \\
        \hline[dashed]
        $6/mmm1^\prime$   &      &   $A_{2u}\oplus E_{1u}$   &   $(A_{1g}, A_{2u}), (A_{1g}, E_{1u}), (A_{2g}, A_{1u}), (A_{2g}, E_{1u}), $ &   $6/mmm$   \\
            &      &      &   $(B_{1g}, B_{2u}), (B_{1g}, E_{2u}), (B_{2g}, B_{1u}), (E_{2g}, E_{2u}), $ &   $ (D_{6h})  $  \\
           &      &      &   $(E_{1g}, A_{1u}), (E_{1g}, A_{2u}), (E_{1g}, E_{1u}), (E_{1g}, E_{2u}), $ &     \\
           &      &      &   $(E_{2g}, B_{1u}), (E_{2g}, B_{2u}), (E_{2g}, E_{1u}), (B_{2g}, E_{2u})$ \\ \hline\hline
    \end{tblr}
    \label{selection_table4_c}
\end{table*}

\begin{table*}
    \centering
    \caption{
    Classification of the Lifshitz invariant based on complex corepresentations of magnetic point groups (MPGs) derived from cubic crystallographic point groups (PGs). The notations and column definitions are the same as in Table~\ref{selection_table1_c}.
    In the second column, we also indicate the cases in which the matrix $\bm N$ in Eq.~(\ref{eq:corepmat_a_b}) becomes non-identity.
    }
    \begin{tblr}{c|l|l|l|c}
    \hline\hline
        MPG   &   (b), (c) reps   &   Corep of $D$   &   Allowed coreps of $(\Psi_i, \Psi_j)$ &   PG  \\
         \hline
        $231^\prime$, $m^\prime\bar{3}^\prime$   &   (c);(${}^{1,2}E$)   &   $T$   &   $(A, T), (E, T), (T, T)$ &   $23$   \\
           &   (c);(${}^{1,2}E$)   &   $T$   &   $(A, T), (E, T), (T, T)$ &   \\
        $\bar{4}^\prime3m^\prime$, $4^\prime32^\prime$  &   $\bm{N}\neq\bm{1}:T$   &   $T$   &   $(A, T), ({}^{1,2}E, T), (T, T)$ &   $(T)$     \\
        \hline[dashed]
        $m\bar{3}1^\prime$   &   (c);(${}^{1,2}E_g$),    &   $T_u$   &   $(A_g, T_u), (A_u, T_g), (E_g, T_u),$ &   $m\bar{3}$  \\
            &  (${}^{1,2}E_u$)  &      &   $(E_u, T_g), (T_g, T_u)$ &   $ (T_h)$  \\
        $m\bar{3}m^\prime$   &   $\bm{N}\neq\bm{1}:T_g,T_u$   &   $T_u$   &   $(A_g, T_u), (A_u, T_g), ({}^{1,2}E_g, T_u),$ &      \\
           &      &      &   $({}^{1,2}E_u, T_g), (T_g, T_u)$ &     \\
        \hline[dashed]
        $4321^\prime$,    &      &   $T_1$   &   $(A_{1}, T_{1}), (A_{2}, T_{2}), (E, T_{1}), (E, T_{2}), $ &   $432$   \\
         $m^\prime\bar{3}m^\prime$  &      &      &   $(T_{1}, T_{1}), (T_{1}, T_{2}), (T_{2}, T_{2})$ &   $(O)$  \\
        \hline[dashed]
        $\bar{4}3m1^\prime$,   &      &   $T_2$   &   $(A_{1}, T_{2}), (A_{2}, T_{1}), (E, T_{1}), (E, T_{2}), $ &  $\bar{4}3m$    \\
         $m^\prime\bar{3}m$    &      &      &   $(T_{1}, T_{1}), (T_{1}, T_{2}), (T_{2}, T_{2})$ &   $ (T_d)$  \\
        \hline[dashed]
        $m\bar{3}m1^\prime$   &      &   $T_{1u}$   &   $(A_{1g}, T_{1u}), (A_{2g}, T_{2u}), (E_{g}, T_{1u}), (E_{g}, T_{2u}), $ &   $m\bar{3}m$   \\
            &      &      &   $(T_{1g}, A_{1u}), (T_{1g}, E_{u}), (T_{1g}, T_{1u}), (T_{1g}, T_{2u}), $ &   $ (O_h)$  \\
           &      &      &   $(T_{2g}, A_{2u}), (T_{2g}, E_{u}), (T_{2g}, T_{1u}), (T_{2g}, T_{2u})$ \\ \hline\hline
    \end{tblr}
    \label{selection_table5_c}
\end{table*}
\begin{table*}
    \centering
    \caption{
    Classification of the Lifshitz invariant based on real corepresentations of magnetic point groups (MPGs) derived from triclinic, monoclinic, and orthorhombic crystallographic point groups (PGs).
    The first column lists the MPGs.
    The second column shows the choice of $R_0$.
    The third column shows parityless corepresentations (coreps) for each MPG.
    The fourth column gives the corep of the covariant derivative $D$.
    The fifth column lists all pairs of corepresentations of the order parameters $(\Psi_i,\Psi_j)$ that allow a Lifshitz invariant for each MPG.
    The sixth column indicates the parent PG in Hermann-Mauguin (Sch\"{o}nflies) notation.
    For each PG, gray MPGs are listed first, followed by black-and-white MPGs. 
    }
    \begin{tblr}{c|l|l|l|l|c}
    \hline\hline
        MPG   &$R_0$&  Parityless coreps   &   Corep of $D$   &   Allowed coreps of $(\Psi_i, \Psi_j)$ &   PG  \\
         \hline
        $11^\prime$   &$E$&      &   $3A^-$   &   $(A^{\pm}, A^{\mp})$ &   $1$  \\
        $\bar{1}^{\prime}$   &$I$&      &   $3A^+$   &   $(A^{\pm}, A^{\pm})$ &   $(C_1)$  \\
        $2^\prime$   &$C_{2z}$&      &   $2A^+ \oplus A^-$   &   $(A^{\pm}, A^{\pm}), (A^{\pm}, A^{\mp})$ &      \\
        $m^\prime$   &$\sigma_z$&      &   $A^+ \oplus 2A^-$   &  $(A^{\pm}, A^{\pm}), (A^{\pm}, A^{\mp})$ &      \\
        \hline[dashed]
        $\bar{1}1^\prime$   &$E$&      &   $3A^-_u$   &   $(A_{g}^{\pm}, A_{u}^{\mp})$ &   $\bar{1}$  \\
        $2^\prime/m^\prime$   &$C_{2z}$&      &   $2A^+_u \oplus A^-_u$   &   $(A_{g}^{\pm}, A_{u}^{\pm}), (A_{g}^{\pm}, A_{u}^{\mp})$ &    $(C_{i},S_{2})$  \\
        \hline[dashed]
        $21^\prime$   &$E$&      &   $A^-\oplus 2B^-$   &   $(A^{\pm}, A^{\mp}), (A^{\pm}, B^{\mp}), (B^{\pm}, B^{\mp})$ &   $2$  \\
        $2/m^\prime$   &$I$&      &   $A^+\oplus 2B^+$   &   $(A^{\pm}, A^{\pm}), (A^{\pm}, B^{\pm}), (B^{\pm}, B^{\pm})$ &   $(C_{2})$  \\
        $2^\prime2^\prime2$   &$C_{2x}$&      &   $A^+\oplus B^+ \oplus B^-$   &   $(A^{\pm}, A^{\pm}), (A^{\pm}, B^{\pm}), (A^{\pm}, B^{\mp}), (B^{\pm}, B^{\pm})$ &      \\
        $m^\prime m^\prime2$   &$\sigma_x$&      &   $A^-\oplus B^+ \oplus B^-$   &   $(A^{\pm}, A^{\mp}), (A^{\pm}, B^{\pm}), (A^{\pm}, B^{\mp}), (B^{\pm}, B^{\mp})$ &      \\
        $4^\prime$   &$C_{4z}$&  $B$   &   $A^-\oplus B$   &   $(A^{\pm}, A^{\mp}), (A^{\pm}, B), (B,B)$ &      \\
        $\bar{4}^\prime$   &$S_{4z}$&  $B$   &   $A^+\oplus B$   &   $(A^{\pm}, A^{\pm}), (A^{\pm}, B), (B,B)$ &      \\
        \hline[dashed]
        $m1^\prime$   &$E$&      &   $2A^{\prime-}\oplus A^{\prime\prime-}$   &   $(A^{\prime\pm}, A^{\prime\mp}), (A^{\prime\pm}, A^{\prime\prime\mp}), (A^{\prime\prime\pm}, A^{\prime\prime\mp})$ &   $m$  \\
        $2^\prime/m$   &$I$&      &   $2A^{\prime+}\oplus A^{\prime\prime+}$   &   $(A^{\prime\pm}, A^{\prime\pm}), (A^{\prime\pm}, A^{\prime\prime\pm}), (A^{\prime\prime\pm}, A^{\prime\prime\pm})$ &   $(C_{s},C_{h})$  \\
        $m^\prime m2^\prime$   &$C_{2y}$&      &   $A^{\prime+}\oplus A^{\prime-}$   &   $(A^{\prime\pm}, A^{\prime\pm}), (A^{\prime\pm}, A^{\prime\mp}), (A^{\prime\pm}, A^{\prime\prime\mp}), $ &      \\
        &&&${}\oplus A^{\prime\prime+}$&$ (A^{\prime\prime\pm},A^{\prime\prime\pm}), (A^{\prime\prime\pm}, A^{\prime\prime\mp})$&\\
        \hline[dashed]
        $2/m1^\prime$   &$E$&      &   $A_u^-\oplus 2B_u^-$   &   $(A_{g}^{\pm}, A_{u}^{\mp}), (A_{g}^{\pm}, B_{u}^{\mp}), (B_{g}^{\pm}, A_{u}^{\mp}), (B_{g}^{\pm}, B_{u}^{\mp})$ &   $2/m$   \\
        $m^\prime m^\prime m$   &$C_{2x}$&      &  $A_u^+\oplus B_u^+ \oplus B_u^-$   &   $(A_{g}^{\pm}, A_{u}^{\mp}), (A_{g}^{\pm}, B_{u}^{\pm}), (A_{g}^{\pm}, B_{u}^{\mp}), (B_{g}^{\pm}, A_{u}^{\pm}), $ &    $(C_{2h})$  \\
        &&&&$(B_{g}^{\pm}, A_{u}^{\mp}), (B_{g}^{\pm}, B_{u}^{\mp})$&\\
        $4^\prime/m$   &$C_{4z}$&  $B_g,B_u$   &   $A_u^-\oplus B_u$   &   $(A_{g}^{\pm}, A_{u}^{\mp}), (A_{g}^{\pm}, B_{u}), (B_{g}, A_{u}^{\pm}), (B_{g}, B_{u})$ &      \\
        \hline[dashed]
        $2221^{\prime}$   &$E$&      &   $B_1^-\oplus B_2^-$   &   $(A^{\pm}, B_{1}^{\mp}), (A^{\pm}, B_{2}^{\mp}), (A^{\pm}, B_{3}^{\mp}), $ &   $222$   \\
           &&       &   ${}\oplus B_3^-$   &   $(B_{1}^{\pm}, B_{2}^{\mp}), (B_{1}^{\pm}, B_{3}^{\mp}), (B_{2}^{\pm}, B_{3}^{\mp})$ &   ($D_{2}$)   \\
        $m^\prime m^\prime m^\prime$   &$I$&      &   $B_1^+\oplus B_2^+$   &   $(A^{\pm}, B_{1}^{\pm}), (A^{\pm}, B_{2}^{\pm}), (A^{\pm}, B_{3}^{\pm}), $ &      \\
           &&      &  ${}\oplus B_3^+$    &   $(B_{1}^{\pm}, B_{2}^{\pm}), (B_{1}^{\pm}, B_{3}^{\pm}), (B_{2}^{\pm}, B_{3}^{\pm})$ &     \\
        $4^\prime22^\prime$   &$C_{2a}$&  $B_{23}$   &   $B_1^+\oplus B_{23}$   &   $(A^{\pm}, B_{1}^{\pm}), (A^{\pm}, B_{23}), (B_1^{\pm}, B_{23}), (B_{23},B_{23})$ &      \\
        $\bar{4}^\prime2m^\prime$   &$\sigma_{da}$&  $B_{23}$   &   $B_1^-\oplus B_{23}$   &   $(A^{\pm}, B_{1}^{\mp}), (A^{\pm}, B_{23}), (B_1^{\pm}, B_{23}), (B_{23},B_{23})$ &      \\
        \hline[dashed]
        $mm21^\prime$   &$E$&      &   $A_1^-\oplus B_1^-$   &   $(A_{1}^{\pm}, A_{1}^{\mp}), (A_{1}^{\pm}, B_{1}^{\mp}), (A_{1}^{\pm}, B_{2}^{\mp}), (A_{2}^{\pm}, A_{2}^{\mp}), $ &   $mm2$  \\
           &&      &   ${}\oplus B_2^-$   &   $(A_{2}^{\pm}, B_{1}^{\mp}), (A_{2}^{\pm}, B_{2}^{\mp}), (B_{1}^{\pm}, B_{1}^{\mp}), (B_{2}^{\pm}, B_{2}^{\mp})$ &   $(C_{2v})$  \\
        $mmm^\prime$   &$I$&      &   $A_1^+\oplus B_1^+$   &   $(A_{1}^{\pm}, A_{1}^{\pm}), (A_{1}^{\pm}, B_{1}^{\pm}), (A_{1}^{\pm}, B_{2}^{\pm}), (A_{2}^{\pm}, A_{2}^{\pm}),$ &      \\
           &&      &   ${}\oplus B_2^+$   &   $ (A_{2}^{\pm}, B_{1}^{\pm}), (A_{2}^{\pm}, B_{2}^{\pm}), (B_{1}^{\pm}, B_{1}^{\pm}), (B_{2}^{\pm}, B_{2}^{\pm})$ &     \\
        $4^\prime mm^\prime$   &$C_{4z}$&  $B_{12}$   &   $A_1^-\oplus B_{12}$   &   $(A_{1}^{\pm}, A_{1}^{\mp}), (A_{2}^{\pm}, A_{2}^{\mp}), (A_{1}^{\pm}, B_{12}),$ &      \\
         && & & $(A_{2}^{\pm}, B_{12}),(B_{12},B_{12})$ & \\
        $\bar{4}^\prime m2^\prime$   &$C_{2a}$&  $B_{12}$   &   $A_1^+\oplus B_{12}$   &   $(A_{1}^{\pm}, A_{1}^{\pm}), (A_{2}^{\pm}, A_{2}^{\pm}), (A_{1}^{\pm}, B_{12}), $ &      \\
         && & & $(A_{2}^{\pm}, B_{12}),(B_{12},B_{12})$ & \\
        \hline[dashed]
        $mmm1^{\prime}$   &$E$&      &   $B_{1u}^-\oplus B_{2u}^-$   &   $(A_{g}^{\pm}, B_{1u}^{\mp}), (A_{g}^{\pm}, B_{2u}^{\mp}), (A_{g}^{\pm}, B_{3u}^{\mp}), (B_{1g}^{\pm}, A_{u}^{\mp}), $ &   $mmm$  \\
           &&      &   ${}\oplus B_{3u}^-$   &   $(B_{1g}^{\pm}, B_{2u}^{\mp}), (B_{1g}^{\pm}, B_{3u}^{\mp}), (B_{2g}^{\pm}, A_{u}^{\mp}), (B_{2g}^{\pm}, B_{1u}^{\mp}),$ &   $(D_{2h})$  \\
           &&      &      &   $ (B_{2g}^{\pm}, B_{3u}^{\mp}), (B_{3g}^{\pm}, A_{u}^{\mp}), (B_{3g}^{\pm}, B_{1u}^{\mp}), (B_{3g}^{\pm}, B_{2u}^{\mp})$ &     \\
        $4^\prime/mmm^\prime$   &$C_{2a}$&  $B_{23g},B_{23u}$   &   $B_{1u}^+\oplus B_{23u}$   &   $(A_g^{\pm}, B_{1u}^{\pm}), (A_g^{\pm}, B_{23u}), (B_{1g}^{\pm}, B_{23u}), $ &      \\
           &&      &      &   $(A_u^{\pm}, B_{1g}^{\pm}), (A_u^{\pm}, B_{23g}), (B_{1u}^{\pm}, B_{23g}),$  &     \\
           &&&&$(B_{23g},B_{23u})$&\\
        \hline\hline
    \end{tblr}
    \label{selection_table1_r}
\end{table*}

\begin{table*}
    \centering
   \caption{
   Classification of the Lifshitz invariant based on real corepresentations of magnetic point groups (MPGs) derived from tetragonal crystallographic point groups (PGs). The notations and column definitions are the same as in Table~\ref{selection_table1_r}.
   }
    \begin{tblr}{c|l|l|l|l|c}
    \hline\hline
        MPG   &$R_0$&  Parityless coreps   &   Corep of $D$   &   Allowed coreps of $(\Psi_i, \Psi_j)$ &   PG  \\
         \hline
        $41^{\prime}$   &$E$&      &   $A^-\oplus E^-$   &   $(A^{\pm}, A^{\mp}), (A^{\pm}, E^{\mp}), (B^{\pm}, B^{\mp}), (B^{\pm}, E^{\mp}), (E^{\pm},E^{\mp})$ &   $4$  \\
        $42^{\prime}2^{\prime}$   &$C_{2x}$&  $E$  &   $A^+\oplus E$   &   $(A^{\pm}, A^{\pm}), (A^{\pm}, E), (B^{\pm}, B^{\pm}), (B^{\pm}, E), (E,E)$  &   $(C_{4})$  \\
        $4/m^\prime$   &$I$&      &   $A^+\oplus E^+$   &   $(A^{\pm}, A^{\pm}), (A^{\pm}, E^{\pm}), (B^{\pm}, B^{\pm}), (B^{\pm}, E^{\pm}), (E^{\pm},E^{\pm})$ &      \\
        $4m^\prime m^\prime$   &$\sigma_{x}$& $E$    &   $A^-\oplus E$   &    $(A^{\pm}, A^{\mp}), (A^{\pm}, E), (B^{\pm}, B^{\mp}), (B^{\pm}, E), (E,E)$ &     \\
        \hline[dashed]
        $\bar{4}1^{\prime}$   &$E$&      &   $B^-\oplus E^-$   &   $(A^{\pm}, B^{\mp}), (A^{\pm}, E^{\mp}), (B^{\pm}, E^{\mp}), (E^{\pm},E^{\mp})$ &   $\bar{4}$  \\
        $4^{\prime}/m^{\prime}$    &$I$&      &   $B^+\oplus E^+$   &   $(A^{\pm}, B^{\pm}), (A^{\pm}, E^{\pm}), (B^{\pm}, E^{\pm}), (E^{\pm},E^{\pm})$ &   $(S_{4})$  \\
        $\bar{4}2^\prime m^\prime$   &$C_{2a}$&  $E$   &   $B^+\oplus E$   &   $(A^{\pm}, B^{\mp}), (A^{\pm}, E), (B^{\pm}, E), (E, E)$ &      \\
        \hline[dashed]
        $4/m1^{\prime}$   &$E$&      &   $A_u^-\oplus E_u^-$   &   $(A_g^{\pm}, A_u^{\mp}), (A_g^{\pm}, E_u^{\mp}), (A_u^{\pm}, E_g^{\mp}), (B_g^{\pm}, B_u^{\mp}), $ &   $4/m$  \\
           &&      &      &   $(B_g^{\pm}, E_u^{\mp}), (B_u^{\pm}, E_g^{\mp}), (E_g^{\pm},E_u^{\mp})$ &   $ (C_{4h})$  \\
        $4/mm^\prime m^\prime$   &$C_{2x}$&  $E_g,E_u$   &   $A_u^+\oplus E_u$   &   $(A_{g}^{\pm}, A_{u}^{\mp}), (A_{g}^{\pm}, E_{u}), (A_{u}^{\pm}, E_{g}), $ &      \\
           &&      &      &   $(B_{g}^{\pm}, B_{u}^{\mp}), (B_g^{\pm}, E_u), (B_u^{\pm}, E_g), (E_g, E_u)$ &     \\
        \hline[dashed]
        $4221^{\prime}$   &$E$&      &   $A_2^-\oplus E^-$   &   $(A_{1}^{\pm}, A_{2}^{\mp}), (A_{1}^{\pm}, E^{\mp}), (A_{2}^{\pm}, E^{\mp}), (B_{1}^{\pm}, B_{2}^{\mp}), $ &   $422$  \\
            &&      &      &   $(B_{1}^{\pm}, E^{\mp}), (B_{2}^{\pm}, E^{\mp}), (E^{\pm}, E^{\mp})$ &   $ (D_4)$  \\
        $4/m^\prime m^\prime m^\prime$   &$I$&      &   $A_2^+\oplus E^+$   &   $(A_{1}^{\pm}, A_{2}^{\pm}), (A_{1}^{\pm}, E^{\pm}), (A_{2}^{\pm}, E^{\pm}), (B_{1}^{\pm}, B_{2}^{\pm}), $ &      \\
           &&      &      &   $(B_{1}^{\pm}, E^{\pm}), (B_{2}^{\pm}, E^{\pm}), (E^{\pm}, E^{\pm})$ &     \\
        \hline[dashed]
        $4mm1^{\prime}$   &$E$&      &   $A_1^-\oplus E^-$   &   $(A_{1}^{\pm}, A_{1}^{\mp}), (A_{1}^{\pm}, E^{\mp}), (A_{2}^{\pm}, A_{2}^{\mp}), $ &   $4mm$  \\
           &&      &      & $(A_{2}^{\pm}, E^{\mp}), (B_{1}^{\pm}, B_{1}^{\mp})$, $(B_{1}^{\pm}, E^{\mp}),$ &   $ (C_{4v})    $  \\
           && & & $ (B_{2}^{\pm}, B_{2}^{\mp}), (B_{2}^{\pm}, E^{\mp}), (E^{\pm}, E^{\mp}) $ & \\
        $4/m^\prime mm$   &$I$&      &   $A_1^+\oplus E^+$   &   $(A_{1}^{\pm}, A_{1}^{\pm}), (A_{1}^{\pm}, E^{\pm}), (A_{2}^{\pm}, A_{2}^{\pm}), $ &      \\
           &&      &      &   $ (A_{2}^{\pm}, E^{\pm}), (B_{1}^{\pm}, B_{1}^{\pm}),(B_{1}^{\pm}, E^{\pm}), $ &     \\
           && & & $ (B_{2}^{\pm}, B_{2}^{\pm}), (B_{2}^{\pm}, E^{\pm}), (E^{\pm}, E^{\pm}) $ & \\
        \hline[dashed]
        $\bar{4}2m1^{\prime}$   &$E$&      &   $B_2^-\oplus E^-$   &   $(A_{1}^{\pm}, B_{2}^{\mp}), (A_{1}^{\pm}, E^{\mp}), (A_{2}^{\pm}, B_{1}^{\mp}), (A_{2}^{\pm}, E^{\mp}), $ &   $\bar{4}2m$  \\
            &&      &      &   $(B_{1}^{\pm}, E^{\mp}), (B_{2}^{\pm}, E^{\mp}), (E^{\pm}, E^{\mp})$ &   $ (D_{2d})$  \\
        $4^\prime/m^\prime m^\prime m$   &$I$&      &   $B_2^+\oplus E^+$   &   $(A_{1}^{\pm}, B_{2}^{\pm}), (A_{1}^{\pm}, E^{\pm}), (A_{2}^{\pm}, B_{1}^{\pm}), (A_{2}^{\pm}, E^{\pm}), $ &      \\
           &&      &      &   $(B_{1}^{\pm}, E^{\pm}), (B_{2}^{\pm}, E^{\pm}), (E^{\pm}, E^{\pm})$ &     \\
        \hline[dashed]
        $4/mmm1^{\prime}$   &$E$&      &   $A_{2u}^-\oplus E_u^-$   &   $(A_{1g}^{\pm}, A_{2u}^{\mp}), (A_{1g}^{\pm}, E_{u}^{\mp}), (A_{2g}^{\pm}, A_{1u}^{\mp}), (A_{2g}^{\pm}, E_{u}^{\mp}), $ &   $4/mmm$  \\
            &&      &      &   $(B_{1g}^{\pm}, B_{2u}^{\mp}),(B_{1g}^{\pm}, E_{u}^{\mp}), (B_{2g}^{\pm}, B_{1u}^{\mp}), $ &   $ (D_{4h})  $  \\
           &&      &      &   $(B_{2g}^{\pm}, E_{u}^{\mp}), (E_{g}^{\pm}, A_{1u}^{\mp}),(E_{g}^{\pm}, A_{2u}^{\mp}), $ \\
           && & & $ (E_{g}^{\pm}, B_{1u}^{\mp}), (E_{g}^{\pm}, B_{2u}^{\mp}), (E_{g}^{\pm}, E_{u}^{\mp}) $ &  \\ \hline\hline 
    \end{tblr}
    \label{selection_table2_r}
\end{table*}

\begin{table*}
    \centering
        \caption{
     Classification of the Lifshitz invariant based on real corepresentations of magnetic point groups (MPGs) derived from trigonal crystallographic point groups (PGs). The notations and column definitions are the same as in Table~\ref{selection_table1_r}.
        }
    \begin{tblr}{c|l|l|l|l|c}
    \hline\hline
        MPG   &     $R_0$    & Parityless coreps   &   Corep of $D$   &   Allowed coreps of $(\Psi_i, \Psi_j)$ &   PG  \\
         \hline
        $31^\prime$   &  $E$  &    &   $A^-\oplus E^-$   &   $(A^{\pm}, A^{\mp}), (A^{\pm}, E^{\mp}), (E^{\pm}, E^{\mp})$ &   $3$  \\
        $32^\prime$   &  $C_{21}^\prime$  &  $E$   &   $A^+\oplus E$   &   $(A^{\pm}, A^{\pm}), (A^{\pm}, E), (E,E)$ &   ($C_{3}$)   \\
        $3m^\prime$   &  $\sigma_{21}$  &  $E$    &   $A^-\oplus E$   &   $(A^{\pm}, A^{\mp}), (A^{\pm}, E), (E,E)$ &      \\
        $\bar{6}^\prime$   &  $\sigma_h$  &     &   $A^+\oplus E^-$   &   $(A^{\pm}, A^{\pm}), (A^{\pm}, E^{\mp}), (E^{\pm}, E^{\pm}), (E^{\pm}, E^{\mp})$ &      \\
        $6^\prime$   &  $C_{2z}$  &      &   $A^-\oplus E^+$   &   $(A^{\pm}, A^{\mp}), (A^{\pm}, E^{\pm}), (E^{\pm}, E^{\pm}), (E^{\pm}, E^{\mp})$ &      \\
        $\bar{3}^\prime$   &  $I$  &     &   $A^+\oplus E^+$   &   $(A^{\pm}, A^{\pm}), (A^{\pm}, E^{\pm}), (E^{\pm}, E^{\pm})$ &      \\
        \hline[dashed]
        $\bar{3}1^{\prime}$   &  $E$  &      &   $A_u^-\oplus E_u^-$   &   $(A_g^{\pm}, A_u^{\mp}), (A_g^{\pm}, E_u^{\mp}),$ &   $\bar{3}$  \\
         &&  & & $(A_u^{\pm}, E_g^{\mp}), (E_g^{\pm}, E_u^{\mp})$ & $ (S_6,C_{3i})$ \\
        $\bar{3}m^\prime$   &  $C_{21}^\prime$  &  $E_g,E_u$    &   $A_u^+\oplus E_u$   &   $(A_g^{\pm}, A_u^{\pm}), (A_g^{\pm}, E_u), (A_u^{\pm}, E_g), $ &    \\
            &   &   &   &  $(E_g,E_u)$ &   \\
        $6^\prime/m^\prime$   &  $C_{2z}$  &      &   $A_u^-\oplus E_u^+$   &   $(A_g^{\pm}, A_u^{\pm}), (A_g^{\pm}, E_u^{\pm}), $ &      \\
         &    & & & $(A_u^{\pm}, E_g^{\pm}), (E_g^{\pm}, E_u^{\pm}), (E_g^{\pm}, E_u^{\mp})$ & \\
        \hline[dashed]
        $321^{\prime}$   &  $E$  &      &   $A_2^-\oplus E^-$   &   $(A_{1}^{\pm}, A_{2}^{\mp}), (A_{1}^{\pm}, E^{\mp}), (A_{2}^{\pm}, E^{\mp}), (E^{\pm}, E^{\mp})$ &   $32$  \\
        $\bar{6}^\prime m^\prime 2$   &  $\sigma_h$  &      &   $A_2^+\oplus E^-$   &   $(A_{1}^{\pm}, A_{2}^{\mp}), (A_{1}^{\pm}, E^{\pm}), (A_{2}^{\pm}, E^{\mp}), (E^{\pm}, E^{\pm}), (E^{\pm}, E^{\mp})$ &   $(D_{3})$   \\
        $\bar{3}^\prime m^\prime$   &  $I$  &      &   $A_2^+\oplus E^+$   &   $(A_{1}^{\pm}, A_{2}^{\pm}), (A_{1}^{\pm}, E^{\pm}), (A_{2}^{\pm}, E^{\pm}), (E^{\pm}, E^{\pm})$ &      \\
        $6^\prime2^\prime2$   &  $C_{2z}$  &      &   $A_2^-\oplus E^+$   &   $(A_{1}^{\pm}, A_{2}^{\mp}), (A_{1}^{\pm}, E^{\pm}), (A_{2}^{\pm}, E^{\pm}), (E^{\pm}, E^{\pm}), (E^{\pm}, E^{\mp})$ &      \\
        \hline[dashed]
        $3m1^\prime$   &  $E$  &      &   $A_1^-\oplus E^-$   &   $(A_{1}^{\pm}, A_{1}^{\mp}), (A_{1}^{\pm}, E^{\mp}), (A_{2}^{\pm}, E^{\mp}), (E^{\pm}, E^{\mp})$ &   $3m$   \\
        $\bar{6}^\prime m2^\prime$   &  $\sigma_h$  &      &   $A_1^+\oplus E^-$   &   $(A_{1}^{\pm}, A_{1}^{\pm}), (A_{1}^{\pm}, E^{\mp}), (A_{2}^{\pm}, E^{\mp}), (E^{\pm}, E^{\pm}), (E^{\pm}, E^{\mp})$ &   $(C_{3v})$  \\
        $\bar{3}^\prime m$   &  $I$  &      &   $A_1^+\oplus E^+$   &   $(A_{1}^{\pm}, A_{1}^{\pm}), (A_{1}^{\pm}, E^{\pm}), (A_{2}^{\pm}, E^{\pm}), (E^{\pm}, E^{\pm})$ &      \\
        $6^\prime m^\prime m$   &  $C_{2z}$  &      &   $A_1^-\oplus E^+$   &   $(A_{1}^{\pm}, A_{1}^{\mp}), (A_{1}^{\pm}, E^{\pm}), (A_{2}^{\pm}, E^{\pm}), (E^{\pm}, E^{\pm}), (E^{\pm}, E^{\mp})$ &      \\
        \hline[dashed]
        $\bar{3}m1^\prime$   &  $E$  &      &   $A_{2u}^-\oplus E_u^-$   &   $(A_{1g}^{\pm}, A_{2u}^{\mp}), (A_{1g}^{\pm}, E_{u}^{\mp}), (A_{2g}^{\pm}, A_{1u}^{\mp}), (A_{2g}^{\pm}, E_{u}^{\mp}), $ &   $\bar{3}m$   \\
            &    &      &      &   $(E_{g}^{\pm}, A_{1u}^{\mp}), (E_{g}^{\pm}, A_{2u}^{\mp}), (E_{g}^{\pm}, E_{u}^{\mp})$ &   $ (D_{3d})$  \\
        $6^\prime/m^\prime m^\prime m$   &  $C_{2z}$  &      &   $A_{2u}^-\oplus E_u^+$   &   $(A_{1g}^{\pm}, A_{2u}^{\mp}), (A_{1g}^{\pm}, E_{u}^{\pm}), (A_{2g}^{\pm}, A_{1u}^{\mp}), (A_{2g}^{\pm}, E_{u}^{\pm}),$ &      \\
           &    &      &      &   $(E_{g}^{\pm}, A_{1u}^{\pm}), (E_{g}^{\pm}, A_{2u}^{\pm}), (E_{g}^{\pm}, E_{u}^{\pm}), (E_{g}^{\pm}, E_{u}^{\mp})$ \\ \hline\hline
    \end{tblr}
    \label{selection_table3_r}
\end{table*}

\begin{table*}
    \centering
    \caption{
    Classification of the Lifshitz invariant based on real corepresentations of magnetic point groups (MPGs) derived from hexagonal crystallographic point groups (PGs). The notations and column definitions are the same as in Table~\ref{selection_table1_r}.
    }
    \begin{tblr}{c|l|l|l|l|c}
    \hline\hline
        MPG   &     $R_0$    & Parityless coreps   &   Corep of $D$   &   Allowed coreps of $(\Psi_i, \Psi_j)$ &   PG  \\
         \hline
        $61^\prime$ & $E$  &    &   $A^-\oplus E_1^-$   &   $(A^{\pm}, A^{\mp}), (A^{\pm}, E_{1}^{\mp}), (B^{\pm}, B^{\mp}), (B^{\pm}, E_{2}^{\mp}), $ &   $6$   \\
           &   &    &      &   $(E_{1}^{\pm}, E_{1}^{\mp}), (E_{1}^{\pm},E_{2}^{\mp}), (E_{2}^{\pm},E_{2}^{\mp})$ &   $(C_{6})$  \\
        $62^\prime2^\prime$   &  $C_{21}$  & $E_1,E_2$    &   $A^+\oplus E_1$   &   $(A^{\pm}, A^{\pm}), (A^{\pm}, E_{1}), (B^{\pm}, B^{\pm}), (B^{\pm}, E_{2}), $ &      \\
           &   &      &    &   $(E_{1}, E_{1}), (E_{1}, E_{2}), (E_{2}, E_{2})$ &     \\
        $6/m^\prime$   &  $I$  &     &   $A^+\oplus E_1^+$   &   $(A^{\pm}, A^{\pm}), (A^{\pm}, E_{1}^{\pm}), (B^{\pm}, B^{\pm}), (B^{\pm}, E_{2}^{\pm}), $ &      \\
           &   &     &      &   $(E_{1}^{\pm}, E_{1}^{\pm}), (E_{1}^{\pm},E_{2}^{\pm}), (E_{2}^{\pm},E_{2}^{\pm})$ &     \\
        $6m^\prime m^\prime$   &  $\sigma_x$  &  $E_1,E_2$    &   $A^-\oplus E_1$   &   $(A^{\pm}, A^{\mp}), (A^{\pm}, E_{1}), (B^{\pm}, B^{\mp}), (B^{\pm}, E_{2}),$ &      \\
           &   &      &     &   $ (E_{1}, E_{1}), (E_{1}, E_{2}), (E_{2}, E_{2})$ &     \\
        \hline[dashed]
        $\bar{6}1^\prime$   & $E$  &    &   $A^{\prime\prime-}\oplus E^{\prime-}$   &   $(A^{\prime\pm}, A^{\prime\prime\mp}), (A^{\prime\pm}, E^{\prime\mp}), (A^{\prime\prime\pm}, E^{\prime\prime\mp}), $ &   $\bar{6}$   \\
            &   &     &      &   $(E^{\prime\pm}, E^{\prime\mp}), (E^{\prime}, E^{\prime\prime\mp}), (E^{\prime\prime\pm}, E^{\prime\prime\mp})$ &   $ (C_{3h})$  \\
        $\bar{6}m^\prime 2^\prime$   &  $C_{21}$  &  $E^\prime,E^{\prime\prime}$   &   $A^{\prime\prime-}\oplus E^{\prime}$   &   $(A^{\prime\pm}, A^{\prime\prime\mp}), (A^{\prime\pm}, E^{\prime}), (A^{\prime\prime\pm}, E^{\prime\prime}), $ &      \\
           &   &      &      &   $(E^{\prime}, E^{\prime}), (E^{\prime}, E^{\prime\prime}), (E^{\prime\prime}, E^{\prime\prime})$ &     \\
        $6^\prime/m$   &  $I$  &      &   $A^{\prime\prime+}\oplus E^{\prime+}$   &   $(A^{\prime\pm}, A^{\prime\prime\pm}), (A^{\prime\pm}, E^{\prime\pm}), (A^{\prime\prime\pm}, E^{\prime\prime\pm}), $ &      \\
           &   &     &      &   $(E^{\prime\pm}, E^{\prime\pm}), (E^{\prime\pm}, E^{\prime\prime\pm}), (E^{\prime\prime\pm}, E^{\prime\prime\pm})$ &     \\
        \hline[dashed]
        $6/m1^\prime$   & $E$  &     &   $A_u^-\oplus E_{1u}^-$   &   $(A_g^{\pm}, A_u^{\mp}), (A_g^{\pm}, E_{1u}^{\mp}), (B_g^{\pm}, B_u^{\mp}), $ &   $6/m$   \\
            &   &     &      &   $(B_g^{\pm}, E_{2u}^{\mp}),(A_u^{\pm}, E_{1g}^{\mp}), (B_u^{\pm}, E_{2g}^{\mp}), $ &   $(C_{6h})$  \\
           &   &   &      &   $(E_{1g}^{\pm}, E_{1u}^{\mp}), (E_{1g}^{\pm},E_{2u}^{\mp}),  (E_{1u}^{\pm},E_{2g}^{\mp}),(E_{2g}^{\pm},E_{2u}^{\mp})$ &     \\
        $6/mm^\prime m^\prime$   &  $C_{21}$  &  $E_{1g},E_{1u},$   & $A_u^+\oplus E_{1u}$  &$(A_g^{\pm}, A_u^{\pm}), (A_g^{\pm}, E_{1u}), (A_u^{\pm}, E_{1g}), (B_g^{\pm}, B_u^{\pm}), (B_g^{\pm}, E_{2u}), $ &      \\
           &   &   $E_{2g},E_{2u}$   &     &  $(B_u^{\pm}, E_{2g}), (E_{1g}, E_{1u}),$   &     \\
           &   &    &   &$(E_{1g},E_{2u}),  (E_{1u},E_{2g}), (E_{2g},E_{2u})$   &   \\
        \hline[dashed]
        $6221^\prime$   & $E$  &      &   $A_2^-\oplus E_1^-$   &   $(A_{1}^{\pm}, A_{2}^{\mp}), (A_{1}^{\pm}, E_{1}^{\mp}), (A_{2}^{\pm}, E_{1}^{\mp}), (B_{1}^{\pm}, B_{2}^{\mp}), (B_{1}^{\pm}, E_{2}^{\mp}), $ &   $622$   \\
            &   &      &      &   $(B_{2}^{\pm}, E_{2}^{\mp}), (E_{1}^{\pm}, E_{1}^{\mp}), (E_{1}^{\pm}, E_{2}^{\mp}), (E_{2}^{\pm}, E_{2}^{\mp})$ &   $(D_{6})$  \\
        $6/m^\prime m^\prime m^\prime$   &  $I$  &      &   $A_2^+\oplus E_1^+$   &   $(A_{1}^{\pm}, A_{2}^{\pm}), (A_{1}^{\pm}, E_{1}^{\pm}), (A_{2}^{\pm}, E_{1}^{\pm}), (B_{1}^{\pm}, B_{2}^{\pm}), (B_{1}^{\pm}, E_{2}^{\pm}), $ &      \\
           &   &      &      &   $(B_{2}^{\pm}, E_{2}^{\pm}), (E_{1}^{\pm}, E_{1}^{\pm}), (E_{1}^{\pm}, E_{2}^{\pm}), (E_{2}^{\pm}, E_{2}^{\pm})$ &     \\
        \hline[dashed]
        $6mm1^\prime$   & $E$  &      &   $A_1^-\oplus E_1^-$   &   $(A_{1}^{\pm}, A_{1}^{\mp}), (A_{1}^{\pm}, E_{1}^{\mp}), (A_{2}^{\pm}, A_{2}^{\mp}), (A_{2}^{\pm}, E_{1}^{\mp}), $ &   $6mm$  \\
           &   &      &      &   $(B_{1}^{\pm}, B_{1}^{\mp}), (B_{1}^{\pm}, E_{2}^{\mp}), (B_{2}^{\pm}, B_{2}^{\mp}), (B_{2}^{\pm}, E_{2}^{\mp}), $ &   $(C_{6v})$  \\
           &   &      &      &   $(E_{1}^{\pm}, E_{1}^{\mp}), (E_{1}^{\pm}, E_{2}^{\mp}), (E_{2}^{\pm}, E_{2}^{\mp})$ &     \\
        $6/m^\prime mm$   &  $I$  &      &   $A_1^+\oplus E_1^+$   &   $(A_{1}^{\pm}, A_{1}^{\pm}), (A_{1}^{\pm}, E_{1}^{\pm}), (A_{2}^{\pm}, A_{2}^{\pm}), (A_{2}^{\pm}, E_{1}^{\pm}), $ &      \\
           &   &      &      &   $(B_{1}^{\pm}, B_{1}^{\pm}), (B_{1}^{\pm}, E_{2}^{\pm}), (B_{2}^{\pm}, B_{2}^{\pm}), (B_{2}^{\pm}, E_{2}^{\pm}), $ &     \\
           &   &      &      &   $(E_{1}^{\pm}, E_{1}^{\pm}), (E_{1}^{\pm}, E_{2}^{\pm}), (E_{2}^{\pm}, E_{2}^{\pm})$ &     \\
        \hline[dashed]
        $\bar{6}m21^\prime$   & $E$  &      &   $A_2^{\prime\prime-}\oplus E^{\prime-}$   &   $(A_{1}^{\prime\pm}, A_{2}^{\prime\prime\mp}), (A_{1}^{\prime\pm}, E^{\prime\mp}), (A_{1}^{\prime\prime\pm}, A_{2}^{\prime\mp}), $ &   $\bar{6}m2$   \\
           &   &      &      &   $(A_{1}^{\prime\prime\pm}, E^{\prime\prime\mp}), (A_{2}^{\prime\pm}, E^{\prime\mp}), (A_{2}^{\prime\prime\pm}, E^{\prime\prime\mp}), $ &   $(D_{3h})$   \\
           &   &      &      &   $(E^{\prime\pm}, E^{\prime\mp}), (E^{\prime\pm}, E^{\prime\prime\mp}), (E^{\prime\prime\pm}, E^{\prime\prime\mp})$ &     \\
        $6^\prime/mmm^\prime$   &  $I$  &      &   $A_2^{\prime\prime+}\oplus E^{\prime+}$   &   $(A_{1}^{\prime\pm}, A_{2}^{\prime\prime\pm}), (A_{1}^{\prime\pm}, E^{\prime\pm}), (A_{1}^{\prime\prime\pm}, A_{2}^{\prime\pm}), $ &      \\
           &   &      &      &   $(A_{1}^{\prime\prime\pm}, E^{\prime\prime\pm}), (A_{2}^{\prime\pm}, E^{\prime\pm}), (A_{2}^{\prime\prime\pm}, E^{\prime\prime\pm}), $ &     \\
           &   &      &      &   $(E^{\prime\pm}, E^{\prime\pm}), (E^{\prime\pm}, E^{\prime\prime\pm}), (E^{\prime\prime\pm}, E^{\prime\prime\pm})$ &     \\
        \hline[dashed]
        $6/mmm1^\prime$   & $E$  &      &   $A_{2u}^-\oplus E_{1u}^-$   &   $(A_{1g}^{\pm}, A_{2u}^{\mp}), (A_{1g}^{\pm}, E_{1u}^{\mp}), (A_{2g}^{\pm}, A_{1u}^{\mp}), (A_{2g}^{\pm}, E_{1u}^{\mp}), $ &   $6/mmm$   \\
            &   &      &      &   $(B_{1g}^{\pm}, B_{2u}^{\mp}), (B_{1g}^{\pm}, E_{2u}^{\mp}), (B_{2g}^{\pm}, B_{1u}^{\mp}), (E_{2g}^{\pm}, E_{2u}^{\mp}), $ &   $ (D_{6h})  $  \\
           &   &      &      &   $(E_{1g}^{\pm}, A_{1u}^{\mp}), (E_{1g}^{\pm}, A_{2u}^{\mp}), (E_{1g}^{\pm}, E_{1u}^{\mp}), (E_{1g}^{\pm}, E_{2u}^{\mp}), $ &     \\
           &   &      &      &   $(E_{2g}^{\pm}, B_{1u}^{\mp}), (E_{2g}^{\pm}, B_{2u}^{\mp}), (E_{2g}^{\pm}, E_{1u}^{\mp}), (B_{2g}^{\pm}, E_{2u}^{\mp})$ \\ \hline\hline
    \end{tblr}
    \label{selection_table4_r}
\end{table*}

\begin{table*}
    \centering
    \caption{
    Classification of the Lifshitz invariant based on real corepresentations of magnetic point groups (MPGs) derived from cubic crystallographic point groups (PGs). The notations and column definitions are the same as in Table~\ref{selection_table1_r}.
    In the third column, we also indicate the cases in which the matrix $\bm N$ in Eq.~(\ref{eq:corepmat_a_b}) becomes non-identity.
    }
    \begin{tblr}{c|l|l|l|l|c}
    \hline\hline
        MPG   &$R_0$&  Parityless coreps   &   Corep of $D$   &   Allowed coreps of $(\Psi_i, \Psi_j)$ &   PG  \\
    \hline
        $231^\prime$   &  $E$  &      &   $T^-$   &   $(A^{\pm}, T^{\mp}), (E^{\pm}, T^{\mp}), (T^{\pm}, T^{\mp})$ &   $23$   \\
        $m^\prime\bar{3}^\prime$   &  $I$  &      &   $T^+$   &   $(A^{\pm}, T^{\pm}), (E^{\pm}, T^{\pm}), (T^{\pm}, T^{\pm})$ &   $(T)$  \\
        $\bar{4}^\prime3m^\prime$   &  $\sigma_x$  &  $E$, $\bm{N}\neq\bm{1}:T$   &   $T^-$   &   $(A^{\pm}, T^{\mp}), (E, T^{\pm}), (T^{\pm}, T^{\mp})$ &      \\
        $4^\prime32^\prime$   &  $C_{2x}$  &  $E$, $\bm{N}\neq\bm{1}:T$   &   $T^+$   &   $(A^{\pm}, T^{\mp}), (E, T^{\pm}), (T^{\pm}, T^{\mp})$ &      \\
        \hline[dashed]
        $m\bar{3}1^\prime$   &  $E$  &      &   $T_u^-$   &   $(A_g^{\pm}, T_u^{\mp}), (A_u^{\pm}, T_g^{\mp}), (E_g, T_u^{\pm}),$ &   $m\bar{3}$  \\
            &    &   &      &   $(E_u, T_g^{\pm}), (T_g^{\pm}, T_u^{\mp})$ &   $ (T_h)$  \\
        $m\bar{3}m^\prime$   &  $C_{2x}$  &  $E_g,E_u$,   &   $T_u^+$   &   $(A_g^{\pm}, T_u^{\mp}), (A_u^{\pm}, T_g^{\mp}), (E_g, T_u^{\pm}),$ &      \\
           &    &   $\bm{N}\neq\bm{1}:T_g,T_u$    &      &   $(E_u, T_g^{\pm}), (T_g^{\pm}, T_u^{\mp})$ &     \\
        \hline[dashed]
        $4321^\prime$   &  $E$  &      &   $T_1^-$   &   $(A_{1}^{\pm}, T_{1}^{\mp}), (A_{2}^{\pm}, T_{2}^{\mp}), (E^{\pm}, T_{1}^{\mp}), (E^{\pm}, T_{2}^{\mp}), $ &   $432$   \\
           &    &      &      &   $(T_{1}^{\pm}, T_{1}^{\mp}), (T_{1}^{\pm}, T_{2}^{\mp}), (T_{2}^{\pm}, T_{2}^{\mp})$ &   $(O)$  \\
        $m^\prime\bar{3}m^\prime$   &  $I$  &      &   $T_1^+$   &   $(A_{1}^{\pm}, T_{1}^{\pm}), (A_{2}^{\pm}, T_{2}^{\pm}), (E^{\pm}, T_{1}^{\pm}), (E^{\pm}, T_{2}^{\pm}), $ &      \\
           &    &      &      &   $(T_{1}^{\pm}, T_{1}^{\pm}), (T_{1}^{\pm}, T_{2}^{\pm}), (T_{2}^{\pm}, T_{2}^{\pm})$ &     \\
        \hline[dashed]
        $\bar{4}3m1^\prime$   &  $E$  &      &   $T_2^-$   &   $(A_{1}^{\pm}, T_{2}^{\mp}), (A_{2}^{\pm}, T_{1}^{\mp}), (E^{\pm}, T_{1}^{\mp}), (E^{\pm}, T_{2}^{\mp}), $ &  $\bar{4}3m$    \\
            &    &      &      &   $(T_{1}^{\pm}, T_{1}^{\mp}), (T_{1}^{\pm}, T_{2}^{\mp}), (T_{2}^{\pm}, T_{2}^{\mp})$ &   $ (T_d)$  \\
        $m^\prime\bar{3}m$   &  $I$  &      &   $T_2^+$   &   $(A_{1}^{\pm}, T_{2}^{\pm}), (A_{2}^{\pm}, T_{1}^{\pm}), (E^{\pm}, T_{1}^{\pm}), (E^{\pm}, T_{2}^{\pm}), $ &      \\
           &    &      &      &   $(T_{1}^{\pm}, T_{1}^{\pm}), (T_{1}^{\pm}, T_{2}^{\pm}), (T_{2}^{\pm}, T_{2}^{\pm})$ &     \\
        \hline[dashed]
        $m\bar{3}m1^\prime$   &  $E$  &      &   $T_{1u}^-$   &   $(A_{1g}^{\pm}, T_{1u}^{\mp}), (A_{2g}^{\pm}, T_{2u}^{\mp}), (E_{g}^{\pm}, T_{1u}^{\mp}), (E_{g}^{\pm}, T_{2u}^{\mp}), $ &   $m\bar{3}m$   \\
            &    &      &      &   $(T_{1g}^{\pm}, A_{1u}^{\mp}), (T_{1g}^{\pm}, E_{u}^{\mp}), (T_{1g}^{\pm}, T_{1u}^{\mp}), (T_{1g}^{\pm}, T_{2u}^{\mp}), $ &   $ (O_h)$  \\
           &    &      &      &   $(T_{2g}^{\pm}, A_{2u}^{\mp}), (T_{2g}^{\pm}, E_{u}^{\mp}), (T_{2g}^{\pm}, T_{1u}^{\mp}), (T_{2g}^{\pm}, T_{2u}^{\mp})$ \\ \hline\hline
    \end{tblr}
    \label{selection_table5_r}
\end{table*}
In this subsection, we discuss the corepresentations of the Lifshitz invariant.
Once a corepresentation of the order parameter $\Psi$ and the covariant derivative $D$ is identified for a specific basis set—denoted as $\mathrm{D}\Gamma^{\Psi}$ and $\mathrm{D}\Gamma^{D}$, respectively—the corepresentation of the Lifshitz invariant is given by 
the Kronecker product $\mathrm{D}\Gamma^{\Psi^*}\otimes \mathrm{D}\Gamma^{D}\otimes \mathrm{D}\Gamma^{\Psi}$.
As in the case of representations, the Lifshitz invariant is allowed under the symmetry if the irreducible decomposition of $\mathrm{D}\Gamma^{\Psi^*}\otimes \mathrm{D}\Gamma^{D}\otimes \mathrm{D}\Gamma^{\Psi}$ contains the trivial corepresentation.

In general, the physical construction of the basis of the corepresentation of the order parameter depends on the details of the system.
For example, let us consider a multiband superconducting system with two distinct configurations of the order parameters, $\psi_i$ and $\phi_i$ ($i$ labels the lattice sites). Suppose that the corresponding basis set $\ket{\psi_i}$ and $\ket{\phi_i}$ transforms under the TR operation as $A\ket{\psi_i}=\theta R_0\ket{\psi_i}=\ket{\phi_i}$.
The corepresentation $\mathrm{D}\Gamma^{\Psi}$ can be constructed in such a way that the order parameters transform in the same way as the coordinates of the corresponding lattice sites $\bm r_i$ under the operation of $G$.
We remark that $\psi_i$ and $\phi_i$ may reside on different unit cells in the case of black-and-white point groups, whose anti-unitary operation $A$ has a spatial component $R_0\neq E$.
A similar kind of construction of order parameter representations is known for the case of point groups \cite{Nagashima2024}, where the obtained representation corresponds to the {\it induced representation}, induced by the trivial representation of the site-symmetry group.

One can see that the corepresentation matrices are block diagonal for unitary operations $R\in G$ and block off-diagonal for anti-unitary operations $B\in AG$, since both $\mathrm{Span}\{\ket{\psi_i}\}$ and $\mathrm{Span}\{\ket{\phi_i}\}$ are invariant under $R$, but are exchanged by $B$. These corepresentation matrices can be written in terms of the permutation matrix $\bm\Delta(R)$ of $\{\ket{\psi_i}\}$ for $R\in G$. 
Since $\bm\Delta(R)$ is a representation matrix, this can be block-diagonalized into irreducible representation matrices as $\bm\Delta(R)=U[\bigoplus_\Gamma\bm\Delta^\Gamma(R)]U^{\dagger}$ for \textit{complex} corepresentation. 
Consequently, the corepresentation matrices can be diagonalized as follows:
\begin{widetext}
\begin{align}    
    \bm D(R)&=
    \begin{pmatrix}
        \bm\Delta(R)&\bm 0\\
        \bm 0&\bm\Delta^*(A^{-1}RA)
    \end{pmatrix}
    \begin{pmatrix}
        U&\bm 0\\
        \bm 0&U^*
    \end{pmatrix}
    \begin{pmatrix}
        \bigoplus_\Gamma\bm\Delta^\Gamma(R)&\bm 0\\
        \bm 0&\bigoplus_\Gamma\bm\Delta^{\Gamma*}(A^{-1}RA)
    \end{pmatrix}
    \begin{pmatrix}
        U^{\dagger}&\bm 0\\
        \bm 0&U^{T}
    \end{pmatrix},\label{corepmat_decomposition_R}\\
\bm D(B)&=
    \begin{pmatrix}
        \bm 0&\bm\Delta(BA)\\
        \bm\Delta^*(A^{-1}B)&\bm 0
    \end{pmatrix}
    \begin{pmatrix}
        U&\bm 0\\
        \bm 0&U^*
    \end{pmatrix}
    \begin{pmatrix}
        \bm 0&\bigoplus_\Gamma\bm\Delta^\Gamma(BA)\\
        \bigoplus_\Gamma\bm\Delta^{\Gamma*}(A^{-1}B)&\bm 0
    \end{pmatrix}
    \begin{pmatrix}
        U^{\dagger}&\bm 0\\
        \bm 0&U^{T}
    \end{pmatrix}^*.\label{corepmat_decomposition_B}
\end{align}
\end{widetext}
One should be careful about the anti-unitary nature of the transformation $B$.
After the transformation, one permutes rows and columns to combine blocks belonging to the same $\Gamma$ and obtain block-diagonalized corepresentation matrices,
\begin{align}
    \tilde{U}^\dagger\bm D(R)\tilde{U}&=\bigoplus_\Gamma
    \begin{pmatrix}
        \bm\Delta^\Gamma(R)&\bm0\\
        \bm0&\bm\Delta^{\Gamma*}(A^{-1}RA)
    \end{pmatrix},\label{eq:blockdiag_r}\\
    \tilde{U}^\dagger\bm D(B)\tilde{U}^*&=\bigoplus_\Gamma
    \begin{pmatrix}
        \bm0&\bm\Delta^\Gamma(BA)\\
        \bm\Delta^{\Gamma*}(A^{-1}B)&\bm0
    \end{pmatrix},\label{eq:blockdiag_b}
   \end{align}
where $\tilde{U}=\mathrm{diag}(U,U^*)\times\text{(permutation matrix)}$.
Then, each block can be transformed into the irreducible corepresentation matrices, described in Eq. \eqref{eq:corepmat_a_r}-\eqref{eq:corepmat_c_b} for the complex corepresentation, depending on the type of $\Gamma$ as described in \cite{Bradley1972}. 

In this way, we can obtain the irreducible decomposition of the complex corepresentation $\mathrm{D}\Gamma^\Psi$.

The corepresentation of the complex conjugate of the order parameter, $\mathrm{D}\Gamma^{\Psi^*}$, can be constructed directly from $\mathrm{D}\Gamma^{\Psi}$. 
Since we can obtain $\mathrm{D}\Gamma^{\Psi^*}$ by taking the complex conjugate of the corepresentation matrices (Eq. \eqref{corepmat_decomposition_R}-\eqref{eq:blockdiag_b}), 
the resulting $\mathrm{D}\Gamma^{\Psi^*}$ depends on the reality properties of $\bm \Delta^\Gamma(R)$.
For $\mathrm{D}\Gamma$ with real $\bm\Delta^\Gamma(R)$, the conjugation does not make any difference in the blocks in Eq.~\eqref{eq:blockdiag_r} and \eqref{eq:blockdiag_b}, so $\mathrm{D}\Gamma$ appears in the decomposition of $\mathrm{D}\Gamma^{\Psi^*}$.

In the case of complex $\bm\Delta^\Gamma(R)$, 
we have to further classify the cases according to the type of the corepresentation. We note that the complex representation always appears in a complex conjugate pair in the context of point groups
(the conjugate of $\Gamma$ is denoted as $\Gamma^*$).
In fact, the complex conjugate pairs are always ${}^1E$ and ${}^2E$ (which correspond to the diagonalized form $E={}^{1}E\oplus{}^{2}E$ of the two-dimensional representation $E$).

If $\bm\Delta^\Gamma(R)$ is complex and $\mathrm{D}\Gamma$ is of type (a) or (b), we can see by conjugating Eq.~\eqref{eq:corepmat_a_r}-\eqref{eq:corepmat_b_b} that $\mathrm{D}\Gamma^{*}$ appears in $\mathrm{D}\Gamma^{\Psi^*}$, corresponding to $\mathrm{D}\Gamma$ in $\mathrm{D}\Gamma^{\Psi}$. We note that $\bm N^{\mathrm{D}\Gamma}$ are real matrices in this case.
On the other hand, for the case of type (c) $\mathrm{D}\Gamma$, we have to consider the relation between $\Gamma^*$ and $\Gamma^\prime$. 
Since the magnetic point groups with type (c) complex corepresentation satisfy the condition $A^{-1}RA=R$ for all $A$ with $A^2=E$, we can see that $\Gamma^*=\Gamma^\prime$:
\begin{align}
    \bm\Delta^{\Gamma'}(R) &= \left[\bm\Delta^{\Gamma}(A^{-1}RA)\right]^{*} \notag \\
    &= \left[ \bm\Delta^{\Gamma}(R)\right]^{*} \notag \\
    &= \bm\Delta^{\Gamma *}(R).
\end{align}
We also see for $B=RA$,
\begin{equation}
    BA = (RA)A = ARA =A^{-1}B,
\end{equation}
where we used $A^{2}=E$.
In addition, the following unitary transformation
\begin{equation}
    U =\bigoplus_\Gamma \begin{pmatrix}
        \bm{0} & \bm{1}_{d(\Gamma)} \\ \bm{1}_{d(\Gamma)} & \bm{0}
    \end{pmatrix},    
\end{equation}
maps the conjugated block $\Gamma^{*}$ into the original block $\Gamma$, where $d(\Gamma)$ is the dimension of $\Gamma$:
\begin{align}
    U^\dagger\bm{D}(R)U 
    =&\bigoplus_\Gamma
    \begin{pmatrix}
        \bm\Delta^{\Gamma*}(A^{-1}RA)&\bm0\\
        \bm0&\bm\Delta^\Gamma(R)
    \end{pmatrix}=\bm{D}^\ast(R),\\
    U^\dagger\bm{D}(B)U^\ast 
    =&\bigoplus_\Gamma
    \begin{pmatrix}
        \bm0&\bm\Delta^{\Gamma*}(A^{-1}B)\\
        \bm\Delta^\Gamma(BA)&\bm0
    \end{pmatrix}=\bm{D}^\ast(R).
\end{align}
Therefore, the Kronecker decomposition of $\mathrm{D}\Gamma^{\Psi*}$ can be obtained by taking the complex conjugate of the corepresentations of the type (a) or (b), while keeping type (c) corepresentations.
This effectively means to consider the case where ${}^{1}E$ and ${}^{2}E$ are swapped.
This decomposition method can be simplified by utilizing the correspondence between the representation and the corepresentation. First, we calculate the set of $\Gamma$, i.e., the irreducible decomposition of the representation of $\Psi$ on $G$. Using the orthogonality of characters of the irreducible representations, the set of $\Gamma$ can be obtained by comparing the characters of the permutation matrices $\bm\Delta(R)$ (the number of sites unchanged by $R$ in the unit cell), and those of the irreducible representations of $G$.

Once we get every $\Gamma$ in the decomposition of the representation, we can use the correspondence between representation $\Gamma$ and corepresentation $\mathrm{D}\Gamma$, which is described in Section~\ref{sec.IIIA}.
First we explain the case of the complex corepresentation.
If $\Gamma$ is of type (a) (here we denote it as $\Gamma_a$), the block characterized by $\Gamma_a$ in Eq.~\eqref{corepmat_decomposition_B} can be further block diagonalized into $2\mathrm{D}\Gamma_a$. If $\Gamma$ is of type (b), the block can be transformed into a type (b) corepresentation $\mathrm{D}\Gamma_b$. If $\Gamma$ is of type (c), the decomposition of the representation includes a pair of $\Gamma_c\oplus\Gamma^\prime_c$, and the paired blocks are transformed into a type (c) corepresentation $2\mathrm{D}\Gamma_c$.
This procedure to derive the decomposition of corepresentation is summarized as 
\begin{equation}\label{eq:rep_to_corep_Psi}
\Gamma_a\rightarrow 2\mathrm{D}\Gamma_a,\ \Gamma_b\rightarrow \mathrm{D}\Gamma_b,\ \Gamma_c\oplus\Gamma_c^\prime\rightarrow 2\mathrm{D}\Gamma_c.
\end{equation}
The corepresentation of the covariant derivative $D$ can be derived similarly. 
Since the spatial derivatives $\partial_{x,y,z}$ and the coordinates $(x,y,z)$ transform identically, the representation of $D$ is given as the representation of the coordinates, which is tabulated for every point group \cite{Atkins1970}, and summarized in the third row of Tables~\ref{selection_table1_c}
-\ref{selection_table5_c}.

For \textit{real} corepresentations, it suffices to replace the unitary matrices $U$ in Eqs.~\eqref{corepmat_decomposition_R}-\eqref{eq:blockdiag_b} with orthogonal matrices $O$. Each block in Eqs.~\eqref{eq:blockdiag_r} and \eqref{eq:blockdiag_b} then corresponds to a real representation $\Gamma_\mathrm{real}$ of $G$.  
Accordingly, the real corepresentations of $M$ can be constructed from a complex representation $\Gamma_\mathrm{comp}$ of $G$ in two steps. First, one determines $\Gamma_\mathrm{real}$ from $\Gamma_\mathrm{comp}$. Second, one further block-diagonalizes the matrices in Eqs. ~\eqref{eq:blockdiag_r} and \eqref{eq:blockdiag_b} into irreducible corepresentation matrices.

The first step is governed by the Frobenius-Schur indicator~\cite{serre1977linear},
\begin{equation}
    \frac{1}{|G|}\sum_{R\in G}\chi^{\Gamma_\mathrm{comp}}(R^2)=
    \begin{cases}
    +1&\text{for real type,}\\
    0&\text{for complex type,}\\
    -1&\text{for quarternionic type,}
    \end{cases}
    \label{frobenius_schur}
\end{equation}
which determines the correspondence between complex and real representations.
A real-type complex representation is identical to its real counterpart, while a complex-type complex representation combines with its conjugate to form a real representation. 
By comparing Eq.~\eqref{wigner_criterion} with Eq.~\eqref{frobenius_schur}, one finds that the real, complex, and quaternionic cases of the Frobenius-Schur indicator for $G$ correspond to types (a), (c), and (b), respectively, for the gray magnetic group $G+\theta G$. In particular, complex-type pairs ${}^{1}E,{}^{2}E$ always merge into a single real representation $E$, while quaternionic types do not occur for crystallographic point groups.
Thus, the real representation $\Gamma_\mathrm{real}$ of the order parameter is obtained from the complex representation by replacing ${}^{1}E\oplus{}^{2}E$ with $E$.

The second step is to construct the real irreducible corepresentation $\mathrm{D\Gamma_{real}}$ of $M$ from $\Gamma_\mathrm{real}$ of $G$. 
If the corresponding corepresentation carries a well-defined parity ($\mathrm{D\Gamma_{real}^\pm}$), the block associated with $\Gamma_\mathrm{real}$ in Eqs.~\eqref{eq:blockdiag_r} and \eqref{eq:blockdiag_b} decomposes as $\mathrm{D\Gamma_{real}^+}\oplus\mathrm{D\Gamma_{real}^-}$. 
This decomposition can be understood as follows. Let $d=\dim \Gamma_\mathrm{real}$.
Each block in Eqs.~\eqref{eq:blockdiag_r} and \eqref{eq:blockdiag_b}, given explicitly in Eq.~\eqref{eq:block}, has dimension $2d$. Since $\mathrm{D\Gamma_{real}^\pm}$ and $\Gamma_\mathrm{real}$ have the same dimension $d$, the $2d$-dimensional block necessarily decomposes into a direct sum of two $\mathrm{D\Gamma_{real}^\pm}$ components,
\begin{equation}
        \begin{pmatrix}
        \bm\Delta^\Gamma(R)&\bm0\\
        \bm0&\bm\Delta^{\Gamma*}(A^{-1}RA)
    \end{pmatrix},
    \begin{pmatrix}
        \bm0&\bm\Delta^\Gamma(BA)\\
        \bm\Delta^{\Gamma*}(A^{-1}B)&\bm0
    \end{pmatrix}.\label{eq:block}
\end{equation}
Here, the upper and lower half columns of the block are spanned by $\{\ket{\psi_i}\}$ and $\{\ket{\phi_i}\}$ with $\ket{\phi_i}:=A\ket{\psi_i}$ ($i=1,\ldots,d$), respectively, 
while the irreducible corepresentations $\mathrm{D\Gamma_{real}^\pm}$ are spanned by $\{\ket{\psi_i}\pm\ket{\phi_i}\}$.
If the block were to be decomposed into $2\mathrm{D\Gamma_{real}^+}$, a direct sum of irreducible corepresentations of the same parity, the decomposed block would have only $d$ independent basis functions. This redundancy of $\{\ket{\psi_1}+\ket{\phi_1},\ldots\}$ contradicts the properties of an orthogonal transformation. Therefore, the block must decompose into $\mathrm{D\Gamma_{real}^+}\oplus\mathrm{D\Gamma_{real}^-}$, a pair of different parity corepresentations.
If the corepresentation associated with $\Gamma_\mathrm{real}$ is parityless $(\mathrm{D\Gamma_{real}})$, the block in Eq.~\eqref{eq:block} is already irreducible and corresponds directly to $\mathrm{D\Gamma_real}$. 

This decomposition procedure is summarized as follows:
\begin{equation}
    \begin{cases}
    \Gamma\rightarrow\mathrm{D\Gamma^+}\oplus\mathrm{D\Gamma^-}&\text{for $\mathrm{D\Gamma}$ with parity,}\\
    \Gamma\rightarrow2\mathrm{D\Gamma}&\text{for  parityless $\mathrm{D\Gamma}$.}
    \end{cases}
\label{real_corep_decomposition}
\end{equation}
This allows us to systematically construct real corepresentations of the order parameters.
Finally, complex conjugation on the order parameter $\Psi^*$ does not alter its real corepresentation, since the corepresentation matrices themselves are real.

Once the irreducible decompositions of the corepresentations $\mathrm{D}\Gamma^\Psi$ and $\mathrm{D}\Gamma^D$ are obtained, we take their Kronecker product and decompose it into irreducible components.
In Appendix \ref{Appendix A}, we explicitly provide the product tables of complex corepresentations of types (b) and (c) for magnetic point groups, as such tables are scarcely available in the literature.
For type (a), we instead refer to the well-established product tables of ordinar point groups.
We also list the product tables for real parityless corepresentations in Appendix \ref{Appendix A}.

\subsection{Classification of the Lifshitz invariant}

By combining the Kronecker-product decomposition rules with the irreducible corepresentations of $\Psi$ and $D$, 
we can systematically classify all Lifshitz invariants (both type I and type II) allowed by the symmetries of magnetic point groups.
The results are summarized in Tables \ref{selection_table1_c}--\ref{selection_table5_r} for all gray and black-and-white magnetic point groups. Tables \ref{selection_table1_c}--\ref{selection_table5_c} present the classification based on the complex corepresentations, whereas Tables~\ref{selection_table1_r}--\ref{selection_table5_r} correspond to real corepresentation.
Tables~\ref{selection_table1_c} and~\ref{selection_table1_r} include magnetic point groups derived from triclinic, monoclinic,
and orthorhombic crystallographic point groups. 
Tables~\ref{selection_table2_c} and~\ref{selection_table2_r}, \ref{selection_table3_c} and~\ref{selection_table3_r}, \ref{selection_table4_c} and~\ref{selection_table4_r}, \ref{selection_table5_c} and~\ref{selection_table5_r} list those derived from tetragonal, trigonal, hexagonal and cubic point groups, respectively.
These tables enumerate all pairs of irreducible corepresentations of the order parameter $\Psi$ that allow a Lifshitz invariant in the GL free energy. 
For the complex-corepresentation tables,
we additionally indicate the type (b) and (c) complex corepresentations (when present). 
For the real-corepresentation tables,
we list the representative anti-unitary operator $A$, identify real corepresentations lacking well-defined $A$-parity, and provide the corresponding corepresentations of $D$, together with the parent point groups for reference.

We emphasize that these classification tables are entirely determined by group-theoretical arguments, indicating that the results are universal. They can be applied not only to superconductors but also to other phases (such as magnetic systems). One can see that many combinations of order-parameter corepresentations are allowed to form the Lifshitz invariant.
In particular, the Lifshitz invariant can appear even in systems with inversion symmetry, as in the case of the point-group-based classification \cite{Nagashima2024}. Thus, absence of inversion symmetry is neither a necessary nor sufficient condition to have the type II Lifshitz invariant.
As discussed in Sec.~\ref{Sec.II}, the TR symmetry must be broken 
for the system to host a type-II Lifshitz invariant.
Hence, one has to consider the black-and-white magnetic point group and the associated classification of the type-II Lifshitz invariant.

Our results provide two complementary classification schemes for the Lifshitz invariant, each with its own advantages and limitations. The first is based on complex corepresentation theory with unitary transformations, under which the $A$-parity cannot be defined consistently. The second utilizes real corepresentation theory with orthogonal transformations, where one can assign the $A$-parity but has to fix the gauge in such a way that the order parameter $\Psi$ takes real values.

We first summarize the results obtained from complex corepresentations.
For the magnetic point groups that admit only type (a) and (b) corepresentations, the classification of the Lifshitz invariant for $M=G+AG$ closely follows that of the underlying point groups $G$ \cite{Nagashima2024}. 
We remark that our distinction between conjugate pairs of type (a) complex corepresentations, ${}^1E$ and ${}^2E$, does not bring additional information, since these pairs always appear together in the corepresentation of $D$ and $\Psi$ in our framework.
By contrast, for type (c) corepresentations, the merging of two representations of $G$ into a single corepresentation of $M$ simplifies the selection rules. Such type (c) corepresentations, e.g., $B_{12}$, $B_{23}$, and $E$ appear in Table~\ref{selection_table1_c} and Tables~\ref{selection_table2_c}-\ref{selection_table5_c}, respectively. 

For real corepresentations, the introduction of $A$-parity imposes stricter selection rules. 
When the irreducible corepresentations of $D$, denoted as $\mathrm{D}\Gamma^{D}$, possess a definite parity, the allowed pairs of order parameters $(\Psi_1^*,\Psi_2)$ must have either the same or opposite parities. 
In groups where $D$ admits corepresentations of mixed parity, the required parity relation between $(\Psi_1^*,\Psi_2)$ depends on the parity of the irreducible corepresentation of $D$ that yields the trivial even-parity corepresentation in the product $\mathrm{D}\Gamma^{\Psi_1^*}\otimes \mathrm{D}\Gamma^{D}\otimes \mathrm{D}\Gamma^{\Psi_2}$.
Furthermore, the presence of parityless corepresentations in some groups relaxes the selection rules, effectively removing the constraint imposed by $A$-parity. 

For the type II Lifshitz invariant to arise, one should confirm that the allowed corepresentation of $\psi_\alpha^\ast D_\lambda \psi_\beta$ becomes symmetric with respect to $\alpha$ and $\beta$.
So far, we have not found any cases where such a symmetric combination is prohibited.

In physical applications, we consider TR-symmetry broken multiband superconductors (see Sec.~\ref{Sec.IV}). In these systems, if the corepresentation of the order parameter ($\mathrm{D}\Gamma^\Psi$) has no parityless corepresentation, different parities ($\mathrm{D}\Gamma^\pm$) always appear in pairs in the irreducible decomposition. Thus, the $A$-parity is not a practical constraint even in the case of the real corepresentation theory.
If one goes beyond classification based on magnetic point groups (such as the double-valued magnetic groups, magnetic space groups, and spin space groups), the $A$-parity may be effective.

We also note that the two-dimensional representation $E$ is expressed as a diagonalized form ${}^{1}E\oplus{}^{2}E$ for the complex corepresentations whenever possible. 
In the standard representation theory, this distinction is often ignored because the two are conjugate and physically equivalent.
We use the notation of ${}^{1,2}E$ to stress that they are either a type (c) pair of $\Gamma$ and $\Gamma^\prime$ corresponding to a corepresentation $E$ or a type (a) pair evolving into corepresentations ${}^1E$ and ${}^2E$.
We note that the pair ${}^1E$ and ${}^2E$ is still conjugate.

Finally, let us mention the cases with non-identity $\bm N$. Here we denote the complex corepresentation theory, but this argument on $\bm N$ is valid even in the real corepresentation theory.
In the magnetic point groups we consider here, a few of the cubic magnetic point groups ($\bar{4}^\prime3m^\prime$, $4^\prime32^\prime$, $m\bar{3}m^\prime$) have corepresentation $T$ ($T_g, T_u$ for $m\bar{3}m^\prime$) with
\begin{align}
\bm{N}^{T}
&=
\begin{pmatrix}
    0&1&0\\
    1&0&0\\
    0&0&-1
\end{pmatrix},
\end{align}
while $\bm N=\bm1$ for the other corepresentations.
The former two groups are generated from the point group $T\ (23)$, while the other is from $T_h\ (m\bar{3})$.

Here we prove that these exceptional cases do not affect the irreducible decomposition of the Kronecker product.
This can be understood by considering a point group with higher symmetries.
For the former two magnetic groups ($\bar{4}^\prime3m^\prime$, $4^\prime32^\prime$), we consider $O\ (432)$.
The matrix $\bm N$ can be identified as a representation matrix of $T_1$ for $O$, $\bm{N}^{T}=\bm\Delta^{T_1}(C_{2xy})$. 
Since the representation $T$ of the point group $T$ and $T_1$ of $O$ shares the same representation matrices, we have  ${\bm\Delta}^T(BA^{-1})\bm{N}^{T}={\bm\Delta}^{T_1}(BA^{-1}C_{2xy})$. 
Consequently, the Kronecker product involving $T$:
\begin{align}
    &\bm D^{T}(B)\otimes\bm D^{T}(B) \notag \\
    &={\bm\Delta}^T(BA^{-1})\bm{N}^{T}\otimes{\bm\Delta}^{T_1}(BA^{-1})\bm{N}^{T} \notag \\
    &={\bm\Delta}^{T_1}(BA^{-1}C_{2xy})\otimes{\bm\Delta}^{T_1}(BA^{-1}C_{2xy}),
\end{align}
can be decomposed as:
\begin{equation}
    T\otimes T=A_1\otimes\!^1E\otimes\!^2E\otimes 2T,
\end{equation}
by the unitary matrix for decomposing $T_1\otimes T_1=A_1\oplus E\oplus T_1\oplus T_2$ on $O$.
For the products between $T$ and the other one-dimensional corepresentations, the effect of $\bm N^{T}$ on the decomposition of $\bm D(B)$ are negligible since the product with one-dimensional corepresentations just adds a phase to the matrices, which can be canceled out by the anti-unitary transformation of $\bm D(B)$ as explained above. 
For $m\bar{3}m^\prime$, a group constructed by adding spatial inversion to $4^\prime32^\prime$, having the same $\bm N^{T_{g,u}}$ can be treated in a similar manner.

\section{
Optical conductivity of time-reversal symmetry broken superconductors}
\label{Sec.IV}

Here we present the microscopic formulation for time-reversal symmetry broken multiband superconductors, which enables the calculation of the linear optical conductivity.
Since the group-theoretical analysis in the previous section yields only necessary conditions for optically active Higgs modes, one should verify that the Higgs mode actually becomes visible in the optical conductivity spectrum from 
the microscopic calculations.
We compute the optical conductivity $\sigma(\omega)$ within the effective action approach in imaginary time, accounting for fluctuations of the order parameters to include collective modes.
We focus on multiband superconductors with orbital loop currents, which can be implemented by inserting a flux $\phi$ into lattice models (the net flux is set to be zero).
We abbreviate the $\bm{k}$ and $\phi$-dependence of functions whenever appropriate.

\subsection{Model}
We begin with a model of a multiband superconductor with on-site ($s$-wave) singlet pairing.
The Hamiltonian is given by $\mathcal{H}=\mathcal{H}_{0}+\mathcal{H}_{\text{int}}$, where $\mathcal H_0$ is the kinetic part,
\begin{equation}
    \mathcal{H}_{0} =\sum_{\bm{k}\alpha\alpha'\sigma}\xi_{\alpha\alpha'}(\bm{k},\phi)c^{\dagger}_{\bm{k}\alpha\sigma}c_{\bm{k}\alpha'\sigma},
\end{equation}
with a flux $\phi$, and $\mathcal H_{\rm int}$ is the interaction part,
\begin{equation}
    \mathcal{H}_{\text{int}} = U\sum_{\bm{k}\bm{k}'\alpha}c^{\dagger}_{\bm{k}\alpha\up}c^{\dagger}_{-\bm{k}\alpha\down}c_{-\bm{k}'\alpha\down}c_{\bm{k}'\alpha\up}.
    \label{Interaction}
\end{equation}
Here, $c$ ($c^{\dagger}$) is the annihilation (creation) operator of electrons, $\alpha=1,2,\cdots, N$ is the band index, $\xi_{\alpha\alpha'}(\bm{k},\phi)$ is the matrix elements of the kinetic term, $\sigma=\up,\down$ represents the spin degrees of freedom, and $U(<0)$ is the attractive interaction.
We set the system volume to one.

To incorporate the influence of an external electromagnetic field described by a vector potential $\bm{A}$, we replace $\bm{k}$ by $\bm{k}-e\bm{A}$ and expand the kinetic part as
\begin{align}
    &\xi_{\alpha\alpha'}(\bm{k}-e\bm{A},\phi) \notag \\
    &= \xi_{\alpha\alpha'}(\bm{k},\phi) - e\nabla_{\bm{k}}\xi_{\alpha\alpha'}(\bm{k},\phi)\cdot\bm{A} + O(\bm{A}^{2}),
\end{align}
where $e(<0)$ is the electric charge.
For the linear optical conductivity, it is enough to include the terms of $\bm{A}$ up to the first order.
We note that the diamagnetic term arising from the $\bm{A}^{2}$ term is omitted, since the linear response of collective modes is encoded in the paramagnetic term from the $\bm{A}^{1}$ term and is insensitive to the diamagnetic term at finite frequencies.

Then the Hamiltonian we use is written as
\begin{equation}
    \mathcal{H} = \mathcal{H}_{0} + \mathcal{H}_{\text{int}} + \mathcal{H}_{\text{EM}},
\end{equation}
with
\begin{equation}
    \mathcal{H}_{\text{EM}} = e\sum_{\bm{k}\alpha\alpha'\sigma}\nabla_{\bm{k}}\xi_{\alpha\alpha'}(\bm{k},\phi)\cdot\bm{A}c^{\dagger}_{\bm{k}\alpha\sigma}c_{\bm{k}\alpha'\sigma}.
\end{equation}
We employ the path integral approach in imaginary time $\tau$~\cite{vanOtterlo1999, Sharapov2002, Benfatto2004, Cea2018}.
The partition function of the whole system $\mathcal{Z}$ is written as $\mathcal{Z}=\int\mathcal{D}(c^{\dagger}d)e^{-S[c^{\dagger},c]}$ with the Euclidean action $S$:
\begin{equation}
    S[c^{\dagger},c] = \int_{0}^{\beta}d\tau\left( \sum_{\bm{k}\alpha\sigma}c^{\dagger}_{\bm{k}\alpha\sigma}\partial_{\tau}c_{\bm{k}\alpha\sigma} + \mathcal{H} \right).
\end{equation}
We perform the Hubbard-Stratonovich transformation to decouple the fermionic interaction~(\ref{Interaction}), introducing a bosonic field $\Delta_{\alpha}(\tau)=\Delta_{0\alpha} + \Delta_{x\alpha}(\tau) - \mathrm{i}\Delta_{y\alpha}(\tau)$ corresponding to the saddle-point contribution ($\Delta_{0\alpha}$), and real ($\Delta_{x\alpha}(\tau)$) and imaginary ($\Delta_{y\alpha}(\tau)$) fluctuations, respectively.

After integrating over the fermionic degrees of freedom, we split the action into a mean-field part and a fluctuation part.
We then integrate out the fluctuations $\Delta_{\mu\alpha}$ ($\mu=x,y$), and perform the analytic continuation for the Matsubara frequency $\mathrm{i}\Omega \to \omega + \mathrm{i}0^{+}$.
Then, the fluctuation part of the effective action $S_{\text{FL}}$ reads
\begin{align}
    &S_{\text{FL}} = \sum_{a,b}\frac{e^{2}}{2}\int\frac{d\omega}{2\pi}A_{a}(\omega)A_{b}(-\omega)\Phi_{ab}(\omega) \notag \\
    &+ \sum_{a,b}\frac{e^{2}}{4}\int\frac{d\omega}{2\pi}A_{a}(\omega)A_{b}(-\omega)Q_{a}^{\text{T}}(\omega)U_{\text{eff}}(\omega)Q_{b}(-\omega),
\end{align}
with $a,b=x,y$ in the case of two-dimensional models.
Here, we have introduced the current-current correlation function $\Phi(\mathrm{i}\Omega)$ by
\begin{align}
    \Phi_{ab}(\mathrm{i}\Omega)&=\frac{1}{\beta}\sum_{n}\int\frac{d^{d}\bm{k}}{(2\pi)^{d}}\text{Tr}[v_{a}G_{0}(\mathrm{i}\omega_{n}+\mathrm{i}\Omega)v_{b}G_{0}(\mathrm{i}\omega_{n})] \notag \\
    &= \int\frac{d^{d}\bm{k}}{(2\pi)^{d}}\sum_{j,l}\frac{f_{jl}v_{a,jl}v_{b,jl}}{\mathrm{i}\Omega - E_{lj}},
\end{align}
where $v_{a}(\bm{k},\phi)$ is the velocity operator:
\begin{equation}
    v_{a}(\bm{k},\phi) = \mqty[ \partial_{a}\xi(\bm{k},\phi) & O \\ O & -\partial_{a}\xi^{\text{T}}(-\bm{k},\phi)],
\end{equation}
with $\partial_{a}=\partial/\partial k_{a}$ and $a$ is the spatial direction ($a=x,y$ in two dimensional systems).
To simplify the notation, we use the generalized Pauli matrices $\tau_{\mu\alpha}$,
\begin{equation}
    \tau_{x\alpha} = \mqty[O & A_{\alpha} \\ A_{\alpha} & O], \quad \tau_{y\alpha} = \mqty[O & -\mathrm{i}A_{\alpha} \\ \mathrm{i}A_{\alpha} & O],
\end{equation}
their band representation
$\tau_{\mu\alpha,jl}=\braket{\varphi_{j}|\tau_{\mu\alpha}|\varphi_{l}}$, and $v_{a,jl}=\braket{\varphi_{j}|v_{a}|\varphi_{l}}$, where $\ket{\varphi_{j}}$ is the $j$th eigenvector with the $j$th largest eigenenergy $E_{j}$, $E_{lj}:=E_{l} - E_{j}$, and $f_{jl}:=f(E_{j}) - f(E_{l})$ with $f(E_{j})$ being the Fermi distribution: $f(E_{j})=1/(e^{\beta E_{j}}+1)$.
We have also introduced the effective interaction $U_{\text{eff}}$ within the random phase approximation:
\begin{equation}
    U_{\text{eff}} = U + U\Pi U + \cdots = (1-U\Pi)U,
\end{equation}
the polarization bubble $\Pi(\mathrm{i}\Omega)$:
\begin{align}
    &[\Pi(\mathrm{i}\Omega)]_{\mu\alpha,\mu'\alpha'} \notag \\
    &= \frac{1}{2\beta}\sum_{n}\int\frac{d^{d}\bm{k}}{(2\pi)^{d}}\text{Tr}[\tau_{\mu\alpha}G_{0}(\mathrm{i}\omega_{n}+\mathrm{i}\Omega)\tau_{\mu'\alpha'}G_{0}(\mathrm{i}\omega_{n})] \notag \\
    &= \frac{1}{2}\int\frac{d^{d}\bm{k}}{(2\pi)^{d}}\sum_{j,l}\frac{f_{jl}\tau_{\mu\alpha}\tau_{\mu'\alpha',lj}}{\mathrm{i}\Omega - E_{lj}},
\end{align}
and the vector $Q_{a}(\mathrm{i}\Omega)$:
\begin{align}
    &[Q_{a}(\mathrm{i}\Omega)]_{\mu\alpha} \notag \\
    &= \frac{1}{\beta}\sum_{n}\int\frac{d^{d}\bm{k}}{(2\pi)^{d}}\text{Tr}[v_{a}G_{0}(\mathrm{i}\omega_{n}+\mathrm{i}\Omega_{n})\tau_{\mu\alpha}G_{0}(\mathrm{i}\omega_{n})] \notag \\
    &= \int\frac{d^{d}\bm{k}}{(2\pi)^{d}}\sum_{j,l}\frac{f_{jl}v_{a,jl}\tau_{\mu\alpha,lj}}{\mathrm{i}\Omega - E_{lj}}.
\end{align}
By performing the analytic continuation $\mathrm{i}\Omega \to \omega + \mathrm{i}0^{+}$, we arrive at the real-frequency form of the effective interaction $U_{\text{eff}}$ and the vector $Q$.

To obtain the linear optical conductivity, we take a functional derivative of $S_{\text{FL}}$ with respect to $A(-\omega)$,
which gives us the current density $j(\omega)$:
\begin{align}
    j(\omega,\phi) &= -\frac{\delta S_{\text{FL}}(\phi)}{\delta A(-\omega)} \notag \\
    &= (\sigma_{\text{QP}}(\omega,\phi)+\sigma_{\text{CM}}(\omega,\phi))E(\omega),
\end{align}
with
\begin{align}
    [\sigma_{\text{QP}}(\omega,\phi)]_{ab} &= \frac{\mathrm{i}e^{2}}{\omega}\Phi_{ab}(\omega,\phi), \notag \\
    [\sigma_{\text{CM}}(\omega,\phi)]_{ab} &= \frac{\mathrm{i}e^{2}}{2\omega}Q_{a}^{\text{T}}(\omega,\phi)U_{\text{eff}}(\omega,\phi)Q_{b}(-\omega,\phi).
\end{align}
The relation $E_{a}(\omega)=\mathrm{i}\omega A_{a}(\omega)$ is used.
Here, $\sigma_{\text{QP}}$ and $\sigma_{\text{CM}}$ correspond to the quasiparticle and collective mode responses, respectively.

\begin{figure}
    \centering
    \includegraphics[scale=1.0]{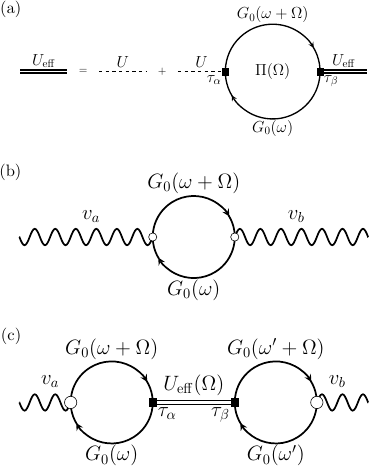}
    \caption{Diagrams of (a) the effective interaction $U_{\text{eff}}$ within the randam phase approximation, and [(b),(c)] the linear optical conductivities in multiband superconductors.
    The diagram (b) [(c)] corresponds to the quasiparticle [collective mode] excitation.
    Here, $U<0$ is the bare interaction and $G_{0}(\omega)$ is the Matsubara Green's function with a frequency $\omega$.
    When the spatial dimension is more than one, the linear optical conductivities depend on two directions $a$ and $b$.}
    \label{Fig_diagrams}
\end{figure}

The diagrams for the linear optical conductivity are described in Fig.~\ref{Fig_diagrams}.
The collective mode response arises from the poles of the effective interaction $U_{\text{eff}}(\omega,\phi)$, which satisfies $1-U\Pi(\omega,\phi)=0$.
One may wonder if the vector $Q$ could also diverge at the frequency of the Higgs mode peak $\omega=2\Delta$, in which case it may not be easy to distinguish the peak arising from the effective interaction $U_{\text{eff}}$ and the vector $Q$.
However, this is not the case: while the pole of the effective interaction $U_{\text{eff}}$ causes the divergence, the seemingly diverging function in the vector $Q$ does not have a singularity, since the singular point is located only at $\omega=2\Delta$ and its singularity is eliminated by the integration with respect to $\bm{k}$.
Hence, if we observe the peak at $\omega=2\Delta$, we can conclude that the peak arises from the pole of the effective interaction.
We can also numerically check the non-divergent behavior of the vector $Q$ at $\omega=2\Delta$.

\subsection{Application of the group-theoretical analysis}

\begin{figure}
    \centering
    \includegraphics[scale=0.8]{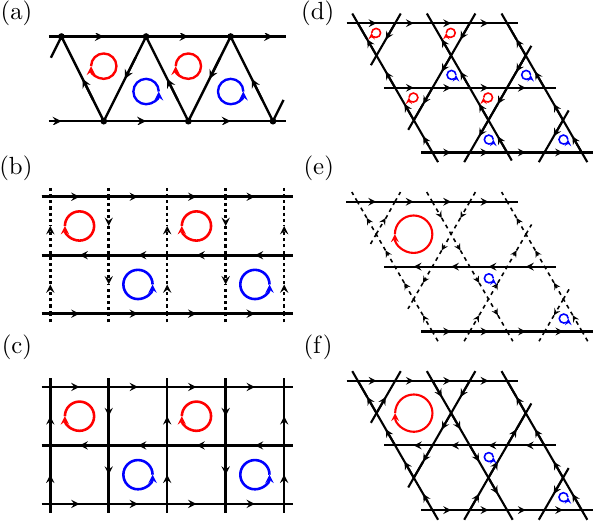}
    \caption{
    Several lattice models of time-reversal symmetry broken superconductors with orbital loop currents, for which we examine whether the type II Lifshitz invariant is allowed or not by the magnetic point-group symmetry:
    (a) The isotropic triangular ladder model,
    (b) the anisotropic square lattice model,
    (c) the isotropic square lattice model,
    (d) the isotropic Kagome lattice model,
    (e) the anisotropic Kagome lattice model,
    and (f) the isotropic Kagome lattice model with a different current pattern from (d).
    The solid and dashed lines represent the hopping strength $t_{1}$ and $t_{2}$ ($t_{1}\neq t_{2}$), respectively.
    When electrons hop along (against) the direction of the arrows on the bonds, they acquire a phase $e^{\mathrm{i}\phi}$ ($e^{-\mathrm{i}\phi}$) in each model.
    The red and blue spiral arrows indicate the loop current patterns.
    The magnetic point group corresponding to each model is shown in Tables~\ref{Models_selectionrule_c} and \ref{Models_selectionrule_r}.
    }
    \label{Fig_lattices}
\end{figure}

Before proceeding to the numerical results, we first present the group-theoretical results for the magnetic point group discussed in the previous section.
We consider several lattice models of time-reversal-symmetry-broken superconductors to demonstrate the effectiveness of our approach.
In the following, we assume $s$-wave superconductivity with a single atomic orbital per lattice site in each model.
We assume the presence of the Aharonov-Bohm flux to break the TR symmetry, 
which implements the effect of orbital loop currents.
Here we analyze an isotropic triangular ladder model, anisotropic and isotropic square lattice models, and Kagome lattice models.
These models are summarized in Fig.~\ref{Fig_lattices}.

For each lattice model, the order parameter $\Psi$ having the sublattice degrees of freedom is decomposed into the irreducible corepresentations.
Then, we can determine whether the type II Lifshitz invariant is allowed or not by the magnetic point-group symmetry 
based on the classification tables in the previous section.
We can choose the Wyckoff position with the highest site symmetry to determine the magnetic point group of the model.
Regarding the permutation matrices of the sublattice degrees of freedom $\psi_i,\phi_i$, a corepresentation matrix allows us to decompose the corepresentation.
The results are summarized in Table~\ref {Models_selectionrule_c} for complex corepresentations and Table~\ref {Models_selectionrule_r} for real corepresentations. 
We observe that a type II Lifshitz invariant is allowed to exist for all the models considered here in both classification schemes (Table~\ref{selection_table1_c}-~\ref{selection_table5_r}) based on complex and real corepresentations.
In particular, the decomposed real corepresentations of the order parameter $\Psi$ always contain both even- $(+)$ and odd-parity $(-)$ irreducible corepresentations under the operation $A$. As a result, the system exhibits an optically active Higgs mode in each lattice model.
In the next subsection, we demonstrate that these group-theoretical considerations are consistent with microscopic numerical calculations.
\begin{table}[]
    \centering
    \caption{
    List of magnetic point groups (MPGs) and the irreducible decomposition of complex corepresentations (coreps) of the order parameter $\Psi$ for each lattice model shown in Fig.~\ref{Fig_lattices}.
    The fourth column shows whether a type-II Lifshitz invariant (LI) is allowed or not for each model.
    }
    \begin{tblr}{cccc}
    \hline\hline
        Model & MPG &Decomposed corep of $\Psi$ & LI \\
        \hline
        (a) &  $2^\prime/m$&$4A^{\prime}$ & \checkmark \\
        (b) &  $m^\prime m^\prime m$&$4(A_g\oplus B_u)$ & \checkmark \\
        (c) &  $4/mm^\prime m^\prime$&$2(A_g\oplus B_g\oplus {}^{1}\!E_u\oplus {}^{2}\!E_u)$ & \checkmark \\
        (d) & $6^\prime/mm m^\prime$ &$2(A_1^{\prime}\oplus E^{\prime})$ & \checkmark \\
        (e) &  $m^\prime m^\prime m$&$12(A_g\oplus B_u)$  & \checkmark\\
        (f) & $6/mm^\prime m^\prime$&$4(A_g\oplus B_u\oplus {}^{1}\!E_{2g}$ & \checkmark\\
        &&$\oplus{}^{2}\!E_{2g}\oplus {}^{1}\!E_{1u}\oplus{}^{2}\!E_{1u})$ \\ \hline \hline
    \end{tblr}
    \label{Models_selectionrule_c}
\end{table}
\begin{table}[]
    \centering
    \caption{
    List of magnetic point groups (MPGs) and the irreducible decomposition of real corepresentations (coreps) of the order parameter $\Psi$ for each lattice model shown in Fig.~\ref{Fig_lattices}.
    The fourth column shows whether a type II Lifshitz invariant (LI) is allowed or not for each model.
    }
    \begin{tblr}{cccc}
    \hline\hline
        Model & MPG &Decomposed corep of $\Psi$ & LI \\
        \hline
        (a) &  $2^\prime/m$&$2A^{\prime+}\oplus 2A^{\prime-}$ & \checkmark \\
        (b) &  $m^\prime m^\prime m$&$2(A_g^+\oplus A_g^-\oplus B_u^+\oplus B_u^-)$ & \checkmark \\
        (c) &  $4/mm^\prime m^\prime$&$A_g^+\oplus A_g^-\oplus B_g^+\oplus B_g^-\oplus 2E_u$ & \checkmark \\
        (d) & $6^\prime/mm m^\prime$ &$A_1^{\prime+}\oplus A_1^{\prime-}\oplus E^{\prime+}\oplus E^{\prime-}$ & \checkmark \\
        (e) &  $m^\prime m^\prime m$&$6(A_g^+\oplus A_g^-\oplus B_u^+\oplus B_u^-)$  & \checkmark\\
        (f) & $6/mm^\prime m^\prime$&$2(A_g^+\oplus A_g^-\oplus B_u^+\oplus B_u^-)$ & \checkmark\\
        &&$\oplus 4(E_{2g}\oplus E_{1u})$ \\
        \hline \hline
    \end{tblr}
    \label{Models_selectionrule_r}
\end{table}
\subsection{Numerical results}

We demonstrate numerical results of the optical conductivity spectrum for the six lattice models shown in Fig.~\ref{Fig_lattices}: the ladder model in one dimension [Fig.~\ref{Fig_lattices}(a)], the square lattice with and without bond-order patterns [(b), (c)], the Kagome lattice with three sites in a unit cell [(d)], and the Kagome lattice with $12$ sites in a unit cell with and without bond-order patterns in two dimensions [(e), (f)].
All the lattice models are assumed to break time-reversal symmetry by a quantum geometric phase carried by electrons, whose effect is included by a flux $\phi$ in the Hamiltonian.
According to the group-theoretical argument presented in the previous section, all the models considered here are permitted to possess the Lifshitz invariant by both complex and real corepresentation analysis (Table~\ref{Models_selectionrule_c} and~\ref{Models_selectionrule_r}), and the Higgs mode response is allowed to emerge in the linear response regime.

The kinetic part of the Hamiltonian of those models $\mathcal{H}_{0}$ is written as
\begin{equation}
    \mathcal{H}_{0} = \sum_{\langle i,j \rangle,\sigma}(-t_{ij}e^{\mathrm{i}\phi_{ij}}\hat{c}^{\dagger}_{i,\sigma}\hat{c}_{j,\sigma} + \text{h.c.} ) - \mu\sum_{i,\sigma}\hat{c}^{\dagger}_{i,\sigma}\hat{c}_{i,\sigma},
\end{equation}
where $\langle i,j\rangle$ denotes a pair of nearest-neighbor sites, $t_{ij}$ is the real hopping strength between sites $i$ and $j$, $e^{\mathrm{i}\phi_{ij}}$ is the phase factor, $\mu$ is the chemical potential, and $\hat{c}_{i,\sigma}$ ($\hat{c}^{\dagger}_{i,\sigma}$) is the annihilation (creation) operator of electrons at site $i$ with spin $\sigma$.
When electrons hop along (against) the direction of the arrows in Fig.~\ref{Fig_lattices} from site $i$ to $j$, they acquire a phase $e^{{\rm i}\phi_{ij}}$ with $\phi_{ij}=\phi$ ($-\phi$).
We fix $\phi=0.01$ and $U=-3.0$ for the numerical calculations.
We numerically verified that all the Higgs-mode peaks in the optical conductivity vanish at $\phi=0$ (i.e., orbital loop currents are absent). 

\begin{figure}
    \centering
    \includegraphics[scale=0.68]{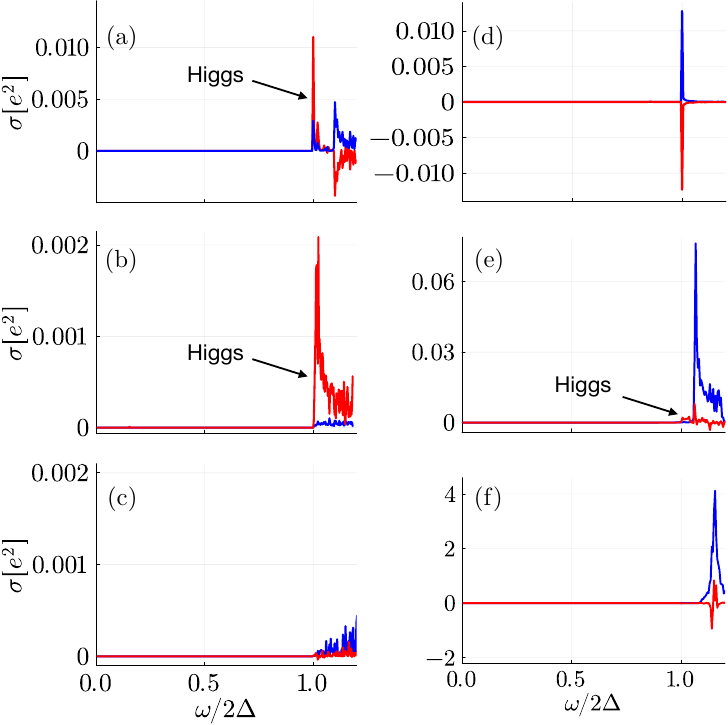}
    \caption{Optical conductivity spectrum for each lattice model shown in Fig.~\ref{Fig_lattices}.
    The blue and red curves correspond to the contributions from the quasiparticles $\sigma_{\text{QP}}$ ($\sigma_{\text{QP}, xx}$) and collective modes $\sigma_{\text{CM}}$ ($\sigma_{\text{CM}, xx}$), respectively.
    $\phi=0.01$ and $U=-3.0$ are used for every model.
    (a) Ladder model (Fig.~\ref{Fig_lattices}(a)) with $t=1.0$ and $\mu=-0.3$.
    (b)[(c)] Anisotropic [Isotropic] square lattice model (Fig.~\ref{Fig_lattices}(b)[(c)]) with $t_{1}=t=1.0$, $t_{2}=0.1$ $[1.0]$, and $\mu=-0.2$.
    (d) Isotropic Kagome lattice model with three sites in the unit cell (Fig.~\ref{Fig_lattices}(d)) and $\mu=0$.
    (e)[(f)] Anisotropic [Isotropic] Kagome lattice model with $12$ sites in the unit cell (Fig.~\ref{Fig_lattices}(e)[(f)]).
    The parameters are $t_{1}=0.5$, $t_{2}=0.05$ [$0.5$], and $\mu=0$.
    }
    \label{Results}
\end{figure}

\subsubsection{Ladder model}
\label{sec: ladder model}

The first example is the triangular ladder model shown in Fig.~\ref{Fig_lattices}(a), where
we fix $t_{ij}=t=1.0$, $\phi=0.01$, $\mu=-0.3$, and $U=-3.0$.
In Fig.~\ref{Results}(a), we show the linear optical conductivity with
blue and red curves corresponding to the response of quasiparticles ($\sigma_{\text{QP}}$) and collective modes ($\sigma_{\text{CM}}$), respectively.
One can see that the collective-mode response has a peak at $\omega/2\Delta=1.0$, which dominates the quasiparticle response at that frequency, suggesting that the Higgs mode becomes visible in the linear optical response.

When we neglect the $\tau_{x}$ (amplitude) channel in the contribution of the optical conductivity, the collective-mode peak at $\omega/2\Delta=1$ changes its sign, and is canceled with the quasiparticle peak, resulting in the absence of the peak at $\omega/2\Delta=1$ in the total optical conductivity.
On the other hand, if we neglect the $\tau_{y}$ (phase) channel, the collective-mode peak survives with a higher weight than that of the quasiparticles.
These results suggest that the peak at $\omega/2\Delta=1$ in the optical conductivity originates from the Higgs mode.
We perform similar analysis for the anisotropic square (Fig.~\ref{Results}(b)) and Kagome (Fig.~\ref{Results}(e)) lattice models, and confirm that the peak at $\omega/2\Delta=1$ mainly comes from the Higgs-mode contribution (see the following subsections).

\subsubsection{Square lattice model}

The second and third examples are the anisotropic and isotropic square lattice models (Fig.~\ref{Fig_lattices}(b) and (c)).
In the anisotropic model, we set the hopping amplitudes $t_{1}=t=1.0$ (solid lines) and $t_{2}=0.1$ (dashed lines), respectively.
In the isotropic model, we fix $t_{1}=t_{2}=t=1.0$.
In both cases, we set $\mu=-0.3$.
The results of the optical conductivity are shown in Fig.~\ref{Results}(b) and (c).
In the anisotropic case, one can see that the collective-mode response has a dominant peak at $\omega/2\Delta=1$. However, this peak vanishes for the isotropic case.
We can understand the absence of the collective-mode peak in the isotropic case from the GL theory; see Appendix~\ref{Appendix.C}.
We also confirm that the peak at $\omega/2\Delta=1$ in the collective-mode response (Fig.~\ref{Results}(b)) arises from the Higgs mode (see the discussion in the previous subsection).

\subsubsection{Kagome lattice model with three sites in the unit cell}
The fourth example is the isotropic Kagome lattice model with three sites in the unit cell (Fig.~\ref{Fig_lattices}(d)).
We set $t_{ij}=t=1.0$ and $\mu=0$.
The calculated linear optical conductivity is shown in Fig.~\ref{Results}(d).
Both the quasiparticle and collective-mode responses have a peak at $\omega/2\Delta=1$ with the same height but with opposite signs. 
We observe that the sum of the two contributions is zero within numerical errors.
The reason for this complete cancellation is not yet clear from a symmetry perspective. At least, the type II Lifshitz invariant is allowed by the magnetic point-group symmetry for this model according to the classification given in the previous section. One possibility is that there might be another accidental symmetry inherent in this particular model that enforces the cancellation of the Higgs mode by the quasiparticles. We leave this issue for future consideration.

\subsubsection{Kagome lattice model with $12$ sites in the unit cell}
The fifth and last examples are the anisotropic and isotropic Kagome lattice models with $12$ lattice sites in the unit cell (Fig.~\ref{Fig_lattices}(e) and (f)).
In the anisotropic model, we set the hopping amplitudes $t_{1}=0.5$ (solid lines) and $t_{2}=0.05$ (dashed lines), respectively.
In the isotropic model, we put $t_{1}=t_{2}=0.5$.
In both cases, we use $\mu=0$.
The optical conductivity for these models is shown in Fig.~\ref{Results}(e) and (f).
In the anisotropic case, the collective-mode response has a tiny peak at $\omega/2\Delta=1$, but this peak vanishes in the isotropic case.
We can understand the absence of the collective-mode peak in the isotropic case within GL theory; see Appendix~\ref{Appendix.C}.
We also confirm that the peak at $\omega/2\Delta=1$ in the collective-mode response (Fig.~\ref{Results}(e)) emerges from the Higgs mode (see the discussion in Sec.~\ref{sec: ladder model}).

\section{Summary and Discussion}
\label{Sec.V}
We have investigated the Lifshitz invariant in time-reversal-symmetry-broken superconductors and its effect on linear optical conductivities.
First, we considered the phenomenological GL theory to show that there are two types of Lifshitz invariant (type I and type II), distinguished by their behavior under the PH transformation (Table~\ref{tab: type I, II}).
When the type II Lifshitz invariant exists in the GL free energy, electromagnetic fields linearly couple to the amplitude fluctuation of the superconducting order parameter.

We then classified all possible combinations of complex and real irreducible corepresentations of order parameters in magnetic point groups that permit a type II Lifshitz invariant in the free energy.
The obtained Tables~\ref{selection_table1_c}--\ref{selection_table5_r}  generalize the results of the previous paper \cite{Nagashima2024}, which implicitly assumed the presence of the time-reversal symmetry  (namely, the type I Lifshitz invariant).
Let us emphasize that the classification is determined entirely by group theory and does not depend on the details of the system. It can be applied not only to superconductors but also to other ordered phases.

The condition under which the existence of the type II Lifshitz invariant is permitted depends on whether the order parameter has a nontrivial corepresentation. One way to realize this is to use sublattice degrees of freedom.
As a demonstration, we considered several lattice models of time-reversal-symmetry-broken superconductors with orbital loop currents (Fig.~\ref{Fig_lattices}) that can possess the type II Lifshitz invariant.
We microscopically calculated the optical conductivity of these models, and found that the Higgs-mode peak can indeed appear at the frequency $\omega/2\Delta=1$ in the optical conductivity spectrum.

We comment that the recent results in Ref.~\cite{Tanaka2025} can be understood within our framework.
In that study, the optical conductivity of multiband $s$-wave superconductors in a magnetic field is calculated with vertex corrections, and the enhancement of the optical conductivity due to the Higgs mode is reported.
Since their model has the symmetry of the magnetic point group $m'=\{E, \theta\sigma_{yz} \}$ in the presence of the magnetic field with the real corepresentation of the order parameter $\rho=A^{+}\oplus A^{-}$, the type II Lifshitz invariant is allowed to exist in the GL free energy according to Table~\ref{selection_table1_r}. 
Hence, one can expect that the Higgs-mode peak appears in the optical conductivity.

Let us mention several open issues to be addressed in future work.
We implicitly assumed that the superconducting order parameters are defined on each lattice site.
This assumption seems to work well since the order parameters reflect the symmetry of the background crystal structure, even though the size of Cooper pairs is much larger than the lattice constant.
It is therefore desirable to check whether this assumption may still be justified when we consider spatially extended Cooper pairs mediated by phonons, where the effective attraction will become nonlocal and frequency dependent.
Another implicit assumption we made is that the sublattice and other degrees of freedom (such as spins or orbitals) are separated as a direct product in the order-parameter corepresentation.
If the system has an intertwining symmetry that cannot be decomposed into such a direct product (e.g., in unconventional superconductors with non-$s$-wave pairing symmetries), more complicated and fruitful phenomena could occur.

Finally, we discuss possible experimental observations of the optically active Higgs mode in time-reversal-symmetry-broken superconductors.
A primary candidate is a kagome superconductor, $\mathrm{CsV}_{3}\mathrm{Sb}_{5}$~\cite{Ortiz2021, Jiang2022, Luo2022, Zheng2022}.
This material is reported to have a CDW phase above the superconducting transition temperature $T_{\text{c}}$, and the phase is thought to coexist with superconductivity below $T_{c}$.
The pairing symmetry of this superconducting phase is considered to be $s$-wave~\cite{Chao2021} and anisotropic~\cite{Roppongi2023}.
Although it has not been clearly understood whether the system breaks the time-reversal symmetry~\cite{Mielke2022, Deng2024, Elmers2025} or not~\cite{Saykin2023, Guo2024}, measuring the optical conductivity in the superconducting phase will provide key information on the underlying symmetry and collective modes of the superconductors.

\section{Acknowledgments}
We thank H. Tanaka for insightful discussions.
This work is supported by JST FOREST (Grant No.~JPMJFR2131) and JSPS KAKENHI (Grant Nos.~JP24H00191, JP25H01246, JP25H01251).
R.N. acknowledges the financial support from Toyota Riken (Toyota Riken Overseas Scholarship).
C.M. acknowledges the financial support from the University of Tokyo (MERIT-WINGS).

\appendix

\twocolumngrid

\section{Product tables of corepresentations for magnetic point groups}
\label{Appendix A}

In this Appendix, we show the product tables of the complex and real corepresentations for the magnetic point groups. 
\subsection{Complex corepresentation}
The product tables for the complex corepresentations of the point group $M=G+AG$ can be obtained from the product tables for the representations of $G$.
Product tables for the crystallographic point groups are widely available. See, for example, Ref.~\cite{Atkins1970}. We remark that some of the literature doesn't distinguish ${}^1E$ and ${}^2E$, the physically equivalent pair of complex representations, and regard them as one two-dimensional representation $E$.

Especially for the magnetic point groups having only type (a) corepresentations, the product tables are completely identical to those of the corresponding $G$. 
An example of such a product table is shown in Table~\ref{only_a_tables}.
Even for magnetic point groups with type (c) corepresentations $E$ generated from representations ${}^1E$ and ${}^2E$ of $G$, the product tables are also identical to the product table of the corresponding $G$ without distinguishing ${}^1E$ and ${}^2E$.

We summarized the product tables, in Tables~\ref{first_product_table_c}-\ref{last_product_table_c} for the other magnetic point groups; namely $4^\prime$, $\bar{4}^\prime$ and $4^\prime/m$ having type (b) corepresentations and $4^\prime22^\prime$, $\bar{4}^\prime2m^\prime$, $4^\prime mm^\prime$, $\bar{4}^\prime m2^\prime$ and $4^\prime/mmm^\prime$ having type (c) corepresentations other than $E$.

\begin{table}[H]
\centering
        \begin{tblr}{c|cccc}
            $\otimes $&$A$&${}^1E$&${}^2E$&$T$  \\
            \hline
            $A$&$A$&${}^1E$&${}^2E$&$T$ \\
            ${}^1E$&&${}^2E$&$A$&$T$ \\
            ${}^2E$&&&${}^1E$&$T$ \\
            $T$&&&&$A\oplus{}^1E\oplus{}^2E\oplus2T$ \\
        \end{tblr}   
    \caption{
    Product table of the complex corepresentations of the magnetic point group $\bar{4}3m^\prime$ and that of the corresponding point group 23 ($T$). Note that $\bar{4}3m^\prime$ has only representations of type (a).
    }
    \label{only_a_tables}
\end{table}    
\subsubsection{$4^\prime$, $\bar{4}^\prime$, $4^\prime/m$}
\begin{table}[H]
 \centering
\begin{tblr}{c|c|[dashed]c}
        &type (a) &type (b)\\
        $\otimes$ & $A$ & $B$  \\
        \hline  
        $\mathrm{A^+}$ & $A$  & $B$ \\
        \hline[dashed]
        $\mathrm{B}$ &    &  $4A$  
\end{tblr}
    \caption{
    Product table of the complex corepresentations of the magnetic point groups $4^\prime$, $\bar{4}^\prime$, and $4^\prime/m$. One should add subscripts $g$ and $u$ for $4^\prime/m$.
    }
    \label{first_product_table_c}
 \end{table}
 
\subsubsection{$4^\prime22^\prime$, $\bar{4}^\prime2m^\prime$, $4^\prime/mmm^\prime$}
\begin{table}[H]
 \centering
\begin{tblr}{c|cc|[dashed]c}
        &\SetCell[c=2]{c}{type (a)}&&type (c)\\
        $\otimes$ & ${A}$ & ${B_1}$ & ${B_{23}}$  \\
        \hline  
        ${A}$ & ${A}$ & ${B_1}$ & ${B_{23}}$ \\
        ${B_1}$ & &${A}$ & ${B_{23}}$ \\
        \hline[dashed]
        ${B_{23}}$ &    &    &${ 2A\oplus 2B_1}$ 
\end{tblr}
    \caption{
    Product table for the magnetic point groups $4^\prime22^\prime$, $\bar{4}^\prime2m^\prime$, and $4^\prime/mmm^\prime$. One should add subscripts $g$ and $u$ for $4^\prime/mmm^\prime$.
    }
    \label{tab:placeholder}
\end{table}

\subsubsection{$4^\prime mm^\prime$, $\bar{4}^\prime m2^\prime$}
\begin{table}[H]
 \centering
\begin{tblr}{c|cc|[dashed]c}
        &\SetCell[c=2]{c}{type (a)}&&type (c)\\
        $\otimes$ & ${A_1}$ & ${A_2}$ & ${B_{12}}$  \\
        \hline  
        ${A_1}$ & ${A_1}$ & ${A_2}$ & ${B_{12}}$ \\
        ${A_2}$ & &${A_1}$ & ${B_{12}}$ \\
        \hline[dashed]
        ${B_{12}}$ &    &    &${ 2A_1\oplus 2A_2}$ 
\end{tblr}
    \caption{
    Product table for the magnetic point groups $4^\prime mm^\prime$ and $\bar{4}^\prime m2^\prime$.
    }
    \label{last_product_table_c}
\end{table}
 
\subsection{Real corepresentation}
The product tables of the real corepresentations can be derived by utilizing the identity between the real corepresentations of $M$ and the real representations of $H$. The real representations of $H$ are, in turn, tabulated from its complex representation~\cite{serre1977linear}.
For the magnetic point groups that do not contain parityless corepresentations, the product tables are constructed by appending the $A$-parity multiplication rules. Namely, $(+)\times(+)=(-)\times(-)=(+),(+)\times(-)=(-)$, to the product tables of the complex corepresentations of the corresponding $G$ , while treating ${}^1E$ and ${}^2E$ as indistinguishable. 
For the remaining magnetic point groups, hosting parityless corepresentations, the product tables are summarized in Tables~\ref{first_product_table_r}-\ref{last_product_table_r}.
In order to feature the $A$-parity of the corepresentation, our notation of the corepresentation is different from the notation used in~\cite{erb2020vector} which using the common notation of the representations of $H$.
\subsubsection{$4^\prime$, $\bar{4}^\prime$, $4^\prime/m$}
\begin{table}[H]
 \centering
\begin{tblr}{c|cc|[dashed]c}
        &\SetCell[c=2]{c}{} &&parityless\\
        $\otimes$ & ${A^+}$ &${ A^-}$ & ${B}$  \\
        \hline  
        ${A^+}$ & ${A^+}$ & ${A^-}$ & ${B}$ \\
        ${A^-}$ &     &${A^+}$ & ${B}$ \\
        \hline[dashed]
        ${B}$ &    &    & ${2A^+\oplus2A^-}$  
\end{tblr}
    \caption{
    Product table for the magnetic point groups $4^\prime$, $\bar{4}^\prime$, and $4^\prime/m$. One should add subscripts $g$ and $u$ for $4^\prime/m$.
    }
    \label{first_product_table_r}
 \end{table}
 
\subsubsection{$4^\prime22^\prime$, $\bar{4}^\prime2m^\prime$, $4^\prime/mmm^\prime$}
\begin{table}[H]
 \centering
\begin{tblr}{c|cccc|[dashed]c}
        &\SetCell[c=4]{c}{}&&&&parityless\\
        $\otimes$ & ${A^+}$ &${A^-}$ & ${B_1^+}$& ${B_1^-}$ & ${B_{23}}$  \\
        \hline  
        ${A^+}$ & ${A^+}$ &${A^-}$ & ${B_1^+}$& ${B_1^-}$ & ${B_{23}}$ \\
        ${A^-}$ &       &${A^+}$ & ${B_1^-}$& ${B_1^+}$ & ${B_{23}}$ \\
        ${B_1^+}$ &    & &${A^+}$ &${A^-}$ &${ B_{23}}$ \\
        ${B_1^-}$ &    & & &${A^+}$ & ${B_{23}}$ \\
        \hline[dashed]
        ${B_{23}}$ &    &    &&    &${ A^+\oplus A^-\oplus B_1^+\oplus B_1^-}$ 
\end{tblr}
    \caption{
    Product table for the magnetic point groups $4^\prime22^\prime$, $\bar{4}^\prime2m^\prime$, and $4^\prime/mmm^\prime$. One should add subscripts $g$ and $u$ for $4^\prime/mmm^\prime$.
    }
    \label{tab:placeholder}
\end{table}

\subsubsection{$4^\prime mm^\prime$, $\bar{4}^\prime m2^\prime$}
\begin{table}[H]
 \centering
    \begin{tblr}{c|cccc|[dashed]c}
        &\SetCell[c=4]{c}{}&&&&parityless\\
        $\otimes$ & ${A_1^+}$ &${A_1^-}$ & ${A_2^+}$& ${A_2^-}$ & ${B_{12}}$  \\
        \hline  
        ${A_1^+}$ & ${A_1^+}$ &${A_1^-}$ & ${A_2^+}$& ${A_2^-}$ & ${B_{12}}$ \\
        ${A_1^-}$ &       &${A_1^+}$ & ${A_2^-}$& ${A_2^+}$ & ${B_{12}}$ \\
        ${A_2^+}$ &    & &${A_1^+}$ &${A_1^-}$ &${ B_{12}}$ \\
        ${A_2^-}$ &    & & &${A_1^+}$ & ${B_{12}}$ \\
        \hline[dashed]
        ${B_{12}}$ &    &    &&    &${ A_1^+\oplus A_1^-\oplus A_2^+\oplus A_2^-}$  
    \end{tblr}
    \caption{
    Product table for the magnetic point groups $4^\prime mm^\prime$ and $\bar{4}^\prime m2^\prime$.
    }
    \label{tab:placeholder}
\end{table}

\subsubsection{$42^\prime2^\prime$, $4m^\prime m^\prime$,$4/mm^\prime m^\prime$}
\begin{table}[H]
 \centering
    \begin{tblr}{c|cccc|[dashed]c}
        &\SetCell[c=4]{c}{}&&&&parityless\\
        $\otimes$ & ${A^+}$ &${A^-}$ & ${B^+}$& ${B^-}$ & ${E}$  \\
        \hline  
        ${A^+}$ & ${A^+}$ &${A^-}$ & ${B^+}$& ${B^-}$ & ${E}$ \\
        ${A^-}$ &       &${A^+}$ & ${B^-}$& ${B^+}$ & ${E}$ \\
        ${B^+}$ &    & &${A^+}$ &${A^-}$ &${ E}$ \\
        ${B^-}$ &    & & &${A^+}$ & ${E}$ \\
        \hline[dashed]
        ${E}$ &    &    &&    &${ A^+\oplus A^-\oplus B^+\oplus B^-}$  
    \end{tblr}
    \caption{
    Product table for the magnetic point groups $42^\prime2^\prime$, $4m^\prime m^\prime$ and $4/mm^\prime m^\prime$. One should add subscripts $g$ and $u$ for $4/mm^\prime$.
    }
    \label{tab:placeholder}
\end{table}
\subsubsection{$\bar{4}2^\prime m^\prime$}
\begin{table}[H]
 \centering
    \begin{tblr}{c|cccc|[dashed]c}
        &\SetCell[c=4]{c}{}&&&&parityless\\
        $\otimes$ & ${A^+}$ &${A^-}$ & ${B^+}$& ${B^-}$ & ${E}$  \\
        \hline  
        ${A^+}$ & ${A^+}$ &${A^-}$ & ${B^+}$& ${B^-}$ & ${E}$ \\
        ${A^-}$ &       &${A^+}$ & ${B^-}$& ${B^+}$ & ${E}$ \\
        ${B^+}$ &    & &${A^+}$ &${A^-}$ &${ E}$ \\
        ${B^-}$ &    & & &${A^+}$ & ${E}$ \\
        \hline[dashed]
        ${E}$ &    &    &&    &${ A^+\oplus A^-\oplus B^+\oplus B^-}$  
    \end{tblr}
    \caption{
    Product table for the magnetic point group $\bar{4}2^\prime m^\prime$.
    }
    \label{tab:placeholder}
\end{table}

\subsubsection{$32^\prime$, $3m^\prime$, $\bar{3}m^\prime$}
\begin{table}[H]
 \centering
    \begin{tblr}{c|cc|[dashed]c}
        &\SetCell[c=2]{c}{}&&parityless\\
        $\otimes$ & ${A^+}$ &${ A^-}$ & ${E}$  \\
        \hline  
        ${A^+}$ & ${A^+}$ & ${A^-}$ & ${E}$ \\
        ${A^-}$ &     &${A^+}$ & ${E}$ \\
        \hline[dashed]
        ${E}$ &    &    & ${A^+\oplus A^-\oplus E}$  
    \end{tblr}
    \caption{
    Product table for the magnetic point groups $32^\prime$, $3m^\prime$, and $\bar{3}m^\prime$. One should add subscripts $g$ and $u$ for $\bar{3}m^\prime$.
    }
    \label{tab:placeholder}
\end{table}

\onecolumngrid
\subsubsection{$62^\prime2^\prime$, $6m^\prime m^\prime$, $6/m m^\prime m^\prime$}
\begin{table}[H]
 \centering
    \begin{tblr}{c|cccc|[dashed]cc}
        &\SetCell[c=4]{c}{}&&&&\SetCell[c=2]{c}{parityless}&\\
 $ \otimes$ & ${A^+}$ & ${A^-}$ & ${B^+}$ & ${B^-}$ & ${E_1}$ & ${E_2}$ \\
        \hline  
        ${A^+}$ & ${A^+}$  & ${A^-}$  & ${B^+}$  & ${B^-}$ & ${E_1}$ & ${E_2}$ \\
        ${A^-}$ &    & ${A^+}$  & ${B^-}$  & ${B^+}$ & ${E_1}$ & ${E_2}$ \\
        ${B^+}$ &    &    & ${A^+}$  & ${A^-}$ & ${E_2}$ & ${E_1}$ \\
        ${B^-}$ &    &    &    & ${A^+}$ & ${E_2}$ & ${E_1}$ \\
        \hline[dashed]
        ${E_1}$ &    &    &    &   & ${A^+\oplus A^-\oplus E_2}$ & ${B^+\oplus B^-\oplus E_1}$ \\
        ${E_2}$ &    &    &    &   &   & ${A^+\oplus A^-\oplus E_2}$ 
    \end{tblr}
    \caption{
    Product table for the magnetic point groups $62^\prime2^\prime$, $6m^\prime m^\prime$ and $6/m m^\prime m^\prime$. One should add subscripts $g$ and $u$ for $6/m m^\prime m^\prime$.
    }
    \label{tab:placeholder}
 \end{table}
\subsubsection{$\bar{4}^\prime3m^\prime$, $4^\prime32^\prime$, $m\bar{3}m^\prime$}
 \begin{table}[H]
 \centering
    \begin{tblr}{c|cccc|[dashed]c}
        &\SetCell[c=4]{c}{}&&&&parityless\\
        $\otimes$ &${A^+}$&${A^-}$&${T^+}$&${T^-}$&${E}$\\
        \hline
         ${A^+}$&${A^+}$&${A^-}$&${T^+}$&${T^-}$&${E}$ \\
         ${A^-}$&&${A^+}$&${T^-}$&${T^+}$&${E}$ \\
         ${T^+}$&&&${A^+\oplus E\oplus 2T^+}$&${A^-\oplus E\oplus 2T^-}$&${T^+\oplus T^-}$\\
         ${T^-}$&&&&${A^+\oplus E\oplus 2T^+}$&${T^+\oplus T^-}$\\
         \hline[dashed]
         ${E}$&&&&&${{A^+\oplus A^-\oplus E}}$ \\
    \end{tblr}
    \caption{
    Product table for the magnetic point groups $\bar{4}^\prime3m^\prime$, $4^\prime32^\prime$ and $m\bar{3}m^\prime$. One should add subscripts $g$ and $u$ for $m\bar{3}m^\prime$.
    }
    \label{last_product_table_r}
 \end{table}

\section{Kronecker products of representations and corepresentations}
\label{app: Kronecker product}

In this Appendix, we review the general correspondence between Kronecker products of representations and those of corepresentations~\cite{Bradley1972}, as summarized in Table~\ref{tab:transform_table} for the complex (co)representations and Table~\ref{tab:transform_table_r} for the real (co)representations.
Using these tables, one can read off the irreducible decomposition of products of corepresentations.

First, we discuss the complex (co)representations.
Here we consider a Kronecker product of representations, $\Gamma_i \otimes \Gamma_j=\oplus_k \Gamma_k$. Each representation ($\Gamma_i, \Gamma_j, \Gamma_k$) is replaced by the corepresentation according to the rule shown in Table~\ref{tab:transform_table}. The rule depends on which type ((a), (b), and (c)) each representation belongs to. After the replacement, one can obtain a Kronecker product of corepresentations. 

For instance, let us consider the Kronecker product between the corepresentations $T$ and $E$ of the magnetic point group $m^\prime3$. 
The corresponding representation of $T$, which is of type (a), is $T$ of the parent point group $23$.
For the type (c) corepresentation $E$, the corresponding representation is ${}^1E,{}^2E$ of $23$.
One can get the irreducible decomposition $T\otimes({}^1E\oplus{}^2E)=2T$ from the product table for representations~\cite{Atkins1970}.
Since the decomposed representation $2T$ is of type (a), one can find a combination,
$\Gamma_i$ of type (a), $\Gamma_j$ of type (c), $\Gamma_k$ of type (a) in  Table~\ref{tab:transform_table}, and obtain $2\Gamma_a=2T\rightarrow 2{D}\Gamma_a=2T$.

\begin{table}[h]
    \centering
    \begin{tblr}{cc|ccc}
    $\otimes$&$\Gamma_j$&$\Gamma_a\rightarrow \mathrm{D}\Gamma_a$&$\Gamma_b\rightarrow \mathrm{D}\Gamma_b$&$\Gamma_c\oplus\Gamma_c^\prime\rightarrow \mathrm{D}\Gamma_c$\\
    $\Gamma_i$&$\Gamma_k$&&&\\
    \hline
    $\Gamma_a$&a&$\Gamma_a\rightarrow \mathrm{D}\Gamma_a$&$\Gamma_a\rightarrow 2\mathrm{D}\Gamma_a$&$2\Gamma_a\rightarrow 2\mathrm{D}\Gamma_a$\\
    $\downarrow$&b&$2\Gamma_b\rightarrow \mathrm{D}\Gamma_b$&$\Gamma_b\rightarrow \mathrm{D}\Gamma_b$&$2\Gamma_b\rightarrow \mathrm{D}\Gamma_b$\\
    $\mathrm{D}\Gamma_a$&c&$\Gamma_c\oplus\Gamma_c^\prime\rightarrow \mathrm{D}\Gamma_c$&$\Gamma_c\oplus\Gamma_c^\prime\rightarrow 2\mathrm{D}\Gamma_c$&$\Gamma_c\oplus\Gamma_c^\prime\rightarrow \mathrm{D}\Gamma_c$\\
    \hline[dashed]
    $\Gamma_b$&a&&$\Gamma_a\rightarrow 4\mathrm{D}\Gamma_a$&$\Gamma_a\rightarrow 2\mathrm{D}\Gamma_a$\\
    $\downarrow$&b&&$\Gamma_b\rightarrow 2\mathrm{D}\Gamma_b$&$\Gamma_b\rightarrow \mathrm{D}\Gamma_b$\\
    $\mathrm{D}\Gamma_b$&c&&$\Gamma_c\oplus\Gamma_c^\prime\rightarrow 4\mathrm{D}\Gamma_c$&$\Gamma_c\oplus\Gamma_c^\prime\rightarrow 2\mathrm{D}\Gamma_c$\\
    \hline[dashed]
    $\Gamma_c\oplus\Gamma_c^\prime$&a&&&$2\Gamma_a\rightarrow 2\mathrm{D}\Gamma_a$\\
    $\downarrow$&b&&&$2\Gamma_b\rightarrow \mathrm{D}\Gamma_b$\\
    $\mathrm{D}\Gamma_c$&c&&&$\Gamma_c\oplus\Gamma_c^\prime\rightarrow \mathrm{D}\Gamma_c$\\
    \end{tblr}
    \caption{
    Correspondence between Kronecker products of complex representations, $\Gamma_i\otimes\Gamma_j=\oplus_k\Gamma_k$, and those of complex corepresentations.
    Depending on which type ((a), (b), or (c)) each representation ($\Gamma_i$, $\Gamma_j$, $\Gamma_k$) belongs to, one can replace the representation to a corepresentation according to the rule shown in the table
    to obtain the Kronecker product of corepresentations. The first column and the first row show the replacement rules for $\Gamma_i$ and $\Gamma_j$, respectively. The second column shows the type of $\Gamma_k$. From the third to fifth column, we show the decomposition rule for $\Gamma_k$. Since the Kronecker product is commutable, the decomposition rule is symmetric against $\Gamma_i \leftrightarrow \Gamma_j$ (and hence the lower left parts of the table are omitted).}
    \label{tab:transform_table}
\end{table}

Next, we turn our attention to the real (co)representations.
Table~\ref{tab:transform_table_r} summarizes the differences between the Kronecker product of real representations and those of real corepresentations, specifically focusing on the assignment of the $A$-parity in the decomposition of the Kronecker product. For brevity, the basis sets of the irreducible corepresentation with parity $\{\ket{\psi_i}+\ket{\phi_i}\}$ and $\{\ket{\psi_i}-\ket{\phi_i}\}$ are abbreviated as $\{\ket{+_i}\}$ and $\{\ket{-_i}\}$, respectively.

First, consider the direct product of two even parity corepresentations. The product space of the two basis, namely $\{\ket{+_i}_1\}$ and $\{\ket{+_i}_2\}$, is given by $\{\ket{+_i}_1\}\otimes\{\ket{+_i}_2\}=\{\ket{+_i+_j}_{12}\}$. Since every basis $\ket{+_i+_j}$ satisfies $A\ket{+_i+_j}=\ket{+_i+_j}$, the Kronecker product should decompose into exclusively even parity or parityless corepresentations. An analogous argument holds for different parity combinations.

Next, consider the product of a corepresentation with even parity and a parityless corepresentation. By choosing an appropriate basis for the second subspace, the combined product space can be expressed as $\{\ket{+_i}_1\}\otimes(\{\ket{\psi_i}_2\}\oplus\{\ket{\phi_i}_2\})=\{\ket{+_i\psi_j}_{12}\}\oplus\{\ket{+_i\phi_j}_{12}\}=\{\ket{+_i+_j}_{12}\}\oplus\{\ket{+_i-_j}_{12}\}$. By applying the argument we discussed in Sec.~\ref{3c} to the even basis $\ket{+_i+_j}$ and odd basis $\ket{+_i-_j}$, we conclude that the product should include pairs of even and odd parity corepresentations or parityless corepresentations.
A similar argument shows that the product of two parityless corepresentations also yields parity pairs or parityless corepresentations.

By systematically applying these parity rules alongside dimensional constraints, we ultimately determine the general correspondence between the Kronecker products of real representations and real corepresentations, as presented in Table~\ref{tab:transform_table_r}.

\begin{table}[h]
    \centering
    \begin{tblr}{cc|cc}
    $\otimes$&$\Gamma_j$&$\Gamma_p\rightarrow \mathrm{D}\Gamma_p^\pm$&$\Gamma_l\rightarrow \mathrm{D}\Gamma_l$\\
    $\Gamma_i$&$\Gamma_k$&&\\
    \hline
    $\Gamma_p$&p&$\Gamma_p\rightarrow \mathrm{D}\Gamma_p^\pm$&$2\Gamma_p\rightarrow \mathrm{D}\Gamma_p^+\oplus\mathrm{D}\Gamma_p^-$\\
    $\rightarrow\mathrm{D}\Gamma_p^\pm$&l&$\Gamma_l\rightarrow \mathrm{D}\Gamma_l$&$\Gamma_l\rightarrow \mathrm{D}\Gamma_l$\\
    \hline[dashed]
    $\Gamma_l$&p&&$2\Gamma_p\rightarrow \mathrm{D}\Gamma_p^+\oplus\mathrm{D}\Gamma_p^-$\\
    $\rightarrow\mathrm{D}\Gamma_l$&l&&$\Gamma_l\rightarrow \mathrm{D}\Gamma_l$\\
    \end{tblr}
    \caption{
    Correspondence between Kronecker products of real representations, $\Gamma_i\otimes\Gamma_j=\oplus_k\Gamma_k$, and those of real corepresentations.
    Depending on whether each real corepresentation ($\mathrm{D}\Gamma_i$, $\mathrm{D}\Gamma_j$, $\mathrm{D}\Gamma_k$) possesses a well-defined parity (denoted by p) or is  parityless (denoted by l), one can replace the representation to a corepresentation according to the rule shown in the table
    to obtain the Kronecker product of corepresentations. The column definitions are the same as in Table~\ref{tab:transform_table}.}
    \label{tab:transform_table_r}
\end{table}

\section{Microscopic description of the coefficient of the Lifshitz invariant
}

\label{Appendix.B}
Here, we present a microscopic description of the coefficient of the Lifshitz invariant, namely $d^{\lambda}_{\alpha\beta}$, where $\lambda$ represents the spatial direction and $\alpha$ and $\beta$ denote the band indices.
We assume that the order parameter of the equilibrium state is spatially homogeneous.
By denoting the phase introduced by an external magnetic field or the Aharonov-Bohm flux as $\phi$, the coefficient $d^{\lambda}_{\alpha\beta}$ is shown to be a linear combination of a real odd function of $\phi$ and an imaginary even function of $\phi$ (where we do not consider the PH symmetry).
When the PH symmetry is taken into account, the coefficient $d^{\lambda}_{\alpha\beta}$ is determined to be either a real odd function of $\phi$ or an imaginary even function of $\phi$ unambiguously.

We begin with the definition of the coefficient $d^{\lambda}_{\alpha\beta}$,
\begin{equation}
d^{\lambda}_{\alpha\beta}(\phi) := \frac{k_{\text{B}}T}{2}\sum_{n}\sum_{\bm{k}}g_{\alpha\beta}(-\mathrm{i}\omega_{n};\bm{k},\phi)\nabla_{\lambda}g_{\alpha\beta}(\mathrm{i}\omega_{n};\bm{k},\phi),
\end{equation}
where $T$ is the temperature, $\mathrm{i}\omega_{n}=(2n+1)\pi k_{\text{B}}T$ is the fermionic Matsubara frequency, $\nabla_{\lambda} = \partial/\partial k_{\lambda}$, and $g(\mathrm{i}\omega_{n};\bm{k},\phi)=(\mathrm{i}\omega_{n}-\xi(\bm{k},\phi))^{-1}$ is the Green's function for the normal state, whose Hamiltonian is given by $\xi(\bm{k},\phi)$.
Since the normal state Hamiltonian $\xi(\bm{k},\phi)$ is hermitian, i.e., $\xi_{\alpha\beta}(\bm{k},\phi)=\xi_{\beta\alpha}^{*}(\bm{k},\phi)$, the Green's function satisfies $g_{\beta\alpha}(\mathrm{i}\omega_{n};\bm{k},\phi)=g^{*}_{\alpha\beta}(-\mathrm{i}\omega_{n};\bm{k},\phi)$:
\begin{align}
g_{\beta\alpha}(\mathrm{i}\omega_{n};\bm{k},\phi) &= [(\mathrm{i}\omega_{n} - \xi(\bm{k},\phi))^{-1}]_{\beta\alpha} \notag \\
&= [(\mathrm{i}\omega_{n} - \xi^{*}(\bm{k},\phi))^{-1}]_{\alpha\beta} \notag \\
&= [(-\mathrm{i}\omega_{n} - \xi(\bm{k},\phi))^{-1}]^{*}_{\alpha\beta} \notag \\
&= g^{*}_{\alpha\beta}(-\mathrm{i}\omega_{n};\bm{k},\phi).
\label{Hermiticity}
\end{align}
The normal state Hamiltonian $\xi(\bm{k},\phi)$ should also satisfy $\xi_{\alpha\beta}(\bm{k},\phi)=\xi_{\beta\alpha}(-\bm{k},-\phi)$ (flipping the direction from site $\alpha$ to site $\beta$), 
leading to the relation
$g_{\beta\alpha}(\mathrm{i}\omega_{n};\bm{k},\phi)=g_{\alpha\beta}(\mathrm{i}\omega_{n};-\bm{k},-\phi)$:
\begin{align}
g_{\beta\alpha}(\mathrm{i}\omega_{n};\bm{k},\phi) &= [(\mathrm{i}\omega_{n} - \xi(\bm{k},\phi))^{-1}]_{\beta\alpha} \notag \\
&= [(\mathrm{i}\omega_{n} - \xi(-\bm{k},-\phi))^{-1}]_{\alpha\beta} \notag \\
&= g_{\alpha\beta}(\mathrm{i}\omega_{n};-\bm{k},-\phi).
\label{Flipping}
\end{align}
These two equations represent general properties of the coefficient of the Lifshitz invariant, which arise from the sublattice structure, and are independent of the system's details.
By using Eqs.~(\ref{Hermiticity},\ref{Flipping}), we obtain the following two properties of the coefficient of the Lifshitz invariant:
\begin{align}
\left[ d^{\lambda}_{\alpha\beta}(\phi) \right]^{*} &= d^{\lambda}_{\beta\alpha}(\phi), \label{Conjugate} \\
d^{\lambda}_{\alpha\beta}(-\phi) &= -d^{\lambda}_{\beta\alpha}(\phi). \label{Phase_reversing}
\end{align}
Equation~(\ref{Conjugate}) can be derived as
\begin{align}
\left[ d^{\lambda}_{\alpha\beta}(\phi) \right]^{*} &= \frac{k_{\text{B}}T}{2}\sum_{n}\sum_{\bm{k}}g^{*}_{\alpha\beta}(-\mathrm{i}\omega_{n};\bm{k},\phi)\nabla_{\lambda}g^{*}_{\alpha\beta}(\mathrm{i}\omega_{n};\bm{k},\phi) \notag \\
&= \frac{k_{\text{B}}T}{2}\sum_{n}\sum_{\bm{k}}g_{\beta\alpha}(\mathrm{i}\omega_{n};\bm{k},\phi)\nabla_{\lambda}g_{\beta\alpha}(-\mathrm{i}\omega_{n};\bm{k},\phi) \notag \\
&= \frac{k_{\text{B}}T}{2}\sum_{n}\sum_{\bm{k}}g_{\beta\alpha}(-\mathrm{i}\omega_{n};\bm{k},\phi)\nabla_{\lambda}g_{\beta\alpha}(\mathrm{i}\omega_{n};\bm{k},\phi) \notag \\
&= d^{\lambda}_{\beta\alpha}(\phi).
\end{align}
Equation~(\ref{Phase_reversing}) can be shown as
\begin{align}
d^{\lambda}_{\alpha\beta}(-\phi) &= \frac{k_{\text{B}}T}{2}\sum_{n}\sum_{\bm{k}}g_{\alpha\beta}(-\mathrm{i}\omega_{n};\bm{k},-\phi)\nabla_{\lambda}g_{\alpha\beta}(\mathrm{i}\omega_{n};\bm{k},-\phi) \notag \\
&= \frac{k_{\text{B}}T}{2}\sum_{n}\sum_{\bm{k}}g_{\beta\alpha}(-\mathrm{i}\omega_{n};-\bm{k},\phi)\nabla_{\lambda}g_{\beta\alpha}(\mathrm{i}\omega_{n};\bm{k},-\phi) \notag \\
&= -\frac{k_{\text{B}}T}{2}\sum_{n}\sum_{\bm{k}}g_{\beta\alpha}(-\mathrm{i}\omega_{n};-\bm{k},\phi)(-\nabla_{\lambda})g_{\beta\alpha}(\mathrm{i}\omega_{n};-\bm{k},-\phi) \notag \\
&= -\frac{k_{\text{B}}T}{2}\sum_{n}\sum_{\bm{k}}g_{\beta\alpha}(-\mathrm{i}\omega_{n};\bm{k},\phi)\nabla_{\lambda}g_{\beta\alpha}(\mathrm{i}\omega_{n};\bm{k},-\phi) \notag \\
&= -d^{\lambda}_{\beta\alpha}(\phi).
\end{align}
Then we have
\begin{equation}
\left[ d^{\lambda}_{\alpha\beta}(\phi) \right]^{*} = -d^{\lambda}_{\alpha\beta}(-\phi).
\label{Conclusive}
\end{equation}
Let us write $d^{\lambda}_{\alpha\beta}(\phi)$ as a sum of odd and even functions of $\phi$:
\begin{equation}
d^{\lambda}_{\alpha\beta}(\phi) = A_{\text{R}}(\phi) + \mathrm{i}A_{\text{I}}(\phi) + B_{\text{R}}(\phi) + \mathrm{i}B_{\text{I}}(\phi),
\end{equation}
where $A_{\text{R}}(\phi)$ and $A_{\text{I}}(\phi)$ are real odd functions of $\phi$, and $B_{\text{R}}(\phi)$ and $B_{\text{I}}(\phi)$ are real even functions of $\phi$, respectively.
Equation~(\ref{Conclusive}) gives us the condition $A_{\text{I}}(\phi)=B_{\text{R}}(\phi)=0$, suggesting that the coefficient of the Lifshitz invariant has to take the following form:
\begin{equation}
d^{\lambda}_{\alpha\beta}(\phi) = A_{\text{R}}(\phi) + \mathrm{i}B_{\text{I}}(\phi).
\end{equation}
This equation is a general consequence of the properties of the normal state Hamiltonian.

Now we consider the PH symmetry.
Since the PH operation changes $\phi$ as $\phi\to-\phi$, we have the dichotomy of the coefficient of the Lifshitz invariant as follows:
\begin{align}
d^{\lambda}_{\alpha\beta}(\phi) &\xrightarrow{\text{PH}} d^{\lambda}_{\alpha\beta}(-\phi) = d^{\lambda}_{\alpha\beta}(\phi) = \mathrm{i}B_{\text{I}}(\phi) \quad (\text{Type I}), \\
d^{\lambda}_{\alpha\beta}(\phi) &\xrightarrow{\text{PH}} d^{\lambda}_{\alpha\beta}(-\phi) = -d^{\lambda}_{\alpha\beta}(\phi) = -A_{\text{R}}(\phi) \quad (\text{Type II}).
\end{align} 
This dichotomy tells us that it is impossible to excite the Leggett and Higgs modes in the linear response regime simultaneously.
We can also conclude that only the linear Leggett mode excitation is allowed when the system preserves the time-reversal symmetry ($\phi=0$), which is consistent with the previous results~\cite{Nagashima2024}.

\section{
Absence of the linear Higgs response in the isotropic square lattice model
}
\label{Appendix.C}
Here we show that the coefficient of the Lifshitz invariant vanishes in the isotropic square lattice case (see Fig.~\ref{Fig_lattices}(c)) due to the $C_{4}$ rotational symmetry.
A similar argument can be applied to the Kagome lattice case with 12 sites in a unit cell (see Fig.~\ref{Fig_lattices}(f)).

Let us consider the isotropic square lattice with $t_{1}=t_{2}$ (Fig.~\ref{Fig_lattices}(c)).
In this case, the system preserves the $C_{4}$ rotational symmetry.
By applying the $C_{4}$ rotation to the system, the coordinate transforms as
\begin{equation}
\mqty[ x \\ y] \mapsto \mqty[ -y \\ x] \quad \Leftrightarrow \quad \mqty[ k_{x} \\ k_{y}] \mapsto \mqty[ -k_{y} \\ k_{x}].
\end{equation}
Following this transformation, we can show that $d^{x}_{\alpha\beta}(\phi) = d^{x}_{\alpha\beta}(-\phi)$:
\begin{align}
d^{x}_{\alpha\beta}(\phi) &= \frac{k_{\text{B}}T}{2}\sum_{n}\sum_{\bm{k}}g_{\alpha\beta}(-\mathrm{i}\omega_{n}; k_{x},k_{y},\phi)\partial_{x}g_{\alpha\beta}(\mathrm{i}\omega_{n}; k_{x},k_{y},\phi) \notag \\
&= \frac{k_{\text{B}}T}{2}\sum_{n}\sum_{\bm{k}}g_{\alpha\beta}(-\mathrm{i}\omega_{n}; -k_{y},k_{x},\phi)(-\partial_{y})g_{\alpha\beta}(\mathrm{i}\omega_{n}; -k_{y},k_{x},\phi) \quad (C_{4}) \notag \\
&= \frac{k_{\text{B}}T}{2}\sum_{n}\sum_{\bm{k}}g_{\alpha\beta}(-\mathrm{i}\omega_{n}; k_{y},k_{x},-\phi)\partial_{y}g_{\alpha\beta}(\mathrm{i}\omega_{n}; k_{y},k_{x},-\phi) \quad ((k_{y},\phi)\to(-k_{y},-\phi)) \notag \\
&= \frac{k_{\text{B}}T}{2}\sum_{n}\sum_{\bm{k}}g_{\alpha\beta}(-\mathrm{i}\omega_{n}; k_{x},k_{y},-\phi)\partial_{x}g_{\alpha\beta}(\mathrm{i}\omega_{n}; k_{x},k_{y},-\phi) \quad ((k_{y},k_{x})\to(k_{x},k_{y})) \notag \\
&= d^{x}_{\alpha\beta}(-\phi).
\end{align}
Since $d^{x}_{\alpha\beta}(\phi)$ is an odd function of $\phi$, we can conclude that $d^{x}_{\alpha\beta}(\phi)=0$.
By considering the symmetry operation $C_{4}^{-1}$, the coordinate transforms as 
\begin{equation}
\mqty[ x \\ y ] \mapsto \mqty[ y \\ -x ] \quad \Leftrightarrow \quad \mqty[ k_{x} \\ k_{y} ] \mapsto \mqty[ k_{y} \\ -k_{y}],
\end{equation}
from which we arrive at $d^{y}_{\alpha\beta}(\phi)=0$ as well:
\begin{align}
d^{y}_{\alpha\beta}(\phi) &= \frac{k_{\text{B}}T}{2}\sum_{n}\sum_{\bm{k}}g_{\alpha\beta}(-\mathrm{i}\omega_{n}; k_{x},k_{y},\phi)\partial_{y}g_{\alpha\beta}(\mathrm{i}\omega_{n}; k_{x},k_{y},\phi) \notag \\
&= \frac{k_{\text{B}}T}{2}\sum_{n}\sum_{\bm{k}}g_{\alpha\beta}(-\mathrm{i}\omega_{n}; k_{y},-k_{x},\phi)(-\partial_{x})g_{\alpha\beta}(\mathrm{i}\omega_{n}; k_{y},-k_{x},\phi) \quad (C_{4}^{-1}) \notag \\
&= \frac{k_{\text{B}}T}{2}\sum_{n}\sum_{\bm{k}}g_{\alpha\beta}(-\mathrm{i}\omega_{n}; k_{y},k_{x},-\phi)\partial_{x}g_{\alpha\beta}(\mathrm{i}\omega_{n}; k_{y},k_{x},-\phi) \quad ((k_{x},\phi)\to(-k_{x},-\phi)) \notag \\
&= \frac{k_{\text{B}}T}{2}\sum_{n}\sum_{\bm{k}}g_{\alpha\beta}(-\mathrm{i}\omega_{n}; k_{x},k_{y},-\phi)\partial_{y}g_{\alpha\beta}(\mathrm{i}\omega_{n}; k_{x},k_{y},-\phi) \quad ((k_{y},k_{x})\to(k_{x},k_{y})) \notag \\
&= d^{y}_{\alpha\beta}(-\phi).
\end{align}
When the square lattice model is anisotropic and lacks the $C_{4}$ rotational symmetry, the above argument cannot be applied, and $d^{\lambda}_{\alpha\beta}(\phi)$ can take a nonzero value.

A similar argument can be applied to the Kagome lattice model.
When the Kagome lattice model has the $C_{6}$ rotational symmetry, $d^{\lambda}_{\alpha\beta}(\phi)$ (or the form suitable for the lattice, such as $d^{x}_{\alpha\beta}(\phi)\pm \sqrt{3}d^{y}_{\alpha\beta}(\phi)$) will be zero.
That is why the Higgs mode peak does not appear in the linear response spectrum of the Kagome lattice model with 12 sites per unit cell and uniform amplitude hopping.

\twocolumngrid
\bibliography{main.bib}

\end{document}